\documentclass[AMA,STIX1COL]{WileyNJD-v2}

\articletype{Article Type}%

\usepackage{amssymb}

\usepackage{pgfplotstable}
\usepackage{pgfplots}
\usepackage{tikz}

\usepackage{amsmath}
\usepackage{amssymb}
\usepackage{epsfig}
\usepackage{hyperref}
\usepackage{stackengine}
\usepackage{booktabs}

\usepackage{color}

\newcommand{\eps}{\varepsilon}
\newcommand{\beq}{\begin{equation}}
\newcommand{\eeq}{\end{equation}}
\newcommand{\bea}{\begin{eqnarray}}
\newcommand{\eea}{\end{eqnarray}}

\newcommand{\half}{\mbox{$\frac{1}{2}$}}
\newcommand{\dif}{{\rm d}}
\newcommand{\dd}{{\rm d}}
\newcommand{\dx}{{\rm d}x}
\newcommand{\dA}{{\rm d}A}
\newcommand{\Eint}{{\cal E}_{int}}
\newcommand{\strf}{\Sigma} % this is the "stress function"
\revised{July 2022}
\newcommand{\added}[1]{{\color{blue}#1}}
\newcommand{\deleted}[1]{}

\begin{document}

\title{Efficient formulation of a two-noded \color{black}{geometrically exact} curved beam element }

\author[1]{Martin  Hor\'{a}k}

\author[2]{Emma La Malfa Ribolla*}

\author[1]{Milan Jir\'{a}sek}

\authormark{HOR\'{A}K \textsc{ET AL}}

\address[1]{\orgdiv{Faculty of Civil Engineering}, \orgname{Czech
Technical University in Prague}, \orgaddress{\state{Prague}, \country{Czech Republic}}}

\address[2]{\orgdiv{Department of Engineering}, \orgname{University of Palermo}, \orgaddress{\state{Palermo}, \country{Italy}}}

\corres{*Emma La Malfa Ribolla, Viale delle Scienze, Ed. 8. 90128 Palermo (PA). \email{emma.lamalfaribolla@unipa.it}}

\abstract[Abstract]{The paper extends the formulation of a 2D geometrically exact beam element  proposed in our previous paper \cite{JLMRH21} to curved elastic beams. This formulation is based on equilibrium equations in their integrated form, combined with the kinematic relations and sectional equations that link the internal forces to sectional deformation variables. The resulting first-order differential equations are approximated by the finite difference scheme and the boundary value problem is converted to an initial value problem using the shooting method. The paper develops the theoretical framework based on the Navier-Bernoulli hypothesis, with a possible extension to shear-flexible beams. Numerical procedures for the evaluation of equivalent nodal forces and of the element tangent stiffness are presented in detail. Unlike standard finite element formulations, the present approach can increase accuracy by refining the integration scheme on the element level while the number of global degrees of freedom is kept constant. The efficiency and accuracy of the developed scheme are documented by seven examples that cover circular and parabolic arches, a spiral-shaped beam, and a spring-like beam with a zig-zag centerline. The proposed formulation does not exhibit any locking. No excessive stiffness is observed for coarse computational grids and the distribution of internal forces is not polluted by any oscillations. It is also shown that a cross effect in the relations between internal forces and deformation variables arises, i.e., the bending moment affects axial stretching and the normal force affects the curvature. This coupling is theoretically explained in the appendix.}


\maketitle

%\footnotetext{\textbf{Abbreviations:} ANA, anti-nuclear antibodies; APC, antigen-presenting cells; IRF, interferon regulatory factor}

\section{Introduction}\label{sec:intro}

Curved beam models are widely used in various engineering applications involving for example arches, pipes and bridge slab structures in civil engineering or lattice metamaterials, tires, and rings in mechanical engineering. Often, these structures are discretized into multiple straight elements that represent the curved geometry only approximately. The majority of contributions to the development of curved beam elements are based on the finite element (FE) method  within the small-displacement theory and address shear and membrane locking caused by the coupling between bending and stretching \cite{Noor1981, Stolarski1982, Stolarski1983, Babu1986, Zhang2003, Ferradi2021}. On the other hand, design and manufacturing of flexible as well as soft metamaterials have opened an area for the application of beams made of polymers or soft materials, for which large displacements
and rotation may arise.

The key contributions to large-deformation analysis of thin beams are the theory of Reissner \cite{reissner1972, reissner1973}, based on the extension of Timoshenko’s assumption to finite deformations, and the FE formulation developed by Simo  and his coworkers \cite{simo1985, simo1986, simo1986pI, simo1986pII}. The geometrically exact beam theory is still attracting researchers, with recent developments in the isogeometric approach (e.g., \cite{borkovic2022geometrically, Marino2019}) or in computational procedures related to the parameterisation of rotations using the rotation vector (\cite{IBRAHIMBEGOVIC1997, MAGISANO2020}). In a recent work \cite{JLMRH21} we have presented a numerical formulation for two-dimensional straight beams under large sectional rotations based on the shooting method: The boundary value problem is converted into an initial value problem handled by a finite difference scheme, and the estimated values used in artificially added initial conditions are iteratively adjusted until the boundary conditions on the opposite beam end are satisfied. On the global (structural) level, the governing equations are assembled in the same way as for a standard FE beam element with six degrees of freedom. It has been demonstrated that the advantage of this approach is a dramatic
reduction of the number of global degrees
of freedom, since the
accuracy of the numerical approximation  can be conveniently increased by refining the integration scheme on the element level instead of introducing additional global unknowns. 

The present paper extends the geometrically exact formulation presented by Jir\'{a}sek et al. \cite{JLMRH21} to curved beams undergoing large displacements and rotations. The theoretical framework is developed in Section~\ref{sec:2} and the corresponding numerical procedures are described in Section~\ref{sec:3}. 
The efficiency and accuracy of the proposed method is illustrated in Section~\ref{sec:numexamples} by five examples,
which treat circular and parabolic arches and a logarithmic spiral. 

\section{Beam with initial curvature}\label{sec:2}

%\subsection{Basic assumptions}\label{sec:basic}

\subsection{Kinematic description}\label{sec:kinematic}

\begin{figure}[h!]
    \centering
    \begin{tabular}{ccc}
    (a) & (b) & (c) 
    \\
    \includegraphics[width=70mm]{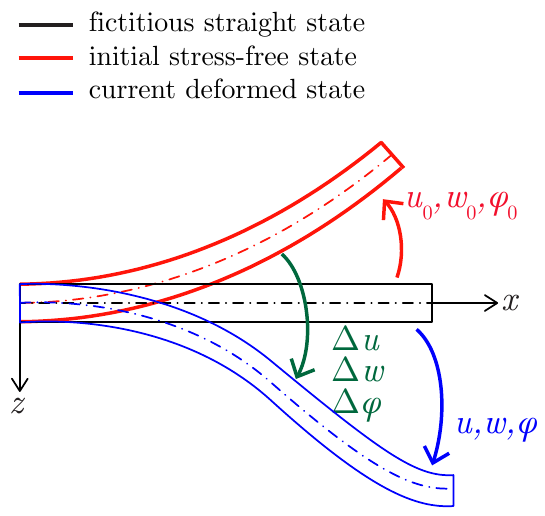}
    &  \includegraphics[width=0.2\linewidth]{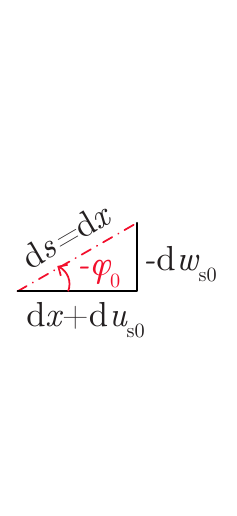}
    & \includegraphics[width=0.2\linewidth]{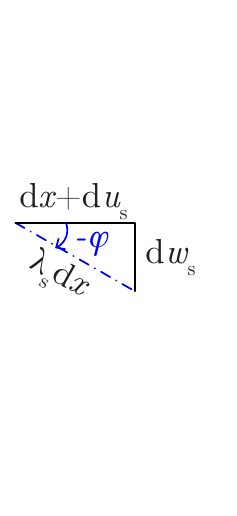}
    \end{tabular}
\caption{(a) Coordinate system $xz$ aligned with the left beam end, fictitious straight configuration (black), initial stress-free configuration (red), and current deformed configuration (blue); (b) infinitesimal triangle with hypotenuse on the centerline in the initial stress-free configuration;
    (c) infinitesimal triangle with hypotenuse on the centerline in the current deformed configuration}
\label{fig1}
\end{figure}

The approach developed by Jir\'{a}sek et al. \cite{JLMRH21} will now be extended to initially curved beams. Consider that the centerline of the undeformed beam is a planar curve of length $L$. An auxiliary curvilinear coordinate
$s$ is defined as the arc length measured along
the undeformed centerline, with $s\in[0,L]$.
For a given shape of the centerline, it is possible to specify function $\varphi_0(s)$
which describes the initial rotation of
section $s$ with respect to the left end
section (this means that $\varphi_0(0)=0)$.
Counterclockwise rotations are considered as positive.
Displacement components $u$ and $w$ will be expressed with respect to 
a local Cartesian coordinate system $xz$ that is attached to the left end section and follows its motion; see Fig.~\ref{fig1}a.

The transformation of the beam from the undeformed state to the current one
is decomposed into 
\begin{itemize}
\item (A) a rigid-body translation and rotation that follows the motion of the left end section, and 
\item (B) a true deformation during which the left end section remains fixed.
\end{itemize}
Phase A can be handled by simple geometrical transformations and does not affect the end forces expressed with respect to the moving local coordinate system. It is therefore
sufficient to focus on phase B and consider the left end of the beam as fixed.  
The black straight beam in Fig.~\ref{fig1}a corresponds to a fictitious state used as a reference, while the initial stress-free configuration of the beam is plotted in red and the current deformed configuration in blue.

%\clearpage

Fictitious displacements that describe the mapping of an arbitrary point $(x,z)$
from its reference position in the fictitious straight state to its actual position in the initial stress-free
configuration are denoted as $u_0$ and $w_0$ and considered as functions of $x$ and $z$. Based on the standard assumption that all cross sections remain planar, 
these functions can be expressed as
\bea\label{equ0} 
u_0(x,z) &=& u_{s0}(x)+z\sin\varphi_0(x) \\
\label{eqw0} 
w_0(x,z) &=& w_{s0}(x)-z(1-\cos\varphi_0(x))
\eea 
where $u_{s0}$ and $w_{s0}$ are functions
of $x$ that correspond to displacements of points
on the centerline. Let us recall that $\varphi_0$ is the already introduced
function defining the angle between a generic section 
and the left end section in the stress-free state. This angle at the same time corresponds
to the rotation from the fictitious straight state to the initial stress-free state.
It is also worth noting that the 
coordinate $s$ measured along the arc 
of the curved centerline in the stress-free
state has the same value as the coordinate
$x$ measured along the Cartesian axis in the
fictitious straight state, and in the subsequent derivations we will write all
functions as dependent on $x$ instead of $s$. 
In the same spirit, Cartesian coordinate $z$
in the fictitious straight state corresponds
to the coordinate that would be measured
in the initial stress-free state in the 
direction normal to the curved centerline.

Functions $u_{s0}$, $w_{s0}$ and  $\varphi_0$
are supposed to be specified in advance but they are not independent. The length of the centerline 
must remain unaffected by the transformation from the fictitious straight reference configuration to the initial
stress-free configuration, which is described by conditions
\bea \label{mj127}
u_{s0}' &=& \cos\varphi_0-1 \\
w_{s0}' &=& -\sin\varphi_0
\label{mj128}
\eea
in which the prime denotes the derivative with respect to $x$. Relations (\ref{mj127})--(\ref{mj128}) follow from the geometry of the infinitesimal triangle depicted in Fig.~\ref{fig1}b. By definition, cross sections in the initial state are perpendicular to the initial centerline, and so the angle
by which the tangent to the centerline deviates
from the $x$ axis is the same as the angle $\varphi_0$ by which the section deviates from the $z$ axis.

In principle it is possible to specify only function
$\varphi_0$ and construct $u_{s0}$ and $w_{s0}$ by integrating equations (\ref{mj127})--(\ref{mj128})
with initial conditions $u_{s0}(0)=0$ and $w_{s0}(0)=0$.
Alternatively, one can characterize the initial shape
by specifying functions $u_{s0}$ and $w_{s0}$, making
sure that they satisfy the constraint $(1+u_{s0}')^2+w_{s0}'^2=1$, and then evaluate
function $\varphi_0=-\arcsin w_{s0}'$.  
However, one should bear in mind that the description based on a given function $\varphi_0$ combined with relations  (\ref{mj127})--(\ref{mj128}) is fully general and permits arbitrary values of the ``rotation'' $\varphi_0$, while the inverse
relation $\varphi_0=-\arcsin w_{s0}'$ is valid
only as long as $\varphi_0\in[-\pi/2,\pi/2]$.
This is always true in the vicinity of the left end section but not necessarily along
the whole beam. For instance, if we consider a circular arch of radius~$R$,
the initial shape is described by
\bea \label{circle1}
\varphi_0(x) &=& \frac{x}{R} \\
\label{circle2}
u_{s0}(x) &=& R\sin\frac{x}{R} - x \\
\label{circle3}
w_{s0}(x) &=& R\left(\cos\frac{x}{R}-1\right)
\eea 
from which 
\bea 
u'_{s0}(x) &=& \cos\frac{x}{R} - 1 \\
w'_{s0}(x) &=& -\sin\frac{x}{R}
\eea
Functions $u_{s0}$ and $w_{s0}$ defined in (\ref{circle2})--(\ref{circle3}) satisfy the constraint $(1+u_{s0}')^2+w_{s0}'^2=1$, but the
corresponding function $\varphi_0$ can be
evaluated as $-\arcsin w_{s0}'$ only for 
$x\le \pi R/2$. If the centerline length
$L$ exceeds $\pi R/2$, one needs to modify 
the formula for the inversion of (\ref{mj128}) accordingly and
set $\varphi_0=\pi+\arcsin w_{s0}'$
for $x\in[\pi R/2,3\pi R/2]$ etc.

The ``total'' centerline displacements, $u_s$ and $w_s$, and the ``total'' rotation, $\varphi$,
are understood as changes between the fictitious straight state
and the final deformed configuration. 
They differ from the initial values by increments
\bea \label{incr1}
\Delta u_s &=& u_s-u_{s0} \\
\Delta w_s &=& w_s-w_{s0} \\
\Delta \varphi &=& \varphi-\varphi_0 
\label{incr3}
\eea 
that represent the actual displacements and rotation.
Similar relations can be written for the  displacements
of an arbitrary point, $u$ and $w$, for which the subscript ''s'' is dropped. In analogy to (\ref{equ0})--(\ref{eqw0}), the displacements
of an arbitrary point can be expressed in terms of the
centerline displacements and sectional rotation as
\bea\label{equ} 
u(x,z) &=& u_{s}(x)+z\sin\varphi(x) \\
\label{eqw} 
w(x,z) &=& w_{s}(x)-z(1-\cos\varphi(x))
\eea 

\subsection{Deformation variables}

Let us proceed to the evaluation of strains. The sections
are assumed to remain perpendicular to the centerline, and so the
shear strains are neglected and it is sufficient to
characterize the stretching in the direction parallel
to the centerline.
Consider a fiber segment parallel to the centerline and located
at section $x$ and at height $z$, which is in the
reference straight configuration represented by
an infinitesimal interval of length $\dd x$. 
In the deformed configuration, this segment is
mapped on the hypotenuse of an orthogonal triangle with catheti
$\dd x + \dd u$ and $\dd w$ and its length is
\beq 
\overline{\dx}=\sqrt{(\dd x + \dd u)^2+\dd w^2} = \dx\sqrt{(1+u')^2+w'^2}
\eeq 
Making use of (\ref{equ0})--(\ref{eqw0}), we express the ratio between the current and reference fiber lengths as
\bea\nonumber 
\frac{\overline{\dx}}{\dx} &=& \sqrt{(1+u')^2+w'^2}
=\sqrt{(1+u_s'+z\varphi'\cos\varphi )^2+(w_s'-z\varphi'\sin\varphi)^2}=
\\
&=& \lambda_s+z\varphi'
\eea 
in which 
\beq\label{eq:lambdas} 
\lambda_s=\sqrt{(1+u_s')^2+w_s'^2}
\eeq 
is the centerline stretch.

When evaluating the actual physical stretch of a generic fiber,
we need to take into account that the length of the 
considered fiber segment in the stress-free configuration
is not $\dx$ but 
\beq 
\dx_0=(\lambda_{s0}+z\varphi_0') \,\dx =(1+z\varphi_0')\, \dx
\eeq 
Here we have taken into account that
\beq 
\lambda_{s0}=\sqrt{(1+u_{s0}')^2+w_{s0}'^2} = 1
\eeq 
due to the constraint on the functions that define the
initial shape (which follows from the assumption that
coordinate $x$ corresponds to the arc length measured
along the centerline).
Based on the above, the stretch of a generic fiber is given by
\beq\label{eq:lambda} 
\lambda= \frac{\overline{\dx}}{\dx_0} = 
\frac{\lambda_s+z\varphi'}{1+z\varphi_0'}
= 
\frac{\lambda_s+z\kappa}{1+z\kappa_0}
\eeq 
where $\kappa_0=\varphi_0'$ is the initial curvature and $\kappa=\varphi'$ is the curvature
in the deformed state.
The fact that for $z=0$ we obtain $\lambda=\lambda_s$
confirms that $\lambda_s$ defined in (\ref{eq:lambdas})
is the stretch evaluated at the centerline.

If we imagine the beam first as straight and then deform it to 
what we later consider as the initial configuration, the stretch
would be given by $1+z\kappa_0$. Measured with respect to the 
fictitious straight shape, the final stretch would be $\lambda_s+z\kappa$. However, since the fiber is actually stress-free
in the initial (but curved) configuration, the effective stretch
that it feels is the ratio $(\lambda_s+z\kappa)/(1+z\kappa_0)$.

\subsection{Internal forces}

The next step is to set up the expression for the strain energy of the deformed beam and identify the internal forces as the variables work-conjugate with the sectional deformation variables $\lambda_s$ and $\kappa$. Since the stress state 
at each material point is considered as uniaxial,
it is sufficient to specify the strain energy density, $\Eint$, as function of the stretch, $\lambda$,
and then integrate it over the volume.
The density is understood here as strain energy
per unit initial volume, i.e., volume in the
stress-free but initially curved state of the beam.
When integrating over the volume of the 
initially curved beam, we have to take into account that an infinitesimal
 segment of length $\dx$ measured along the centerline contains fibers whose length $(1+z\kappa_0)\,\dx$ varies as function of their
 distance from the centerline. The strain energy is
therefore written  as
\beq \label{eq:strainenergy}
E_{int} = \int_0^L \int_A (1+z\kappa_0)
\Eint(\lambda) \,\dA\,\dx
\eeq 
and its variation is
\beq 
\delta E_{int} = \int_0^L \int_A (1+z\kappa_0)
\frac{\partial\Eint}{\partial\lambda}\delta\lambda \,\dA\,\dx
\eeq
where $A$ is the cross section and 
\beq
\delta\lambda = \frac{\delta\lambda_s+z\delta\kappa}{1+z\kappa_0}
\eeq 
is the variation of stretch.
The derivative $\partial\Eint/\partial\lambda=\sigma$ is the
stress work-conjugate with the Biot strain, because differentiation
with respect to $\eps=\lambda-1$ gives the same result as differentiation with respect to $\lambda$. This stress can be interpreted as the
normal component of the back-rotated first Piola-Kirchoff stress (where the back rotation
eliminates the effects of the cross-sectional rotation). 

The expression for the strain energy variation can be further converted into
\bea \label{eq:deltaEint}
\delta E_{int} &=& \int_0^L \int_A (1+z\kappa_0)\,
\sigma\,\frac{\delta\lambda_s+z\delta\kappa}{1+z\kappa_0} \,\dA\,\dx
= \int_0^L \int_A 
\sigma\,(\delta\lambda_s+z\delta\kappa) \,\dA\,\dx =\int_0^L (N\,\delta\lambda_s+M\,\delta\kappa) \,\dx
\eea
in which 
\bea \label{eq:25}
N&=&\int_A \sigma\,\dA \\
\label{eq:26}
M&=&\int_A z\sigma\,\dA
\eea 
are identified as the normal force and bending moment, playing the role of stress resultants
that are work-conjugate with the centerline stretch
and curvature.

If the expression for strain energy density
is taken as quadratic, given by
\beq\label{ss27}
\Eint(\lambda) = \half E(\lambda-1)^2
\eeq 
where $E$ is the Young modulus, the resulting
stress-strain relation
\beq \label{eq:stress-strain}
\sigma = \frac{\dd\Eint(\lambda)}{\dd\lambda}=
E(\lambda-1)=E\eps
\eeq 
is linear in terms of
the Biot strain $\eps=\lambda-1$ and the (back-rotated) first Piola-Kirchhoff stress $\sigma$.
However,
since the distribution of stretches across the height of the section 
\added{of an initially curved beam}
is not linear
but is given by the rational function (\ref{eq:lambda}),
the distribution of stresses across the section height is nonlinear
even for \color{black}{this type of} a linear stress-strain law,
and it is given by
\beq\label{eq:29} 
\sigma =  E\left(\frac{\lambda_s+z\kappa}{1+z\kappa_0}-1\right) = E\,\frac{\lambda_s-1+z(\kappa-\kappa_0)}{1+z\kappa_0} = E\,\frac{\eps_s+z\Delta\kappa}{1+z\kappa_0}
\eeq
where 
\begin{equation}\label{eq:epss}
    \eps_s=\lambda_s-1
\end{equation}
is the strain at the centerline and
\begin{equation}\label{eq:dkap}
    \Delta\kappa=\kappa-\kappa_0
\end{equation}
is the difference between the curvatures in the
deformed state and in the stress-free state.

Substituting the stress expressed from (\ref{eq:29}) into the integral formulae for internal forces, eqs.~(\ref{eq:25})--(\ref{eq:26}), we obtain
%Despite the nonlinearity, the
%integrals over the section can be evaluated analytically:
\bea \nonumber
N&=&\int_A \sigma\,\dA = E\int_A \frac{\eps_s+z\Delta\kappa}{1+z\kappa_0}\,\dA
= E\int_A \frac{\dA}{1+z\kappa_0}\,\eps_s + E\int_A \frac{z\,\dA}{1+z\kappa_0}\,\Delta\kappa\\
\\
\nonumber
M&=&\int_A z\sigma\,\dA = E\int_A \frac{z\eps_s+z^2\Delta\kappa}{1+z\kappa_0}\,\dA
= E\int_A \frac{z\,\dA}{1+z\kappa_0}\,\eps_s + E\int_A \frac{z^2\dA}{1+z\kappa_0}\,\Delta\kappa\\
\label{mj75}
\eea 
The resulting \color{black}{sectional equations, i.e.,} relations between internal forces
and \color{black}{sectional} deformation variables\color{black}{,} can be written as
\bea \label{eq:n}
N&=& EA_{\kappa_0}\eps_s + ES_{\kappa_0}\Delta\kappa
\\
\label{eq:m}
M&=& ES_{\kappa_0}\eps_s + EI_{\kappa_0}\Delta\kappa
\eea 
where
\bea \label{e:A0}
A_{\kappa_0} &=& \int_A \frac{\dA}{1+z\kappa_0} \\
S_{\kappa_0} &=& \int_A \frac{z\,\dA}{1+z\kappa_0} \\
\label{e:I0}
I_{\kappa_0} &=& \int_A \frac{z^2\dA}{1+z\kappa_0} 
\eea 
are modified sectional characteristics, dependent on the initial curvature $\kappa_0$.
They remain constant during the simulation
and thus can be computed in advance, so their evaluation does not represent any problem even for general sections. 
\color{black}{Moreover, by manipulating the integrals
it is easy to show that $S_{\kappa_0}=-\kappa_0 I_{\kappa_0}$ and $A_{\kappa_0}=A-\kappa_0 S_{\kappa_0}=A+\kappa_0^2 I_{\kappa_0}$ where
$A$ is the standard sectional area. Therefore,
it is sufficient to evaluate the modified moment
of inertia, $I_{\kappa_0}$, and the modified area and modified static moment are then obtained
in a straightforward way.}

For a rectangular section,
\color{black}{the modified moment of inertia} can be expressed analytically:
\beq
%\nonumber
%A_{\kappa_0} &=& b_s\int_{-h_s/2}^{h_s/2} \frac{\dA}{1+z\kappa_0} = \frac{b_s}{\kappa_0}\ln\frac{2+h_s\kappa_0}{2-h_s\kappa_0}=\\
%&=& b_sh_s\left(1+\frac{h_s^2\kappa_0^2}{12}+\frac{h_s^4\kappa_0^4}{80}+\frac{h_s^6\kappa_0^6}{448} +\ldots\right) \\
%\nonumber
%S_{\kappa_0} &=& b_s\int_{-h_s/2}^{h_s/2} \frac{z\,\dA}{1+z\kappa_0} =\frac{b_s}{\kappa_0^2}\left(h_s\kappa_0-\ln\frac{2+h_s\kappa_0}{2-h_s\kappa_0}\right) =\\
%&=& -b_sh_s^2\left(\frac{h_s\kappa_0}{12}+\frac{h_s^3\kappa_0^3}{80}+\frac{h_s^5\kappa_0^5}{448}+\ldots\right)\\
%\nonumber
I_{\kappa_0} = b_s\int_{-h_s/2}^{h_s/2} \frac{z^2\dA}{1+z\kappa_0} 
= \frac{b_s}{\kappa_0^3}\left(\ln\frac{2+h_s\kappa_0}{2-h_s\kappa_0}-h_s\kappa_0\right) = b_sh_s^3\left(\frac{1}{12}+\frac{h_s^2\kappa_0^2}{80}+\frac{h_s^4\kappa_0^4}{448}+\ldots\right)
\label{eq39}
\eeq 
%Note that $S_{\kappa_0}=-\kappa_0 I_{\kappa_0}$
%and $A_{\kappa_0}=b_sh_s+\kappa_0^2 I_{\kappa_0}$,
%and so it is sufficient to evaluate  $I_{\kappa_0}$.
The dimensionless product $h_s\kappa_0$ is equal to the ratio $h_s/R_0$ where $R_0=1/\kappa_0$ is the initial radius of curvature. If this ratio is 1:10,
the relative difference between the \color{black}{modified
moment of inertia}
$I_{\kappa_0}$ 
%and the standard area $A=b_sh_s$ is less than $10^{-3}$, and for 
\color{black}{and the standard moment of inertia $I=b_sh_s^3/12$ is just $0.1503~\%$.}
The approximation
\beq\label{eq:Ikappa} 
I_{\kappa_0} \approx b_sh_s^3\left(\frac{1}{12}+\frac{h_s^2\kappa_0^2}{80}\right)=\frac{b_sh_s^3}{12}\left(1+0.15\frac{h_s^2}{R_0^2}\right)
\eeq 
is then sufficiently accurate.

\color{black}{Sectional equations (\ref{eq:n})--(\ref{eq:m}) have been derived
using a rigorous procedure that combines (i) the definition of internal forces
as stress resultants (\ref{eq:25})--(\ref{eq:26}), (ii) the kinematic assumption of planar cross sections
remaining planar, which leads to (\ref{eq:lambda}), and (iii) the linear uniaxial stress-strain law (\ref{eq:stress-strain}).
If no further approximations are made, the consistently derived equations 
reflect a certain coupling between axial stretching and bending,
in the sense that the normal force depends not only on the axial strain
but also on the change of curvature, and the bending moment depends
not only on the change of curvature but also on the axial strain.
Of course, this is true for a beam with nonzero initial curvature.
For a straight beam, sectional characteristics $A_{\kappa_0}$ and
$I_{\kappa_0}$ are respectively equal to the standard area $A$ and moment of inertia
$I$, and $S_{\kappa_0}$ vanishes because it corresponds to the
static sectional moment $S=0$. In this case, sectional equations 
(\ref{eq:n})--(\ref{eq:m}) reduce to their simple form
\bea \label{eq:ns}
N&=& EA\,\eps_s 
\\
\label{eq:ms}
M&=&  EI\,\Delta\kappa
\eea 
in which the axial stretching and bending are decoupled.
For curved beams, sectional equations in the decoupled form 
(\ref{eq:ns})--(\ref{eq:ms}) need to be considered as a simplified
version of the consistently derived equations (\ref{eq:n})--(\ref{eq:m}),
which is based on an approximation and introduces a certain error.

It is worth noting that the coupling between axial stretching and bending for 
curved beams is a phenomenon described in the literature but often ignored. 
The consistent sectional equations were, in a different notation, derived for instance 
by Bauer et al.\cite{Bauer2016}; see their equations (72)--(73).
Similar issues arise also in shell theory.
Detailed discussion of the appropriate form of sectional equations
for curved beams is provided in Appendix~\ref{appA}.
}

%%%%%%%%%%%%%%%%%%%%%%%%%%%%%%%%%%%%%%%%%%%%%%%%%%%%%%%%%%% 

\subsection{Equilibrium equations}

The equilibrium equations in their differential form could be derived from the stationarity conditions of the total potential energy functional using the standard variational approach.
This procedure is described in detail in \color{black}{our previous paper}\cite{JLMRH21} and it will not be repeated
here because the resulting equations
\bea \label{eq26}
-(N\,\cos\varphi)'-\left(\frac{M'}{\lambda_s}\sin\varphi\right)' &=& 0
\\
(N\,\sin\varphi)'-\left(\frac{M'}{\lambda_s}\cos\varphi\right)'&=&0
\label{eq27}
\eea 
are exactly the same as equations (26)--(27) in \cite{JLMRH21}. It is important to note that
$\varphi$ needs to be properly understood as the
rotation of the section with respect to the
fictitious straight configuration and not as 
the actual rotation with respect to the initial
curved state, which would be $\Delta\varphi=\varphi-\varphi_0$.
Furthermore, equations (\ref{eq26})--(\ref{eq27}) can be integrated
in closed form using the same approach as in 
\cite{JLMRH21}, which finally leads to 
\bea \label{e149x}
N(x) &=& -X_{ab}\cos\varphi(x) + Z_{ab}\sin\varphi(x) \\
M(x) &=& -M_{ab}+X_{ab} w_s(x) - Z_{ab}(x+u_s(x)) \label{e150x}
\eea 
where $X_{ab}$, $Z_{ab}$ and $M_{ab}$ are integration constants that physically correspond
to the end forces and end moment acting on the
left end of the beam.
Again, $\varphi$, $u_s$ and $w_s$ are the rotation
and displacements with respect to the fictitious straight configuration,
and the components of end forces are
expressed with respect to the local $xz$ coordinate system attached to the left end.
Of course, equations (\ref{e149x})--(\ref{e150x})
could be set up directly as equilibrium conditions 
deduced from a free-body diagram, but it is reassuring
that they can be derived variationally and that
the  internal forces (primarily
defined as work conjugates of the deformation 
variables) have indeed their usual meaning. 

\subsection{Treatment of the governing equations}\label{sec:treat}

The approach used here when setting up the relations between the generalized end forces (i.e., end forces and moments) and the generalized end displacements (i.e., end displacements and rotations) follows the main idea
described in \cite{JLMRH21}. Instead
of approximating the centerline displacement functions by
a linear combination of pre-selected functions
(e.g., polynomials) and enforcing equilibrium in the weak sense, we consider the integrated equilibrium
equations (\ref{e149x})--(\ref{e150x}) and
combine them with the sectional constitutive
equations (\ref{eq:n})--(\ref{eq:m}) and with a set of three first-order differential
equations 
\bea \label{mj142x}
\varphi' &=& \kappa \\
\label{mj143x}
u_s' &=& \lambda_s\cos\varphi-1 \\
w_s' &=& -\lambda_s\sin\varphi
\label{mj144x}
\eea
which link the centerline displacement 
functions $u_s$ and $w_s$ and the sectional rotation function $\varphi$ to the deformation
variables---centerline stretch $\lambda_s$ and curvature $\kappa$.
Equation (\ref{mj142x}) directly follows from the definition of the curvature, specified in the text after equation (\ref{eq:lambda}), while equations  (\ref{mj143x})--(\ref{mj144x}) follow from
the geometry of an orthogonal triangle with
hypotenuse given by a deformed centerline segment
of length $\lambda_s\,\dx$ inclined by $\varphi$
with respect to the horizontal axis; see Fig.~\ref{fig1}c. 

One component of the outlined approach is the evaluation of
the deformation variables from the internal forces,
which is based on the inverted form
of consistent \color{black}{sectional} equations (\ref{eq:n})--(\ref{eq:m}),
\bea \label{mj140z}
\eps_s &=& \frac{I_{\kappa_0}N-S_{\kappa_0} M}{E(A_{\kappa_0}I_{\kappa_0}-S_{\kappa_0}^2)}\\
\Delta\kappa &=&  \frac{-S_{\kappa_0}N+A_{\kappa_0} M}{E(A_{\kappa_0}I_{\kappa_0}-S_{\kappa_0}^2)}
\label{mj141z}
\eea 
%In the special case of a rectangular section,
%one can further exploit the fact that
\color{black}{Exploiting previously mentioned relations}
$S_{\kappa_0}=-\kappa_0 I_{\kappa_0}$
and $A_{\kappa_0}=A+\kappa_0^2 I_{\kappa_0}$,
\color{black}{this can be rewritten as}\footnote{\color{black}{According to equation (\ref{mj140}), the centerline strain is not proportional to the 
standard normal force $N$ but to a certain combination of the normal
force and bending moment, $N+\kappa_0 M$. This may evoke the notion
of ``effective membrane stress resultant'' introduced by Simo and Fox\cite{Simo1989}
in the context of shell analysis.
However, it turns out that our corrective term $\kappa_0 M$ has a different origin.
The relation between the present theory and the framework used by Simo and Fox\cite{Simo1989} is discussed in detail in Appendix~\ref{appA}.}}
\bea \label{mj140}
\eps_s &=& \frac{N+\kappa_0 M}{EA}\\
\Delta\kappa &=& \frac{\kappa_0 N}{EA} + \frac{M}{EI_{\kappa_0}} + \frac{\kappa_0^2M}{EA} = \frac{M}{EI_{\kappa_0}} + \kappa_0\eps_s
\label{mj141}
\eea 
\color{black}{If the consistent sectional equations (\ref{eq:n})--(\ref{eq:m})
are replaced by their simplified form (\ref{eq:ns})--(\ref{eq:ms}),
the inverted relations read
\bea \label{mj140s}
\eps_s &=& \frac{N}{EA}\\
\Delta\kappa &=& \frac{M}{EI} 
\label{mj141s}
\eea 
}

One could now express the internal forces
on the right-hand sides of (\ref{mj140})--(\ref{mj141})
using the integrated equilibrium equations (\ref{e149x})--(\ref{e150x}), and then transform the deformation variables
$\eps_s$ and $\Delta\kappa$ into $\lambda_s=1+\eps_s$ and $\kappa=\kappa_0+\Delta\kappa$ and
substitute
the resulting expressions into the right-hand
sides of (\ref{mj142x})--(\ref{mj144x}).
This would lead to a set of three ordinary
differential equations for functions
$u_s$, $w_s$ and $\varphi$, with all the other
unknown functions eliminated.
However,
for the purpose of numerical
treatment, it is preferred to keep the equations
separate and process them one by one, because
the procedure will be more transparent and 
individual operations will retain a clear physical
meaning. 

\color{black}{
Before we proceed with the details of numerical implementation, let us note
that the proposed approach shares some common features with the technique
developed by Saje~\cite{Saje1991} in a more general context of the Reissner beam
model\cite{reissner1972}, i.e., with shear distortion taken into account. Saje~\cite{Saje1991} started from
a variational formulation based on the Hu-Washizu principle. 
The potential energy was written as the sum of (i) the strain energy
dependent on functions that characterize deformation of individual beam
segments (axial strain, curvature and shear
distortion) and (ii) the load energy dependent on the displacement and
rotation functions and on the generalized displacements at both end sections. Compatibility constraints that link the deformation and
displacement functions were enforced by an additional term that contains two Lagrange multiplier
functions, and the curvature was expressed directly as the derivative of rotation. The resulting functional was then reduced  by 
imposing relations that correspond to some of the stationarity conditions (Euler-Lagrange equations). In particular, 
the axial strain and shear distortion were expressed in terms of the rotation and Lagrange
multipliers, and the Lagrange multiplier functions were expressed
in terms of their values at the left end and prescribed loads.
The reduced functional was dependent on only one function---the rotation, and on eight discrete values---six generalized end displacements and two left-end values of Lagrange multiplier, which
physically correspond to the left-end forces. The functional was then discretized by approximating the rotation function by
a polynomial of degree $M-1$ with $M$ unknown coefficients. 
The discretized model has $M+6$ degrees of freedom, because
the unknown end rotations directly depend on the coefficients of
the polynomial rotation approximation, and thus only four end displacements and two
end forces need to be considered as additional degrees of freedom.
}

\color{black}{
In the present context (of an Euler-Bernoulli beam model), 
the approach of Saje~\cite{Saje1991} would correspond to 
the second-order differential equation for the rotation function
that can be constructed in the following way: Equation (\ref{mj142x})
is differentiated with respect to $x$, $\kappa'$ is replaced by
$\varphi_0''+\Delta\kappa'$, $\Delta\kappa'$ is expressed as $M'/EI$
using the differentiated form of  (\ref{mj141s}), 
$M'$  is then evaluated according to (\ref{e150x}) with the derivatives
$u_s'$ and $w_s'$ expressed by (\ref{mj143x})--(\ref{mj144x}), and finally $\lambda_s$
is written as $1+\eps_s$ where $\eps_s$ is set to $N/EA$ according
to  (\ref{mj140s}) and $N$ is expressed using (\ref{e149x}). 
After a simple rearrangement, the resulting equation reads 
\beq 
EI\varphi'' +\left(1+\frac{Z_{ab}\sin\varphi-X_{ab}\cos\varphi}{EA}\right)\left(X_{ab}\sin\varphi+Z_{ab}\cos\varphi\right)=EI\varphi_0''
\eeq 
Instead of solving this equation in its weak form by constructing 
a polynomial approximation of function $\varphi$, in our approach
we treat the original equations (\ref{e149x})--(\ref{mj144x}) and  (\ref{mj140})--(\ref{mj141}) separately,
replace the first derivatives in (\ref{mj142x})--(\ref{mj144x}) by finite differences, and
handle the problem by marching from the left end to the right end
without the need for solving a coupled set of equations. 
Of course, the left-end forces and moment still have to be determined
iteratively, but the resulting set always consists
of only three nonlinear algebraic equations, no matter how
fine discretization we use along the beam. In contrast to that,
the approach of Saje~\cite{Saje1991} leads to a coupled set of $M+6$ nonlinear
algebraic equations if the rotation function is approximated
by a polynomial of order $M-1$. We also use the fully
consistent inverted sectional equations (\ref{mj140})--(\ref{mj141}) instead of their simplified
form (\ref{mj140s})--(\ref{mj141s}), which would follow
from the formulation adopted by Saje.~\cite{Saje1991} 

Another potential advantage
of our approach is that it can handle (with high precision) cases
when the curvature varies along the beam in a non-smooth way,
which would be quite hard to approximate by a polynomial. 
An example of such a problem will be presented in section~\ref{sec:springexample}. 
}

\section{Numerical procedures}\label{sec:3}

\subsection{Choice of primary unknown functions}

Based on the theoretical description developed in the preceding section, 
a natural choice of the primary unknown functions
would be the centerline displacements 
$u_s$ and $w_s$ and the sectional rotation $\varphi$. However, we should bear in mind that
these kinematic variables describe changes of the current beam state with
respect to the fictitious straight state.
A potential disadvantage of the  approach based on such ``total'' displacements is that
even the initial stress-free shape would be computed numerically with some error,
even though functions $u_{s0}$, $w_{s0}$ and
$\varphi_0$ that characterize the stress-free
state are assumed to be known.
One also needs to take into account that
since the computed ``total'' displacements and rotation
at the right end of the beam, $u_s(L)$,
$w_s(L)$ and $\varphi(L)$, are referred
to the fictitious straight shape, they do
not correspond to the actual displacements 
and rotation of the joint to which the 
right end is attached. The joint
displacements and rotation are in fact
$\Delta u_s(L)$,
$\Delta w_s(L)$ and $\Delta\varphi(L)$.
For instance, if the beam ends do not move at all, the target values that should be obtained by the shooting method (to be described in Section~\ref{sec:shoot})
would not be zero but $u_{s0}(L)$,
$w_{s0}(L)$ and $\varphi_0(L)$.
These corrections would need to be included
in the expressions for the residual used
by the shooting method.

The above considerations motivate an alternative
choice of  displacements and rotation with respect to the stress-free configuration, 
$\Delta u_s$, $\Delta w_s$ and $\Delta \varphi$,
as the primary unknown functions. 
In terms of these ``true'' displacements and rotation, equations (\ref{mj142x})--(\ref{mj144x}) can be rewritten as
\bea \label{mj145}
\Delta\varphi' &=& \Delta\kappa \\
\label{mj146}
 \Delta u_s' &=& (1+\eps_s)\cos(\varphi_0+\Delta\varphi)-\cos\varphi_0 \\
 \Delta w_s' &=& -(1+\eps_s)\sin(\varphi_0+\Delta\varphi)+\sin\varphi_0
 \label{mj147}
\eea
These equations have been obtained by combining
(\ref{mj142x})--(\ref{mj144x}) with (\ref{incr1})--(\ref{incr3}), (\ref{mj127})--(\ref{mj128}) and (\ref{eq:epss})--(\ref{eq:dkap}).
Function $\varphi_0$ is known (it specifies the initial geometry) and the deformation variables $\eps_s$ and $\Delta\kappa$ are directly evaluated from $N$ and $M$ using
(\ref{mj140})--(\ref{mj141}). Of course,
when the internal forces are computed
based on equations (\ref{e149x})--(\ref{e150x}), the displacements and rotation must be substituted in their total form,
i.e., $\Delta u_s$, $\Delta w_s$ and $\Delta \varphi$ must be increased by the known values of
$u_{s0}$, $w_{s0}$ and $\varphi_0$.

\subsection{Shooting method}\label{sec:shoot}

From the mathematical point of view, the problem that we need to solve looks like an initial value problem, because equations (\ref{mj145})--(\ref{mj147}) are first-order
differential equations for $\Delta\varphi$, $\Delta u_s$ and $\Delta w_s$ and the initial
values $\Delta\varphi(0)=0$, $\Delta u_s(0)=0$
and $\Delta w_s(0)=0$ are known (due to the definition of the local coordinate system that remains firmly attached to the left end section). 
However, in order to proceed with the integration,
one also needs to know the left-end forces $X_{ab}$ and $Z_{ab}$ and the left-end moment $M_{ab}$, which are used when expressing the
internal forces according to (\ref{e149x})--(\ref{e150x}). 

In the context of structural analysis,
the beam under consideration is attached to
joints that link it to other beams, and the
joint displacements and rotations play the role
of basic global unknowns. Therefore, a typical
task at the beam element level is to evaluate
the end forces and moments (at both ends) 
generated by prescribed displacements and rotations
of the joints.
Numerical treatment of this task can be based
on a special version of the shooting method,
i.e., of the method that converts a boundary value
problem into an initial value problem with 
an iterative modification of those initial values
that are not known. These values are first
guessed and then repeatedly corrected until the
boundary conditions on the opposite end of the
interval are satisfied with sufficient accuracy. In the present setting,
the left-end forces $X_{ab}$ and $Z_{ab}$ and the left-end moment $M_{ab}$ need to be adjusted until
the numerically computed 
displacements and rotation of the right end 
(with respect to the coordinate system attached to the left end)
become equal to
the target values determined from the prescribed
joint displacements and rotations.
This iterative process at the beam element level is embedded in the global iteration of joint
displacements and rotations leading to satisfaction of joint equilibrium conditions.

To formalize the computational procedure outlined above,
let us introduce the column matrix of the left end generalized forces,
\beq\label{eq:fab}
\boldsymbol{f}_{ab} = \left(\begin{array}{c} X_{ab} \\ Z_{ab} \\ M_{ab} \end{array}\right)
\eeq 
and the column matrix of the right end generalized displacements,
\beq \label{eq:ub}
\boldsymbol{u}_b = \left(\begin{array}{c}  u_b \\  w_b \\ \varphi_b \end{array}\right)
\eeq
Numerical integration along the beam,
starting from zero values of generalized displacements at the left end and using
generalized forces $\boldsymbol{f}_{ab}$,
leads to the values
of generalized displacements at the right end.
This is formally described by the mapping
\beq\label{e241}
 \boldsymbol{g}(\boldsymbol{f}_{ab})=\left(\begin{array}{c}  \Delta u_s(L;X_{ab},Z_{ab},M_{ab}) \\  \Delta w_s(L;X_{ab},Z_{ab},M_{ab}) \\ \Delta\varphi(L;X_{ab},Z_{ab},M_{ab}) \end{array}\right)
\eeq 
where for instance $\Delta u_s(L;X_{ab},Z_{ab},M_{ab})$ means the value
of function $\Delta u_s$ at $x=L$ determined 
with the left end forces set to $X_{ab}$ and $Z_{ab}$ and the left end moment to $M_{ab}$.

For given values of $\boldsymbol{u}_b$, condition
\beq\label{e241w}
 \boldsymbol{g}(\boldsymbol{f}_{ab}) = \boldsymbol{u}_b 
\eeq 
represents a set
of three nonlinear equations for unknowns $\boldsymbol{f}_{ab}$. The solution is found by the
Newton-Raphson method, using the recursive formula
\beq \label{eq:NR}
\boldsymbol{f}_{ab}^{(k+1)}=\boldsymbol{f}_{ab}^{(k)}+\boldsymbol{G}^{-1}\left(\boldsymbol{f}_{ab}^{(k)}\right)\left(\boldsymbol{u}_b-\boldsymbol{g}\left(\boldsymbol{f}_{ab}^{(k)}\right)\right), \hskip 10mm k=0,1,2,\ldots
\eeq 
where 
\beq 
\boldsymbol{G} = \frac{\partial \boldsymbol{g}}{\partial \boldsymbol{f}_{ab}}
\eeq 
is the Jacobi matrix of mapping $\boldsymbol{g}$.

Note that
the target values of $\Delta u_s(L)$, $\Delta w_s(L)$ and $\Delta \varphi(L)$
are the true displacements and rotation
of the joint attached to the right end of the beam. Of course, the displacement
components must be taken with respect to the
local beam coordinate system aligned with the left end.
The advantage of the approach based on displacements and rotation with respect to the 
initial shape
is that, for this choice, $\boldsymbol{g}(\boldsymbol{0})=\boldsymbol{0}$
holds exactly, even when the mapping $\boldsymbol{g}$ is evaluated numerically,
and so for zero prescribed displacements of the joints, leading to $\boldsymbol{u}_b=\boldsymbol{0}$, equation (\ref{e241w}) yields
zero end forces, $\boldsymbol{f}_{ab}=\boldsymbol{0}$.
This would not be the case if the numerical 
integration used the ``total displacements''
as primary unknowns, because the initial shape
would not be captured exactly by the numerical
approximation.

\subsection{Algorithms}
\label{sec:algo}

A numerical algorithm for evaluation
of function $\boldsymbol{g}$ will be developed
in Section~\ref{sec:algo1}%--\ref{sec:algo2} 
and the corresponding algorithm
for evaluation of the Jacobi matrix $\boldsymbol{G}$ will be described
in Section~\ref{sec:algo3}. 
Approximation of the governing equations is based on finite difference expressions.
The interval $[0,L]$ is divided into $N$ segments of equal length $h=L/N$, which connect
the grid points $x_i=ih$, $i=0,1,2,\ldots N$.
The approximate values of various quantities
at point $x_i$ will be denoted by subscript $i$. To increase the accuracy, in some cases
it is beneficial to deal with approximate values at the midpoint of segment number $i$,
which will be denoted by subscripts $i-1/2$.
\color{black}{Of course, the coordinates of the midpoints
are given by $x_{i-1/2}=(i-1/2)h$, $i=1,2,\ldots N$.}

\subsubsection{\color{black}{Evaluation of right-end displacements and rotation}}\label{sec:algo1}

For a beam with arbitrary geometry, the initial shape 
is supposed to be described by given functions 
$u_{s0}$, $w_{s0}$ and $\varphi_0$, from which
it is possible to derive the initial
curvature function, $\kappa_0=\varphi_0'$. 
\color{black}{In the special case of a circular arch,
we have $\kappa_0=1/R_0=$ const., where
$R_0$ is the initial radius of curvature
of the centerline. More precisely, 
$\kappa_0=1/R_0$ if the shape of the arch is convex (center of curvature above the arch),
while $\kappa_0=-1/R_0$ should be used if the shape is concave.

Based on the shape and dimensions of the
cross section, one can determine the
modified moment of inertia, $I_{\kappa 0}$,
as function of the initial curvature.
For a rectangular section, the exact expression
(\ref{eq39}) or its approximation (\ref{eq:Ikappa}) can be used. For other sectional shapes, 
an appropriate analytical formula can be derived or,
as the last resort, numerical evaluation
of the integral in (\ref{e:I0}) can be performed.
The algorithm also works with the standard
sectional area, $A$.}

%To clearly show the structure of the algorithm,
%let us first consider the special case of a circular arch of radius $R$ with a rectangular cross section of width $b_s$ and depth $h_s$.
%In this case, the initial curvature $\kappa_0=1/R$ (convex shape) or  $\kappa_0=-1/R$ (concave shape) is constant
%and sectional parameters are constant as well,
%so they do not need to be recomputed in each step. %Let us denote the initial curvature as $\kappa^*_0$, to avoid confusion with the  value of total curvature at $x=0$, which is denoted as $\kappa_0$.

\color{black}{Evaluation of right-end displacements
and rotation  for given left-end
forces and moment and for left-end displacements and rotation set to zero can be described as follows:}

\begin{enumerate}
    %\item 
    %Evaluate sectional characteristics %$A=b_sh_s$, $I=b_sh_s^3/12$, $I_{\kappa 0} = %(1+0.15h_s^2/R^2)\,I$.
    \item
    Set initial values 
    \bea \label{ee61}
    \Delta u_0&=&0\\ \Delta w_0&=&0 \\ \Delta\varphi_0&=&0\label{ee63}
   \\
    M_0&=&-M_{ab}
    \\ 
    \kappa_{0,0} &=& \kappa_0(0)
    \\ \label{ee71}
    \eps_0&=&(-X_{ab}+\kappa_{0,0}M_0)/EA\\
    \label{ee72}
    \Delta\kappa_0&=&\kappa_{0,0}\eps_0+M_0/EI_{\kappa 0}(\kappa_{0,0})
    \eea
    \item For $i=1,2,\ldots N$ evaluate
\bea \label{e59}
\Delta\varphi_{i-1/2} &=& \Delta\varphi_{i-1} + \Delta\kappa_{i-1}{\Delta x}/{2} \\
\label{e60}
\varphi_{i-1/2} &=& \varphi_0(x_{i-1/2})+ \Delta\varphi_{i-1/2}\\
\label{e61}
N_{i-1/2} &=&  -X_{ab}\cos\varphi_{i-1/2} + Z_{ab}\sin\varphi_{i-1/2} \\
\kappa_{0,i-1/2} &=& \kappa_0(x_{i-1/2})
\\
\label{e62}
\eps_{i-1/2} &=& 
(N_{i-1/2}+\kappa_{0,i-1/2}M_{i-1})/EA\\
\Delta u_i &=& \Delta u_{i-1}+\left[(1+\eps_{i-1/2})\cos\varphi_{i-1/2}-\cos\varphi_0(x_{i-1/2})\right] \,\Delta x \label{e63} \\
\Delta w_i &=& \Delta w_{i-1}+\left[\sin\varphi_0(x_{i-1/2})-\left(1+\eps_{i-1/2}\right)\sin\varphi_{i-1/2}\right]\,\Delta x \label{e64} \\
%\label{e64b}
%\tilde{x}_i &=& R\sin(x_i/R) + \Delta u_i
%\\
%\label{e64c}
%w_i &=& R(\cos(x_i/R)-1) + \Delta w_i
%\\
\label{e65}
M_i &=& -M_{ab}+X_{ab}(w_0(x_i)+\Delta w_i)-Z_{ab}(x_i+u_0(x_i)+\Delta u_i)
\\
\kappa_{0,i} &=& \kappa_0(x_i)
\\
\label{e66}
\eps_{i} &=& (N_{i-1/2}+\kappa_{0,i}M_{i})/EA\\
\label{e67}
\Delta\kappa_{i} &=& \kappa_{0,i}\eps_{i} +M_{i}/EI_{\kappa 0}(\kappa_{0,i})\\
\label{e68}
\Delta\varphi_i &=& \Delta\varphi_{i-1/2} + \Delta\kappa_i\,{\Delta x}/{2}\label{e240a}
\eea 
\item The resulting values of right-end displacements and rotation are $u_b=\Delta u_N$, $w_b=\Delta w_N$ and $\varphi_b=\Delta\varphi_N$.
\end{enumerate}

This algorithm has a similar structure to the
procedure developed in \cite{JLMRH21} for a straight beam. It is fully explicit and at the same time second-order accurate because time derivatives
are replaced either by central differences,
or by a sequence of two half-steps that use
a forward difference combined with a backward difference.
For instance, equation (\ref{mj145}) is integrated in two
half-steps described by (\ref{e59}) and (\ref{e68}), the first
one being based on a forward difference and
the second on a backward difference.
For an initially straight beam, the axial strain depends only on the normal force, which can be determined from the sectional rotation without  
knowing the displacements, and the curvature
depends only on the bending moment, which
can be determined from the displacements without
knowing the rotation. In that case, integration
of equation (\ref{mj145})  in two
half-steps described by (\ref{e59}) and (\ref{e68}) corresponds
to the trapezoidal rule, and integration
of equations (\ref{mj146})--(\ref{mj147}) described by (\ref{e63})--(\ref{e64}) corresponds to the midpoint rule. 
The initial curvature introduces a slight coupling
but the effects of the normal force on the change of curvature and of the bending moment on the axial strain remain small.
This is why the accuracy is not substantially compromised 
if 
\begin{itemize}
    \item 
    the
evaluation of the axial strain at midpoint described by (\ref{e62})
is based on the normal force
at midpoint and the bending moment at the beginning
of the segment, and
\item 
the evaluation of the curvature change at the end of the segment described by (\ref{e67}) is based on
the bending moment at the end of the segment combined with the normal force at midpoint;
see (\ref{e66}).
\end{itemize}
Using this scheme, it is sufficient to evaluate
the normal force only at the midpoint
and the bending moment only at the end of the segment. 
One can avoid the evaluation of
the bending moment and displacements at the midpoint
and of the normal force at the end of the segment,
which would involve
additional calculations without a substantial
increase in accuracy. 
\color{black}{Note that $\eps_{i-1/2}$ and $\eps_i$ computed in (\ref{e62}) and (\ref{e66}) are auxiliary variables, which are directly used in (\ref{e63})--(\ref{e64}) and (\ref{e67}) but do not need to be stored. On the other hand,
$\Delta\kappa_i$ evaluated in (\ref{e67}) is used
not only in (\ref{e68}) but also
in the next segment in (\ref{e59}) under the name of $\Delta\kappa_{i-1}$ (with the value of $i$ incremented). In fact, one could
proceed from $\Delta\varphi_{i-1/2}$ to $\Delta\varphi_{i+1/2}$ in one single step of length $\Delta x$ and skip the evaluation
of $\Delta\varphi_i$. A half-step is needed
only in the first segment and the last one,
to obtain $\Delta\varphi_N$ as one of the basic output variables.}

\subsubsection{Jacobi matrix}\label{sec:algo3}

The foregoing algorithm
defines the mapping $\boldsymbol{g}$
of the generalized left-end forces $\boldsymbol{f}_{ab}$ on the generalized right-end displacements $\boldsymbol{g}(\boldsymbol{f}_{ab})$,
which is needed for the evaluation of the 
left-hand side of (\ref{e241w}). For iterative
solution of equations (\ref{e241w}) by the Newton-Raphson method, one also needs the Jacobi matrix of mapping $\boldsymbol{g}$, i.e., the matrix
\beq 
\boldsymbol{G} = \frac{\partial \boldsymbol{g}}{\partial \boldsymbol{f}_{ab}}
\eeq 
which contains derivatives of the right-end
displacements and rotation with respect to the left-end forces and moment.

The entries of the Jacobi matrix are evaluated numerically
using the linearized version of the computational scheme.
Suppose that the input values of left-end forces $X_{ab}$, $Z_{ab}$ and $M_{ab}$ are perturbed by infinitesimal increments
$\dif X_{ab}$, $\dif Z_{ab}$ and $\dif M_{ab}$.
The corresponding infinitesimal changes of displacements and rotations along the beam can be
computed from the
linearized form of equations (\ref{e59})--(\ref{e68}), which reads
\bea \label{e234b}
\dif\varphi_{i-1/2} &=& \dif\varphi_{i-1} + \dif\kappa_{i-1}{\Delta x}/{2} \\
\dif N_{i-1/2} &=&  -\dif X_{ab}\cos\varphi_{i-1/2} + \dif Z_{ab}\sin\varphi_{i-1/2}+X_{ab}\sin\varphi_{i-1/2}\dif\varphi_{i-1/2} + Z_{ab}\cos\varphi_{i-1/2}\dif\varphi_{i-1/2}
\\
\dif\eps_{i-1/2} &=& 
(\dif N_{i-1/2}+\kappa_{0,i-1/2}\dif M_{i-1})/EA
\\
\dif u_i &=& \dif u_{i-1}+\left[\dif\eps_{i-1/2}\cos\varphi_{i-1/2}-(1+\eps_{i-1/2})\sin\varphi_{i-1/2}\dif\varphi_{i-1/2}\right] \,\Delta x\\
\dif w_i &=& \dif  w_{i-1}-\left[\dif\eps_{i-1/2}\sin\varphi_{i-1/2}+\left(1+\eps_{i-1/2}\right)\cos\varphi_{i-1/2}\dif\varphi_{i-1/2}\right]\,\Delta x \\
\dif M_i &=& -\dif M_{ab}+\dif X_{ab}(w_0(x_i)+\Delta w_i)-\dif Z_{ab}(x_i+u_0(x_i)+\Delta u_i)+X_{ab}\dif w_i-Z_{ab}\dif u_i
\\ 
\dif\eps_{i} &=& (\dif N_{i-1/2}+\kappa_{0,i}\dif M_{i})/EA\\
\dif\kappa_{i} &=& \kappa_{0,i}\dif\eps_{i} +\dif M_{i}/EI_{\kappa 0}(\kappa_{0,i})\\
\dif\varphi_i &=& \dif\varphi_{i-1/2} + \dif\kappa_i\Delta x/{2}\label{e240b}
\eea 

The values of $\dif u_0$, $\dif w_0$ and $\dif\varphi_0$ are set to zero, because the zero values of $\Delta u_0$,
$\Delta w_0$ and $\Delta\varphi_0$ are fixed;
see (\ref{ee61})-(\ref{ee63}).
The value of $\dif M_0$ is set to $-\dif M_{ab}$, and
$\dif\kappa_0$ is obtained as
$\kappa_{0,0}\dif\eps_0-\dif M_{ab}/EI_{\kappa 0}(\kappa_{0,0})$ where $\dif\eps_0=-(\dif X_{ab}+\kappa_{0,0}\dif M_{ab})/EA$.

If we set $\dif X_{ab}=1$ and $\dif Z_{ab}=\dif M_{ab}=0$,
the resulting values of $\dif u_N$, $\dif w_N$ and $\dif\varphi_N$ will correspond to the first column of the Jacobi matrix $\boldsymbol{G}$. They are evaluated using the adapted scheme
with $\dif\kappa_0=-\kappa_{0,0}/EA$ and
\bea \label{e234c}
\dif\varphi_{i-1/2} &=& \dif\varphi_{i-1} + \dif\kappa_{i-1}{\Delta x}/{2} \\
\dif N_{i-1/2} &=&  -\cos\varphi_{i-1/2} +X_{ab}\sin\varphi_{i-1/2}\dif\varphi_{i-1/2}+ Z_{ab}\cos\varphi_{i-1/2}\dif\varphi_{i-1/2} \\
\dif\eps_{i-1/2} &=& 
(\dif N_{i-1/2}+\kappa_{0,i-1/2}\dif M_{i-1})/EA\\
\dif u_i &=& \dif u_{i-1}+\left[\dif\eps_{i-1/2}\cos\varphi_{i-1/2}-(1+\eps_{i-1/2})\sin\varphi_{i-1/2}\dif\varphi_{i-1/2}\right] \,\Delta x \\
\dif w_i &=& \dif  w_{i-1}-\left[\dif\eps_{i-1/2}\sin\varphi_{i-1/2}+\left(1+\eps_{i-1/2}\right)\cos\varphi_{i-1/2}\dif\varphi_{i-1/2}\right]\,\Delta x \\
\dif M_i &=& w_0(x_i)+\Delta w_i+X_{ab}\dif w_i-Z_{ab}\dif u_i \\ 
\dif\eps_{i} &=& (\dif N_{i-1/2}+\kappa_{0,i}\dif M_{i})/EA\\
\dif\kappa_{i} &=& \kappa_{0,i}\dif\eps_{i} +\dif M_{i}/EI_{\kappa 0}(\kappa_{0,i})\\
\dif\varphi_i &=& \dif\varphi_{i-1/2} + \dif\kappa_i\Delta x/{2}\label{e240c}
\eea 
The other two columns are obtained in an analogous fashion. In practice, the evaluation
of (\ref{e234c})--(\ref{e240c}) in a loop over $i=1,2,\ldots N$ is performed simultaneously with the evaluation of  (\ref{e59})--(\ref{e68}),
so that various auxiliary values such as
$\cos\varphi_{i-1/2}$, $\sin\varphi_{i-1/2}$ 
or $\eps_{i-1/2}$ can be reused.

\subsection{Transformation to global coordinates}

Suppose that the algorithms described in Section~\ref{sec:algo}  have been implemented
%with initial conditions $\Delta\boldsymbol{u}(0)=\boldsymbol{0}$. 
and equations (\ref{e241w}) can be solved numerically based on the iterative scheme (\ref{eq:NR}).
The prescribed values of right-end displacements $\boldsymbol{u}_b$ on the
right-hand side of (\ref{e241w}) as well as
the resulting left-end forces $\boldsymbol{f}_{ab}$ 
are expressed in a local coordinate system $xz$, with the origin located at the left end of the beam in the deformed configuration 
%(point $\tilde a$ in Fig.~\ref{f:xtilde})
and with the $x$ axis in the direction of the tangent to the
deformed centerline at the left end. Now we need
to link the local components of forces and displacements to the components expressed with respect
to the global coordinate system used for the whole structural model, which
will be denoted by a superscript $G$. 

The initial geometry is described by global coordinates
of the joints connected by the beam, i.e., 
$x_{a}^{G}$ and $z_{a}^{G}$
at the left end and $x_{b}^{G}$ and $z_{b}^{G}$ at the right end,
and also by the beam length measured along the centerline, $L$, and by functions $u_{s0}$, $w_{s0}$, 
$\varphi_0$ and $\kappa_0$
that specify the curved shape of the beam. These functions
are not independent, and in principle it would be sufficient
to specify the curvature, $\kappa_0$,
because the rotation, $\varphi_0$, could be computed by integrating $\kappa_0$ and imposing initial condition
$\varphi_0(0)=0$,
and the functions describing the centerline,  $u_{s0}$ and $w_{s0}$,
could be computed by integrating equations (\ref{mj127})--(\ref{mj128}) and imposing
initial conditions $u_{s0}(0)=0$ and $w_{s0}(0)=0$. Instead of performing these operations
numerically or developing complicated rules for symbolic 
integration, the implementation in OOFEM \cite{patzak2001, patzak2012}
leaves it up to the user
to prepare all needed functions and specify them on input,
making sure that they are consistent.

For the purpose of transformation between local and global
coordinate systems, we need to characterize the angle
$\alpha_{ab}$ by which the local axes are rotated
(clockwise) with respect to the global axes. 
%Fig.~\ref{} shows that this angle can be expressed as
Suppose that $\alpha_{0ab}$ is the value
of this angle in the initial stress-free state,
which can be determined from the given geometrical data. During the deformation process,
the local coordinate system rotates with the
left end joint, and so 
\beq 
\alpha_{ab} = \alpha_{0,ab}-\varphi_{a}^{G}
%=\gamma_{0ab}-\beta_{0ab}-\varphi_{a}^{G}
\eeq 
where 
%$\alpha_{0ab}=\gamma_{0ab}-\beta_{0ab}$ is determined by the initial geometry and 
$\varphi_{a}^{G}$ is the rotation of the left joint (positive counterclockwise).
Another quantity that will play a role in the transformations is the angle by which the
beam chord deviates from the local $x$-axis.
The value of this angle in the initial stress-free state, denoted as $\beta_{0ab}$,
can easily be deduced from the description of the initial beam shape.

As already explained in Section~\ref{sec:kinematic}, the total transformation of the beam can be conceptually decomposed into two parts: (A) rigid-body motion and (B) pure deformation. 
We can imagine that during stage A the beam moves as a rigid body
such that it gets translated by $u_{a}^{G}$ and $w_{a}^{G}$
and then rotated about the left end by $\varphi_{a}^{G}$ counterclockwise. This part of the overall motion does not affect the deformation state
of the beam and has no effect on the end forces and moments, provided that their components
are expressed with respect to a coordinate system that rotates with the beam. 

During stage B, the left end remains fixed while the 
right end is moved to its actual position in the
deformed configuration and the right
end section is rotated by $\varphi_{b}^{G}-\varphi_{a}^{G}$. 
The displacements of the right end experienced
during this second stage and expressed with respect to
the local coordinate system attached to the left end are
\bea\nonumber 
u_b &=& (u_{b}^{G}-u_{a}^{G})\cos\alpha_{ab} + (w_{b}^{G}-w_{a}^{G})\sin\alpha_{ab}+L_{ab}(\cos(\beta_{0ab}+\varphi_{a}^{G})-\cos\beta_{0ab})\\
\label{mj215} \\
\nonumber
w_b &=& -(u_{b}^{G}-u_{a}^{G})\sin\alpha_{ab} + (w_{b}^{G}-w_{a}^{G})\cos\alpha_{ab}+L_{ab}(\sin(\beta_{0ab}+\varphi_{a}^{G})-\sin\beta_{0ab})
\\
\eea 
where $L_{ab}$ is the length of the initial chord,
see (\ref{eq:Lab}),
and the rotation is
\beq \label{mj217}
\varphi_b = \varphi_{b}^{G}-\varphi_{a}^{G}
\eeq 
Therefore, if the global displacements and rotations are prescribed,
they can be transformed into
the local displacements and rotation of the right
end with respect to the left end, which are assembled into the column matrix ${\boldsymbol{u}}_b$.
The corresponding column matrix of left-end forces ${\boldsymbol{f}}_{ab}$,
obtained by solving equations (\ref{e241}) and
formally denoted as ${\boldsymbol{g}}^{-1}({\boldsymbol{u}}_b)$,
has components ${X}_{ab}$, ${Z}_{ab}$ and
${M}_{ab}$. Here, ${M}_{ab}$ is directly
the end moment acting at the left end, while the end forces must be transformed into the global components
\bea 
X_{ab}^{G} &=& {X}_{ab}\cos{\alpha}_{ab} - {Z}_{ab}\sin{\alpha}_{ab} \\
Z_{ab}^{G} &=& {X}_{ab}\sin{\alpha}_{ab} + {Z}_{ab}\cos{\alpha}_{ab} 
\label{mj219}
\eea 
Finally, equilibrium equations written for the whole beam
lead to expressions for the right-end forces 
\bea\label{e:122} 
X_{ba}^{G} &=& -X_{ab}^{G} \\
\label{e:123}
Z_{ba}^{G} &=& -Z_{ab}^{G}
\eea 
and the right-end moment 
\beq \label{e348}
{M}_{ba} = -{M}_{ab} + {X}_{ab}(w_{s0}(L)+ w_b) - {Z}_{ab}(L+u_{s0}(L)+u_b)
\eeq 

\begin{figure}[h!]
\centering
\begin{tabular}{c}
\includegraphics[width=0.3\linewidth]{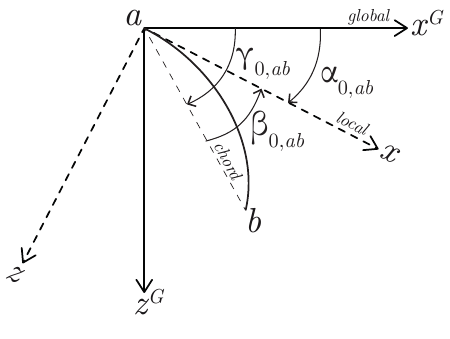}
\end{tabular}
\caption{Local and global coordinate axes in the initial state and definition of angles $\alpha_{0,ab}$, $\beta_{0,ab}$ and $\gamma_{0,ab}$}
\label{FigRS}
\end{figure}

The angle characterizing the initial deviation of the local axes from the global ones can be expressed as
\beq 
\alpha_{0ab}=\gamma_{0ab}-\beta_{0ab}
\eeq 
where 
\beq\label{mj208} 
\gamma_{0ab} = \arctan \frac{z_{b}^{G}-z_{a}^{G}}{x_{b}^{G}-x_{a}^{G}}
\eeq 
is the angle between
the beam chord (straight line connecting the end joints) and
the global axis $x^G$, and $\beta_{0ab}$ is the angle 
between
the beam chord and the tangent to the centerline at the
left end; see Fig.~\ref{FigRS}.
Recall that $(x_{a}^{G},z_{a}^{G})$ and $(x_{b}^{G},z_{b}^{G})$ are the global
coordinates of the end joints $a$ and $b$ in the initial state.

Strictly speaking, formula (\ref{mj208}) gives the correct result
only if $x_{b}^{G}>x_{a}^{G}$ and, to make it general, the rule for evaluation
of $\gamma_{ab}$ would need to be split into several cases. However, for evaluation of the transformation formulae we will
not really need the angle $\gamma_{0ab}$ as such but rather
its sine and cosine, which are conveniently expressed as
\bea\label{e:cosgam} 
\cos\gamma_{0ab} &=& \frac{x_{b}^{G}-x_{a}^{G}}{L_{ab}}
\\ \label{e:singam}
\sin\gamma_{0ab} &=& \frac{z_{b}^{G}-z_{a}^{G}}{L_{ab}}
\eea 
where
\beq \label{eq:Lab}
L_{ab} = \sqrt{(x_{b}^{G}-x_{a}^{G})^2+(z_{b}^{G}-z_{a}^{G})^2}
\eeq
is the initial chord length (distance between the end joints). These equations give the correct values of $\cos\gamma_{0ab}$ and $\sin\gamma_{0ab}$ including the signs for arbitrary inclinations of the chord.

In a similar fashion, the sine and cosine
of angle $\beta_{0ab}$ between the beam chord and the local axis (in the initial state)
can be expressed as
\bea\label{e:cosbet} 
\cos\beta_{0ab} &=& \frac{L+u_{s0}(L)}{L_{ab}}
\\ \label{e:sinbet}
\sin\beta_{0ab} &=& \frac{w_{s0}(L)}{L_{ab}}
\eea 
Consistency requires that
\beq 
(L+u_{s0}(L))^2 + w^2_{s0}(L) = L_{ab}^2
\eeq 
Recall that $L$ is the length of the beam measured along the curved centerline  while the chord length $L_{ab}$ represents the distance between the end joints, both in the initial stress-free state.

\subsection{Matrix formalism and stiffness matrix}

It is convenient to rewrite relations (\ref{mj215})--(\ref{mj219}) in the matrix notation:
\bea\label{mj223}
{\boldsymbol{u}}_b &=& \boldsymbol{T}(\varphi_{a}^{G})(\boldsymbol{u}_{b}^{G}-\boldsymbol{u}_{a}^{G}) + \boldsymbol{l}(\varphi_{a}^{G})\\
\boldsymbol{f}_{ab}^{G} &=& \boldsymbol{T}^T(\varphi_{a}^{G}){\boldsymbol{f}}_{ab}
\label{mj224}
\eea
where
\bea
\boldsymbol{T}(\varphi_{a}^{G}) &=& \left(\begin{array}{ccc}
\cos(\alpha_{0,ab}-\varphi_{a}^{G}) & \sin(\alpha_{0,ab}-\varphi_{a}^{G}) & 0 \\
-\sin(\alpha_{0,ab}-\varphi_{a}^{G}) & \cos(\alpha_{0,ab}-\varphi_{a}^{G}) & 0 \\
0 & 0 & 1
\end{array}\right)
\\
\boldsymbol{l}(\varphi_{a}^{G}) &=& L_{ab}\left(\begin{array}{ccc}
\cos(\beta_{0ab}+\varphi_{a}^{G})-\cos\beta_{0ab} \\ \sin(\beta_{0ab}+\varphi_{a}^{G})-\sin\beta_{0ab} \\ 0 \end{array} \right)
\eea
Recall that ${\boldsymbol{f}}_{ab}$ and ${\boldsymbol{u}}_b$ are the
column matrices of local components defined in (\ref{eq:fab})--(\ref{eq:ub}). The column matrices of global components are
defined as
\beq\label{eq:fabg}
\boldsymbol{u}_{a}^G = \left(\begin{array}{c} u_{a}^G \\ w_{a}^G \\ \varphi_{a}^G \end{array}\right),
\hskip 10mm
\boldsymbol{u}_{b}^G = \left(\begin{array}{c} u_{b}^G \\ w_{b}^G \\ \varphi_{b}^G \end{array}\right),
\hskip 10mm
\boldsymbol{f}_{ab}^G = \left(\begin{array}{c} X_{ab}^G \\ Z_{ab}^G \\ M_{ab} \end{array}\right)
\eeq 

For computing purposes, the coefficients in
matrices $\boldsymbol{T}$ and $\boldsymbol{l}$
can be expanded into
\bea
\cos(\beta_{0ab}+\varphi_{a}^{G})&=&\cos \beta_{0ab} \cdot \cos \varphi_{a}^{G} - \sin \beta_{0ab} \cdot \sin \varphi_{a}^{G}\\
\sin(\beta_{0ab}+\varphi_{a}^{G})&=&\sin \beta_{0ab} \cdot \cos \varphi_{a}^{G} + \cos \beta_{0ab} \cdot \sin \varphi_{a}^{G}
\\
\cos(\alpha_{0,ab}-\varphi_{a}^{G})&=&\cos\alpha_{0,ab}\cdot \cos\varphi_{a}^{G} + \sin\alpha_{0,ab}\cdot \sin \varphi_{a}^{G}
\\
\sin(\alpha_{0,ab}-\varphi_{a}^{G})&=&\sin\alpha_{0,ab}\cdot \cos\varphi_{a}^{G} - \cos\alpha_{0,ab}\cdot \sin \varphi_{a}^{G}
\eea
where $\cos\beta_{0ab}$ and $\sin\beta_{0ab}$
are pre-computed constants given by (\ref{e:cosbet})--(\ref{e:sinbet})
and  
\bea
\cos\alpha_{0,ab}&=& \cos\gamma_{0ab}\cdot\cos\beta_{0ab} + \sin\gamma_{0ab}\cdot\sin\beta_{0ab}
\\
\sin\alpha_{0,ab}&=& \sin\gamma_{0ab}\cdot\cos\beta_{0ab} - \cos\gamma_{0ab}\cdot\sin\beta_{0ab}
\eea 
are pre-computed constants obtained from the
constants given by (\ref{e:cosgam})--(\ref{e:sinbet}).

Combining (\ref{mj223})--(\ref{mj224}) with 
the inverse form of (\ref{e241w}),
\beq \label{e:143}
{\boldsymbol{f}}_{ab} = {\boldsymbol{g}}^{-1}({\boldsymbol{u}}_b)
\eeq 
we get
\beq 
\boldsymbol{f}_{ab}^{G} = \boldsymbol{T}^T(\varphi_{a}^{G})\,{\boldsymbol{g}}^{-1}(\boldsymbol{T}(\varphi_{a}^{G})(\boldsymbol{u}_{b}^{G}-\boldsymbol{u}_{a}^{G}) + \boldsymbol{l}(\varphi_{a}^{G}))
\eeq 
This formula summarizes the process of evaluation of the left-end forces and moment from the end displacements and rotations. 
For better clarity, let us rewrite it in the
simplified form
\beq 
\boldsymbol{f}_{ab}^{G} = \boldsymbol{T}^T{\boldsymbol{g}}^{-1}(\boldsymbol{T}(\boldsymbol{u}_{b}^{G}-\boldsymbol{u}_{a}^{G}) + \boldsymbol{l})
\eeq 
bearing in mind that matrices $\boldsymbol{T}$ and $\boldsymbol{l}$ depend on the left-end rotation, $\varphi_{a}^{G}$, which is at the same time the last component of column matrix $\boldsymbol{u}_{a}^{G}$.

The dependence of  $\boldsymbol{T}$ and $\boldsymbol{l}$ on $\varphi_{a}^{G}$ needs to be taken into account when developing
the relation between infinitesimal
increments of end displacements and end forces,
which will provide an appropriate formula for the tangent element stiffness matrix.

The linearized form of equations (\ref{mj223})--(\ref{mj224}) reads
\bea
{\rm d}{\boldsymbol{u}}_b &=& \boldsymbol{T}({\rm d}\boldsymbol{u}_{b}^{G}-{\rm d}\boldsymbol{u}_{a}^{G}) + \left(\boldsymbol{T}'(\boldsymbol{u}_{b}^{G}-\boldsymbol{u}_{a}^{G}) +\boldsymbol{l}'\right)\,{\rm d}\varphi_{a}^{G}\\
{\rm d}\boldsymbol{f}_{ab}^{G} &=& \boldsymbol{T}^T{\rm d}{\boldsymbol{f}}_{ab} + \boldsymbol{T}'^T{\boldsymbol{f}}_{ab}\,{\rm d}\varphi_{a}^{G}
\eea
where 
\bea
\boldsymbol{T}' &=&\frac{\partial\boldsymbol{T}(\varphi_a^{G})}{\partial\varphi_a^{G}} = \left(\begin{array}{ccc}
\sin(\alpha_{0,ab}-\varphi_{a}^{G}) & -\cos(\alpha_{0,ab}-\varphi_{a}^{G}) & 0 \\
\cos(\alpha_{0,ab}-\varphi_{a}^{G}) & \sin(\alpha_{0,ab}-\varphi_{a}^{G}) & 0 \\
0 & 0 & 0
\end{array}\right) 
\\
\boldsymbol{l}' &=&\frac{\partial\boldsymbol{l}(\varphi_a^{G})}{\partial\varphi_a^{G}} = L_{ab}\left(\begin{array}{ccc}
-\sin(\beta_{0ab}+\varphi_{a}^{G}) \\ \cos(\beta_{0ab}+\varphi_{a}^{G}) \\ 0 \end{array} \right)
\eea
Combining this with the linearized version of (\ref{e:143}),
\beq 
{\rm d}{\boldsymbol{f}}_{ab} = {\boldsymbol{G}}^{-1}({\boldsymbol{u}}_b)\,{\rm d}{\boldsymbol{u}}_b
\eeq 
we get
\bea\nonumber 
{\rm d}\boldsymbol{f}_{ab}(\varphi_a^{G})  &=& \boldsymbol{T}^T{\boldsymbol{G}}^{-1}[\boldsymbol{T}({\rm d}\boldsymbol{u}_{b}^{G}-{\rm d}\boldsymbol{u}_{a}^{G}) + \left[\boldsymbol{T}'(\boldsymbol{u}_{b}^{G}-\boldsymbol{u}_{a}^{G}) +\boldsymbol{l}'\right]\,{\rm d}\varphi_a] + \boldsymbol{T}'^T{\boldsymbol{f}}_{ab}\,{\rm d}\varphi_a = \\
\nonumber
&=& \boldsymbol{T}^T{\boldsymbol{G}}^{-1}\boldsymbol{T}({\rm d}\boldsymbol{u}_{b}^{G}-{\rm d}\boldsymbol{u}_{a}^{G})+
\left[\boldsymbol{T}^T{\boldsymbol{G}}^{-1}\left[\boldsymbol{T}'(\boldsymbol{u}_{b}^{G}-\boldsymbol{u}_{a}^{G}) +\boldsymbol{l}'\right]+ \boldsymbol{T}'^T{\boldsymbol{f}}_{ab}\right]\,{\rm d}\varphi_a\\
\label{eq:stif1}
\eea
where $\boldsymbol{G}^{-1}$ is the inverse of Jacobi matrix $\boldsymbol{G}$ evaluated at
$\boldsymbol{f}_{ab}=\boldsymbol{g}^{-1}(\boldsymbol{u}_b)$ where
$\boldsymbol{u}_b=\boldsymbol{T}(\boldsymbol{u}_{b}^{G}-\boldsymbol{u}_{a}^{G})+\boldsymbol{l}$.

Based on (\ref{eq:stif1}), we can set up the first three rows
of the $6\times 6$ element tangent stiffness matrix (in global coordinates). In view of (\ref{e:122})--(\ref{e:123}),
the fourth row is minus the first row,
and the fifth row is minus the second row,
because ${\rm d}X^G_{ba}=-{\rm d}X^G_{ab}$
and ${\rm d}Z^G_{ba}=-{\rm d}Z^G_{ab}$.
To determine the sixth row,
one needs to linearize the expression for ${M}_{ba}$.
Instead of using (\ref{e348}), it is convenient to set up an equivalent formula written in terms of the global components.
From the moment equilibrium condition written with respect to the
centroid of the right end section in the deformed state,
we get
\beq 
{M}_{ba} = -{M}_{ab} + X_{ab}^{G}(z_{b}^{G}-z_{a}^{G}+w_{b}^{G}-w_{a}^{G}) - Z_{ab}^{G}(x_{b}^{G}-x_{a}^{G}+u_{b}^{G}-u_{a}^{G})
\eeq 
and the infinitesimal increment of the right-end moment 
can be expressed as
\bea\nonumber
{\rm d}{M}_{ba} &=& -{\rm d}{M}_{ab} + (z_{b}^{G}-z_{a}^{G}+w_{b}^{G}-w_{a}^{G}){\rm d}X_{ab}^{G} - (x_{b}^{G}-x_{a}^{G}+u_{b}^{G}-u_{a}^{G}){\rm d}Z_{ab}^{G}+
\\ &&
+ X_{ab}^{G}({\rm d}w_{b}^{G}-{\rm d}w_{a}^{G}) - Z_{ab}^{G}({\rm d}u_{b}^{G}-{\rm d}u_{a}^{G}) 
\eea
Consequently, the sixth row of the stiffness matrix can be constructed as a linear
combination of the first three rows with coefficients
$(z_{b}^{G}-z_{a}^{G}+w_{b}^{G}-w_{a}^{G})$, $-(x_{b}^{G}-x_{a}^{G}+u_{b}^{G}-u_{a}^{G})$ and $-1$, resp.,
added to the row  $(Z_{ab}^{G},-X_{ab}^{G},0,-Z_{ab}^{G},X_{ab}^{G},0)$.
However, this does not even have to be done, since the
stiffness matrix must be symmetric and we already have its sixth
column, except for the last (i.e., diagonal) entry.
So it is sufficient to mirror the entries from the sixth column into
the sixth row and put 
\beq 
k_{66} = (z_{b}^{G}-z_{a}^{G}+w_{b}^{G}-w_{a}^{G})\,k_{16} - (x_{b}^{G}-x_{a}^{G}+u_{b}^{G}-u_{a}^{G})\,k_{26} - k_{36}
\eeq 
on the diagonal.

%%%%%%%%%%%%%%%%%%%%%%%%%%%%%%%%%%%%%%%%%%%%%%%%
%\clearpage
\section{Numerical examples}
\label{sec:numexamples}
%%%%%%%%%%%%%%%%%%%%%%%%%%%%%%%%%%%%%%%%%%%%%%%%

A nonlinear curved beam element based on the proposed approach has been implemented into OOFEM \cite{patzak2001, patzak2012}, an object-oriented finite element code.
Even though the present formulation is not based on an interpolation of displacements and rotations using fixed shape functions multiplied by unknown nodal values, the element
can still be incorporated into a structural model 
in the same way as conventional beam elements. 
For given input values of nodal (joint) displacements and rotations, the corresponding end forces and moments are evaluated and assembled into nodal equilibrium equations.
Also, the corresponding $6\times 6$ element tangent stiffness matrix is constructed and assembled (by standard procedures) into
the structural tangent stiffness matrix used
in global equilibrium iterations.
The accuracy and efficiency 
of the suggested approach will now be demonstrated using \color{black}{seven} examples.

The first \color{black}{four} examples treat arches of a circular shape, for which  
functions $\varphi_0$, $u_{s0}$ and $w_{s0}$ 
that characterize the
initial geometry in terms of the arc-length coordinate
are given  by (\ref{circle1})--(\ref{circle3}).
The \color{black}{fifth} example deals with a parabolic arch,
and it shows that analogous closed-form expressions are in this case
not available; therefore, it is explained how
the description of the initial geometry can
be handled numerically. \color{black}{The sixth} example
analyzes a logarithmic spiral, for which functions
(\ref{spiral1})--(\ref{spiral3}) describing the initial geometry are
derived in Appendix~\ref{appB}.
\color{black}{Finally, the last example shows that the present formulation
can handle not only beams with a smooth centerline, but also cases
in which the centerline exhibits kinks, e.g., a beam consisting of multiple straight segments. Even in this case, the whole beam can be 
treated as one element, and no intermediate nodes hosting additional
global degrees of freedom need to be introduced.
}

%%%%%%%%%%%%%%%%%%%%%%%%%%%%%%%%%%%%
\subsection{Symmetric circular arch, initial stiffness}\label{sec:4.1}
%%%%%%%%%%%%%%%%%%%%%%%%%%%%%%%%%%%%

It is well-known that curved finite elements suffer by excessive stiffness unless the axial displacement interpolation is of a sufficiently high order. 
When low-order axial displacement approximations are used, 
\color{black}{membrane locking makes it difficult for the elements
to bend without stretching, and parasitic oscillating stresses may appear.}
%membrane locking causes a bending-dominated response. 
This locking \color{black}{effect} can be eliminated by using  selectively reduced integration of the stiffness terms \cite{Stolarski1982,Stolarski1983}. 

\begin{figure}[h!]
\centering
\begin{tabular}{c}
\includegraphics[width=0.2\linewidth]{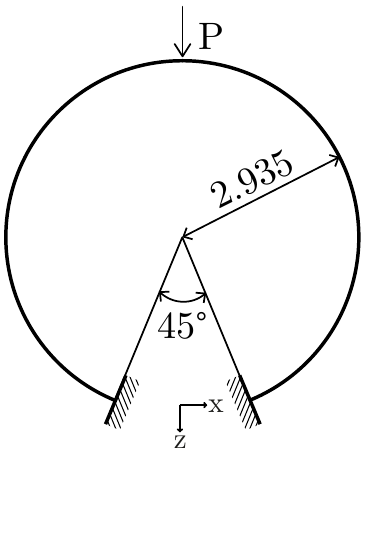}
\end{tabular}
\caption{Symmetric circular arch: geometry}
\label{ring}
\end{figure}

The present formulation exploits the equilibrium equations in their strong form and does not work with a priori chosen shape functions for the kinematic approximation, unlike standard displacement-based finite elements. 
\color{black}{In each integration segment, the values of curvature
and axial strain are completely independent.}
The membrane locking effect is therefore not expected to occur, and accuracy can be increased simply by increasing the number of integration segments.
This  can be verified by analyzing the circular arch in Fig.~\ref{ring}a. The same problem was studied by
Stolarski and Belytschko, who constructed a 
geometrically nonlinear model in the spirit of the shallow-shell approximation of Marguerre \cite{Marguerre1938},
first for a curved Euler-Bernoulli beam \cite{Stolarski1982}
and then for a curved beam with shear distortion
\cite{Stolarski1983}.
%geometrically nonlinear FE solution was presented by  \cite{Stolarski1982}, who also evaluated the performance of various types  of finite elements in the geometrically linear range \cite{Stolarski1983}. 

To facilitate the comparison,
let us take the
geometrical and constitutive parameters from the original papers. They were specified in British
units (inches and psi), which will be omitted here. The cross section is a rectangle of width $b_s=1.2$  and depth $h_s=0.125$, the radius of the
undeformed centerline is $R_0=2.935$, and the elastic modulus is set to $E=1.05 \times 10^7$
(which is the value for aluminum in psi). 
Symmetry is exploited, and so the numbers of elements reported below always correspond to one half of the structure. On the other hand, the applied force $P$ is considered
as the full load on the whole structure. 

The complete load-deflection curve is nonlinear
but locking phenomena can be detected already 
in the geometrically linear range, where the 
structural response is conveniently characterized
by the load-deflection ratio, $P/w$, playing the role of the initial structural
stiffness (i.e., the initial slope of the nonlinear load-deflection diagram). Here, $w$ denotes the vertical deflection
under the loading force, $P$.
Stolarski and Belytschko \cite{Stolarski1982} reported
the analytical value of the $P/w$ 
ratio for the Euler-Bernoulli model to be equal to 471.09, but this value was actually
based on the load acting on one half of the structure
and later \cite{Stolarski1983} it was corrected to 942.2.
However, our analytical calculation based on the force method
indicates that, for a model that neglects the shear distortion and takes into account the axial and flexural deformation, the initial stiffness is 943.73.
In this calculation, and also in the numerical simulations, the interaction between bending and
axial deformation is neglected,   \color{black}{so that the underlying theory remains the same as}
 in the original papers by
Stolarski and Belytschko\cite{Stolarski1982,Stolarski1983}
\color{black}{and the results can be directly compared.}
This means that the
\color{black}{consistent sectional equations}  (\ref{eq:n})--(\ref{eq:m}) are replaced by \color{black}{their simplified form} (\ref{eq:ns})--(\ref{eq:ms}),
\color{black}{which is reflected by appropriate modifications in lines
(\ref{ee71})--(\ref{ee72}),
(\ref{e62})
and (\ref{e66})--(\ref{e67}) of the algorithm (in fact, it is sufficient to set $\kappa_{0,i}=0$, $i=0,1,2 \ldots N$).}
If the effect of shear distortion
is added, 
the stiffness decreases to 941.11,
provided that the
Poisson ratio $\nu=0.3$ and the effective shear area is
equal to the actual area, as assumed by Stolarski and Belytschko.\cite{Stolarski1983} With the shear area reduction
factor set to the standard value for a rectangle, i.e., $5/6$,
the resulting stiffness would be 940.59.

The results presented by Stolarski and Belytschko \cite{Stolarski1982} showed that a simulation on a mesh composed of eight curved finite elements with a linear approximation of the
displacement component in the direction of the chord
and a cubic approximation of the displacement perpendicular to the chord
leads to an excessive structural stiffness
if the integration scheme uses 
4 or 3 Gauss integration points.
The resulting $P/w$ ratios were 1396.6 and
1405.1, respectively, which corresponds to relative
errors of 48~\% and 49~\%.
Reduced integration with 2 integration points
per element 
resulted into $P/w=900.26$, which is by 4.6~\% lower than the
exact value.

Stolarski and Belytschko \cite{Stolarski1983} analyzed the same problem using a slightly adjusted model with
shear distortion taken into account and with the
effective shear area considered as equal to the
actual area. In this case, the theoretical stiffness
is 941.11. The displacement-based element used 
a cubic interpolation for transverse displacements,
quadratic for sectional rotations and linear for
axial displacements. Eight elements with full integration would again
lead to an excessive stiffness (1389.8, 48~\% above the exact value) while a 2-point integration
(which corresponds to reduced integration of the membrane terms and full integration of the shear terms) gives a substantial
improvement (946.6, 0.6~\% above). Various
hybrid and mixed formulations were tested as well but
none of them gave a closer approximation of the
analytical result.

In our analyses, we used discretizations by one or two curved elements. 
It is remarkable that, for the one-element mesh,
no global degrees of freedom were needed and the
problem was solved simply by prescribing the 
end displacements and rotations and evaluating the
corresponding end forces and moments. For the 
two-element mesh, only three global unknowns had
to be introduced.
Accuracy was increased by refining the integration segments and the
full nonlinear model was used, but the initial load-deflection ratio was evaluated
from the stiffness matrix computed in the undeformed
configuration. 

\begin{center}
 \begin{table}[h!]
     \centering
          \begin{tabular}{|llr|llr|}
\toprule
\multicolumn{3}{|c|}{\textbf{ 1 element}} & \multicolumn{3}{c|}{\textbf{ 2 elements}} \\
        \textbf{NIS} & \textbf{$P/w$} & \textbf{error $[\%]$} &\textbf{NIS} & \textbf{$P/w$} & \textbf{error $[\%]$} \\
\midrule
        Exact & 943.73 & &&&\\
         4 & 632.01 &  33.031 & 2 &  633.33 & 32.891 \\
         8 & 839.37 &   11.058 & 4 & 839.82 & 11.010   \\
        16 & 915.23 &  3.020 & 8 & 915.36 & 3.007    \\
       32 & 936.44  & 0.773 & 16 &  936.47 & 0.769  \\
        64 & 941.90 & 0.194 & 32 &  941.91 & 0.193   \\
        128 & 943.27 &  0.049 & 64 &  943.28 & 0.048  \\
        256 & 943.62 &  0.012 & 128 & 943.62 & 0.012   \\
        $\to\infty$ &  943.73 & 0.000 &  $\to\infty$ & 943.73 & 0.000 \\
         \bottomrule
     \end{tabular}
            \caption{Symmetric circular arch: initial load-deflection ratios and the corresponding errors caused by numerical integration.}
 \end{table}
     \label{tabSR}
 \end{center}
The results in Tab.~\ref{tabSR} show that
as the number of integration segments (NIS) is increased, the load-deflection ratio converges to the exact value, 943.73. Convergence is monotonic (from below) and fully regular.
Asymptotically, quadratic convergence is observed,
in the sense that the error is inversely proportional to the square of the NIS
(i.e., when the NIS is doubled, the error is reduced by a factor close to 4). Solutions obtained
on the one-element and two-element meshes are
very similar, provided that the total number of
integration segments is kept
the same. The relative error obtained with 
one element divided into 16 integration segments
(or for two elements, each divided into 8 segments)
is about 3~\% and it is lower than the error
reported by Stolarski and Belytschko \cite{Stolarski1982} for 8 finite
elements with reduced two-point integration,
which required 18 global unknowns.

%%%%%%%%%%%%%%%%%%%%%%%%%%%%%%%%%%%%
\color{black}{\subsection{Simply supported circular arch, internal forces}\label{sec:ex431}}

\begin{figure}[h!]
\centering
\begin{tabular}{c}
\includegraphics[width=0.3\linewidth]{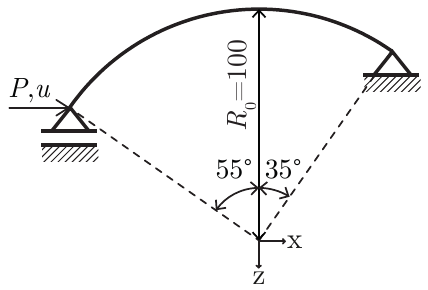}
\end{tabular}
\caption{\color{black}{Simply supported circular arch: geometry and loading} }
\label{fig:newexample1}
\end{figure}

\color{black}{
The second example corresponds to the cylindrical shell strip analyzed by
Bieber et al.~\cite{Bieber2018} in their section 4.3.1. Since the Poisson ratio was considered as zero, the problem is equivalent
to the circular arch schematically
shown in Fig.~\ref{fig:newexample1}. The radius of the undeformed centerline is $R_0=100$
while the sectional depth is just $h_s=0.25$, which corresponds to a very slender arch.
Bieber et al.~\cite{Bieber2018} analyzed the structural response in an extremely large range of support displacements $u$ up to 300,
using 10 isogeometric NURBS elements of 
polynomial orders $p$ varying between 2 and 15. 
They showed that the load-displacement curve becomes reasonably accurate already
for $p=4$ while the distribution of normal force exhibits wild oscillations even for $p=10$, provided that the simplified formulation is used. With an improved formulation, Bieber et al.~\cite{Bieber2018} were able to get
substantially better results using low-order polynomials.
}

\begin{figure}[h!]
\centering
\begin{tabular}{cc}
(a) & (b)
\\
\includegraphics[width=0.45\linewidth]{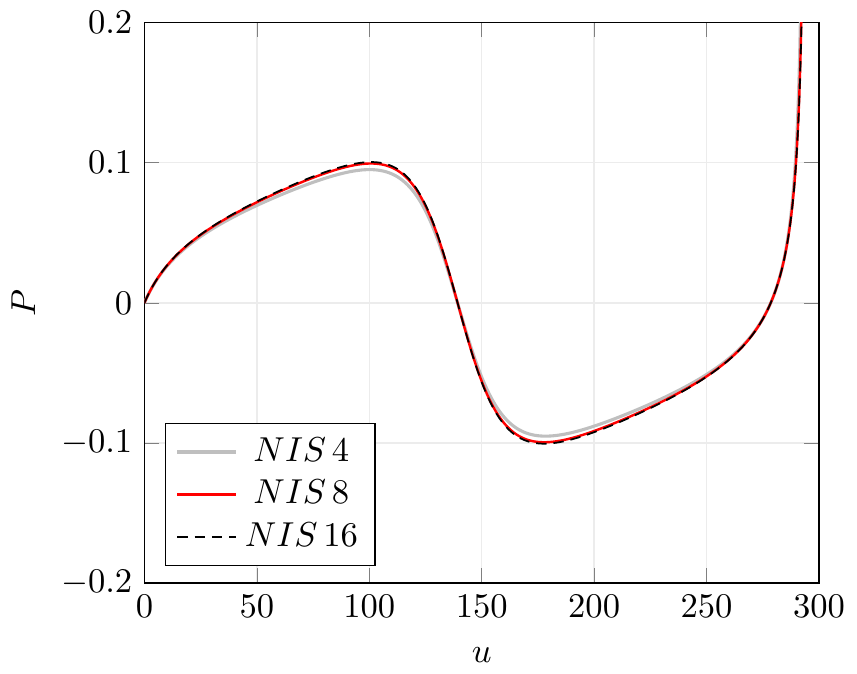}
&
\includegraphics[width=0.45\linewidth]{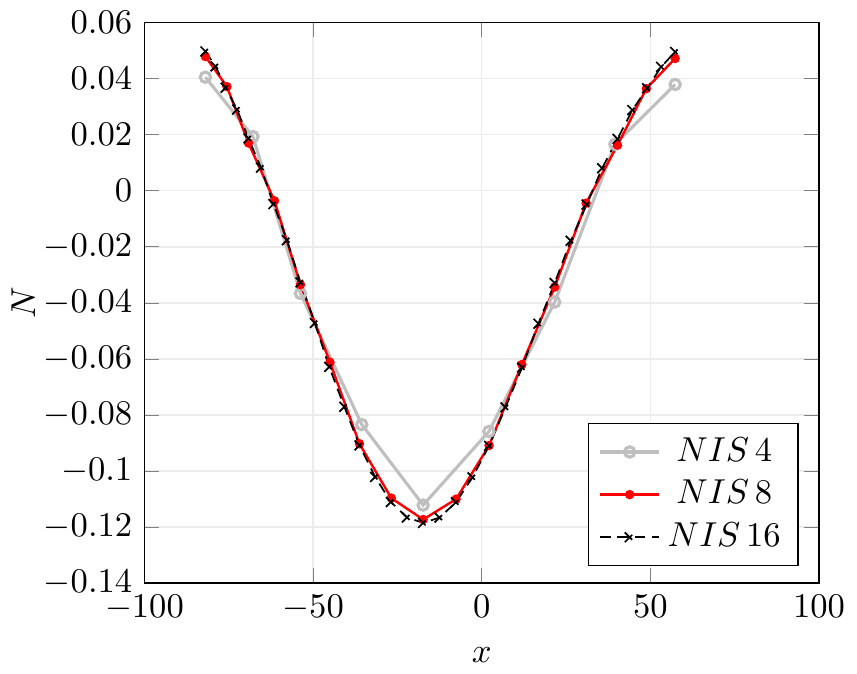}
\end{tabular}
\caption{\color{black}{Simply supported circular arch: (a) load-displacement diagram, (b) distribution of normal force in the state characterized by $u=100$, plotted as function of the global coordinate $x$ marked in Fig.~\ref{fig:newexample1}.} }
\label{fig:newexample2}
\end{figure}

\color{black}{
We can analyze the same problem using again just one element. 
If the simulation is run under displacement control (prescribed displacement $u$ at the left support), only two global unknowns are needed.
To be consistent with 
example 4.3.1 in Bieber et al.~\cite{Bieber2018}, the elastic modulus
is set to $E=21000$, the sectional width 
to $b_s=10$, and simplified sectional equations (\ref{eq:ns})--(\ref{eq:ms}) are adopted.  The load-displacement diagram
plotted in Fig.~\ref{fig:newexample2}a 
is in good agreement\footnote{\color{black}{It is worth noting that we report here
the total force $P$ acting on the beam of width $b_s=10$ and the (total)
normal force $N$, while Bieber et al.~\cite{Bieber2018} reported 
the load multiplier $\lambda$ and the specific normal force $n_{11}=N/b_s$. This is why our normal forces are 10 times larger than those in Figs.~27-29 in Bieber et al.~\cite{Bieber2018}
According to their Fig.~25, the load intensity was set to $0.1\lambda$ on a strip of width 10, and so the load multiplier $\lambda$ should correspond to our load resultant, $P$. However, the values of $\lambda$ in Fig.~26 in Bieber et al.~\cite{Bieber2018} are 10 times smaller than the values of $P$ in our Fig.~\ref{fig:newexample2}a, which
means that Bieber et al.~\cite{Bieber2018} probably plotted the
load intensity $0.1\lambda$ instead of the load multiplier.}} 
with the results
of Bieber et al.~\cite{Bieber2018}, and
reasonable accuracy
is attained already for 8 integration segments. Moreover, the distribution of normal forces shown in Fig.~\ref{fig:newexample2}b 
is not polluted by any oscillations and exhibits fast convergence.
To illustrate the convergence rate, 
the value of the normal force at the right
support in the state at which the left support is displaced by $u=100$ (i.e., near the first peak of the load-displacement diagram)
is taken as a representative example. 
Values of this normal force computed
for different numbers of integration segments (NIS) are reported in Tab.~\ref{tab:newexample}.
Convergence is seen to be quadratic,
i.e., of the same order as convergence in terms of displacements.
}

\begin{table}[h!]
    \centering
    \begin{tabular}{lrr}
    \hline
    {\bf NIS} & {\bf normal force} & {\bf error [\%]}
    \\ \hline
      4   & 0.03793  & 24.66 \\
      8   & 0.04723  & 6.20\\
    16   &  0.04956 & 1.56\\
    32   & 0.05015 & 0.39\\
$\to\infty$   & 0.05035   & 0.00\\
                  \hline
    \end{tabular}
    \caption{\color{black}{Simply supported circular arch: normal force at the right support  in the state when the displacement of the left support is $u=100$, obtained numerically with different values of the number of integration segments (NIS), and the corresponding relative errors.}}
    \label{tab:newexample}
\end{table}

%}
%\\[5mm]

%%%%%%%%%%%%%%%%%%%%%%%%%%%%%%%%%%%%

%%%%%%%%%%%%%%%%%%%%%%%%%%%%%%%%%%%%%%%%%%%%%%%%%%%%
\subsection{Unfolding of a circular cantilever beam}\label{sec:unfolding}
%%%%%%%%%%%%%%%%%%%%%%%%%%%%%%%%%%%%%%%%%%%%%%%%%%%%

In the previous examples, the \color{black}{sectional equations that link the} internal forces to deformation variables were
used in \color{black}{their simplified form (\ref{eq:ns})--(\ref{eq:ms})}, so that the results
could be directly compared to those reported in
\color{black}{the literature}. For linear elastic beams, 
it is usually assumed that the bending moment
is not affected by the axial strain and the normal
force is not affected by the curvature. However,
this can be rigorously proven only if the beam
is initially straight. For beams with an initial
curvature, we have derived \color{black}{consistent} relations
(\ref{eq:n})--(\ref{eq:m}), which contain
modified sectional characteristics and cross-coupling
terms. 
Let us now explain the physical origin of these
modifications and illustrate the resulting effects
on the structural behavior. 
For this purpose, we will consider a curved cantilever which has the initial form of a full
circle, cut at a certain section. One side of the
cut is clamped and the other is
loaded by an increasing bending moment, which
gradually unfolds the beam to a straight configuration and subsequently folds it again to a circular shape with the opposite curvature.

Even though this fictitious test might seem to be the direct analog of the simple example of a straight cantilever folded to a circle by an end moment \cite{JLMRH21,simo1986}, the case of the initially curved beam is actually more involved. For pure bending (i.e., zero normal force) and a rectangular cross section of width $b_s$ and depth $h_s$, \color{black}{the consistent inverted sectional} equations (\ref{mj140})--(\ref{mj141}) reduce to
\bea \label{mj140x}
\eps_s &=& \frac{\kappa_0 M}{EA} = \kappa_0\,\frac{M}{Eb_sh_s}
\\ \label{mj141x}
\Delta\kappa &=&   \left(\frac{1}{EI_{\kappa_0}} + \frac{\kappa^2_0}{EA}\right) M
=\frac{12+h_s^2\kappa_0^2+0.15h_s^4\kappa_0^4}{1+0.15h_s^2\kappa_0^2}\,\frac{M}{Eb_sh_s^3}
\eea
in which $\kappa_0=1/R_0$ is the initial curvature.
Here we have taken into account that, for the rectangular section, the modified moment of inertia
$I_{\kappa_0}$ can be approximated by formula (\ref{eq:Ikappa}) and the standard sectional area is $A=b_sh_s$.

For an initially straight cantilever of length $L=2\pi R_0$, the moment needed to fold the cantilever into a full circle would be $M_0=EI\kappa_0=EI/R_0$, and the radius of that circle would be $R_0$.
On the other hand, if the cantilever in its initial
stress-free shape has the form of a circle of radius $R_0$, application of moment $-M_0$
(the negative sign means that the moment acts clockwise)
leads to the change of curvature
\beq 
\Delta\kappa =   -\frac{12+h_s^2\kappa_0^2+0.15h_s^4\kappa_0^4}{1+0.15h_s^2\kappa_0^2}\,\frac{EI\kappa_0}{Eb_sh_s^3} =
-\frac{12+h_s^2\kappa_0^2+0.15h_s^4\kappa_0^4}{12+1.8\,h_s^2\kappa_0^2}\,\kappa_0
\eeq
and the resulting curvature
\beq 
\kappa = \kappa_0+\Delta\kappa = \frac{0.8\,h_s^2\kappa_0^2-0.15h_s^4\kappa_0^4}{12+1.8\,h_s^2\kappa_0^2}\,\kappa_0
\eeq 
is not zero. For instance, for $h_s=0.4$ and $R_0=1$ we obtain $h_s\kappa_0=h_s/R_0=0.4$ and $\kappa=0.0101$. To get precisely zero curvature,
the applied moment needs to be set  to $-M_1$ where
\beq 
M_1 = \frac{12+1.8\,h_s^2\kappa_0^2}{12+h_s^2\kappa_0^2+0.15h_s^4\kappa_0^4}\,EI\kappa_0
=\frac{12+1.8\,h_s^2\kappa_0^2}{12+h_s^2\kappa_0^2+0.15h_s^4\kappa_0^4}\,M_0
\eeq
Under this load, the initially curved cantilever becomes 
perfectly straight but its length does not remain
equal to the initial centerline length $L=2\pi R_0$ because the
axial strain 
\beq 
\eps_{s1}=\frac{\kappa_0}{Eb_sh_s}\cdot (-M_1)
=-
\frac{h_s^2\kappa_0^2+0.15\,h_s^4\kappa_0^4}{12+h_s^2\kappa_0^2+0.15h_s^4\kappa_0^4}
\eeq 
is not zero. For our example with $h_s\kappa_0=0.4$,
we obtain $\eps_{s1}=-0.0135$.

%\begin{figure}[t]
%    \centering
%    \begin{tabular}{cc}
%    (a) & (b)
%    \\
%    \includegraphics[width=0.5\linewidth]{figures/circle_curved_vs_straight.pdf}
%    &
%    \includegraphics[width=0.46\linewidth]{figures/circle_curved_vs_straight_final.pdf}
%    \end{tabular}
%\caption{Pure bending of a circular cantilever beam subjected to an increasing end moment ranging from 0 to $-2 M_1$.}
%\label{fig:circular_unfolding}
%\end{figure}

\begin{figure}[t]
    \centering
    \begin{tabular}{c}
    \includegraphics[width=0.7\linewidth]{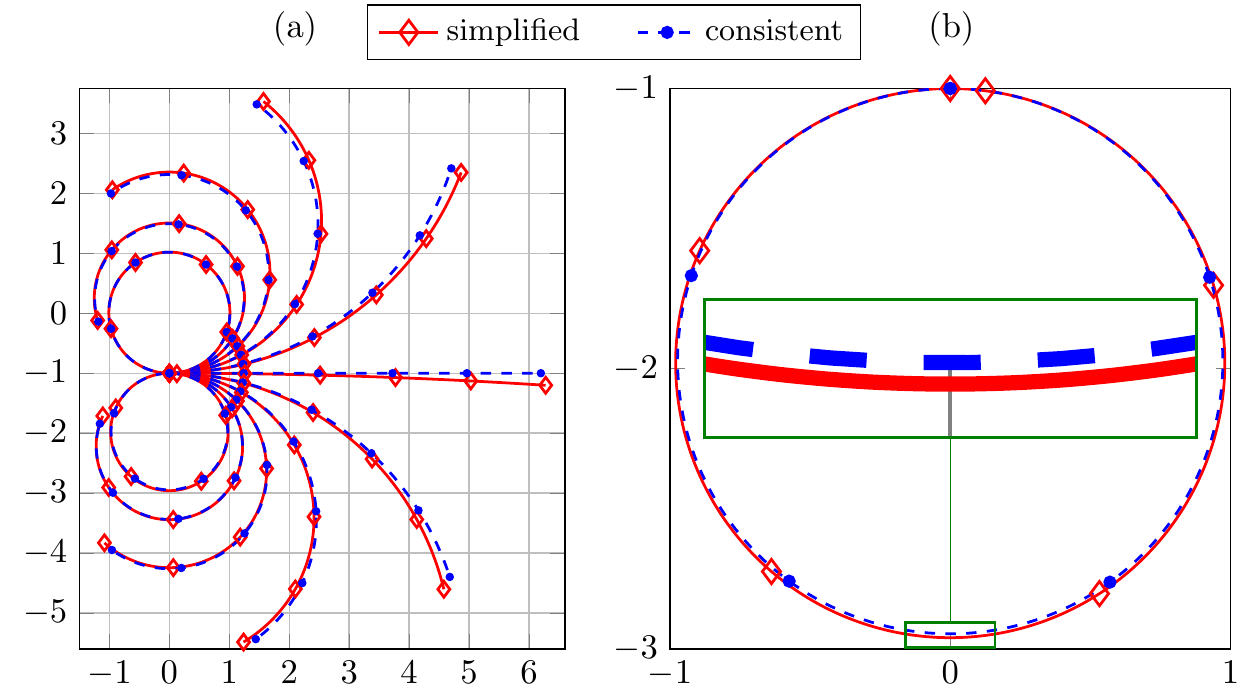}
    \end{tabular}
%\caption{Pure bending of a circular cantilever beam subjected to an increasing end moment ranging from 0 to $-2 M_1$.}
\caption{Comparison of the \color{black}{consistent and simplified} theories for pure bending of a circular cantilever beam: (a) deformed configurations for an increasing end moment ranging from 0 to $-2 M_1$, (b) closer look at the final stage. }
\label{fig:circular_unfolding}
\end{figure}

%\begin{figure}[t]
%    \centering
%    \begin{tabular}{c}
%    \includegraphics[width=0.6\linewidth]{figures/circle_curved_vs_straight_final.pdf}
%    \end{tabular}
%\caption{Final configuration of the beam, i.e., configuration of the cantilever loaded by an end moment equal to $-2M_1$.}
%\label{fig:circular_unfolding_final}
%\end{figure}

The physical reason why the circular arch unfolded
by a uniform moment into a straight beam tends to shrink is that if it did not, the compressive strain
in the fibers that were initially on the outer 
surface and had initial length $2\pi(R_0+h_s/2)$
would be $-h_s/(2R_0+h_s)$ while the tensile strain
in the fibers that were initially on the inner 
surface and had initial length $2\pi(R_0-h_s/2)$
would be $h_s/(2R_0-h_s)$, i.e., it would be higher
in magnitude than the compressive strain. 
If the material law is linear elastic \color{black}{(in terms of the Biot strain and its work-conjugate stress)}, the same
would hold for the stresses and the resulting
normal force would be tensile. To get zero normal
force, the centerline length must be reduced.

The case of a fully unfolded circle
that still remains linear elastic in terms of 
the engineering strain and nominal stress may look purely
academic, but the same argument applies even to 
much lower strain levels and thus the centerline
would shrink at least during an initial stage of the unfolding
process \color{black}{(see Appendix~\ref{appA} for deeper analysis)}. This phenomenon is neglected by the \color{black}{widely used simplified}
theory, which is in many cases justified because
 the effect is
indeed weak provided that $h_s\kappa_0\ll 1$. 
The difference between $M_1$ and $M_0$ 
for $h_s\kappa_0=0.4$, 0.2, 0.1 and 0.05 is about
1~\%, 0.3~\%, 0.07~\% and 0.02~\%, respectively. However, note
that the case of $h_s\kappa_0=0.4$ is not outside
the range of what is normally considered as slender
beams, because the total length of the circular centerline is
$L=2\pi R_0=2\pi/\kappa_0$ and thus the span-to-depth ratio of the unfolded cantilever is $L/h_s=2\pi/(h_s\kappa_0)\approx 15.7$. 
In fact, the unfolding of a circular cantilever
was studied in \cite{IBRAHIMBEGOVIC1995} using the \color{black}{simplified} theory, with dimensions set to $L=10$ and $h_s=1$, which corresponds to $h_s\kappa_0=2\pi h_s/L\approx 0.628$.
In this case, the described modification would play
a significant role and the difference between
$M_1$ and $M_0$ would be about 2.4~\%.

Another (closely related) consequence of the coupling between membrane and bending effects is that if the applied moment
is $-2M_1$, the curvature changes from $\kappa_0$
to $-\kappa_0$ and one may think that the centerline is located
on a circle of radius $R_0$ (i.e., of the same radius as in the initial state), but in reality 
the radius is different. The reason is that,
\color{black}{according to the definition adopted here},
the curvature \color{black}{is equal} to the derivative
of rotation with respect to the arc length coordinate measured along the initial
centerline, but the arc length along the
deformed centerline is different,
since the axial strain is not zero. The actual radius of curvature
under loading by constant moment $-2M_1$
leading to curvature $-\kappa_0$ and axial
strain $2\eps_{s1}$
is evaluated as 
\beq \label{eq:162}
R = \left\vert\frac{(1+2\eps_{s1})\,\dx}{{\rm d}\varphi}\right\vert = 
\frac{1+2\eps_{s1}}{\kappa_0} = (1+2\eps_{s1})\,R_0
\approx 0.973\,R_0
\eeq 
This means that if the \color{black}{consistent} theory is used and moment $-2M_1$ is applied, the resulting shape 
is a perfectly closed circle of a smaller radius
than the initial one, with the radius reduced
by approximately 2.7~\% (here we consider again our specific example with $h_s\kappa_0=0.4$).

The deformed shapes of the initially circular beam that correspond to the applied moment ranging from 0 to $-2M_1$ with step $-0.2\,M_1$ are shown in Fig.~\ref{fig:circular_unfolding}. The solid red curves 
have been obtained using the \color{black}{simplified} relations
$N=EA\,\eps_s$ and $M=EI\,\Delta\kappa$ while the dashed
purple curves have been obtained with the \color{black}{consistent}
relations (\ref{eq:n})--(\ref{eq:m}).
In the middle of the deformation process, at applied moment $M=-M_1$, the current shape 
computed using the \color{black}{consistent} theory
is straight and the centerline length is slightly reduced as compared to the initial one.
On the other hand, according to the \color{black}{simplified} theory
the centerline length would remain constant, and the 
current shape would be straight at applied moment 
$M=-M_0$ while at $M=-M_1$ it would be slightly curved.
The markers plotted in  Fig.~\ref{fig:circular_unfolding} indicate the
position of points that are initially regularly spaced on the centerline
at $s=iL/5$ where $i=0,1,\ldots 5$. 

The simulations also confirm the theoretical
solution that corresponds to the ultimate stage
of loading, with applied moment $-2M_1$; see the ``lower'' circle
in Fig.~\ref{fig:circular_unfolding}a,
which is also shown in more detail in Fig.~\ref{fig:circular_unfolding}b.
For the \color{black}{consistent} theory, the deformed shape plotted
by the dashed blue line corresponds to a full circle of radius close to 0.973, as predicted in (\ref{eq:162}).
On the other hand, the \color{black}{simplified} theory would
give a perfectly closed circle of the same radius
as the initial one only if moment $-2M_0$ was applied.
Under moment $-2M_1$ it gives a deformed centerline located on a circle of radius approximately 
$0.98\,R_0$, plotted in Fig.~\ref{fig:circular_unfolding} by the solid red line. As seen in Fig.~\ref{fig:circular_unfolding}b,
this circle is ``more than closed'', i.e., the free end of the cantilever initially located at the origin does not end up at the origin but 
is slightly shifted along the circle clockwise. 

To provide not only a visual but also a quantitative assessment, the distances between the positions of the end point computed using the \color{black}{consistent and simplified} theories at different levels of loading 
are evaluated in Table~\ref{sometable}.
The distance (i.e., the Cartesian norm
of the difference between displacement vectors) is normalized here by the 
initial radius $R_0$, so the values
around 0.2 represent a considerable error.
    \begin{center}
 \begin{table}[htb]
     \label{tab_cb}
     \centering
          \begin{tabular}{cc}
\toprule
        $\mathbf{-M/M_{1}}$ & \textbf{normalized error}  \\
\midrule             
       0.2  & 0.0228\\
       0.4  & 0.0660\\
       0.6  & 0.1224\\
       0.8  & 0.1780\\
       1.0  & 0.2168\\
       1.2  & 0.2279\\
       1.4  & 0.2095 \\
       1.6  & 0.1726\\
       1.8  & 0.1388 \\
       2.0  & 0.1256\\    
       \bottomrule
     \end{tabular}
           \caption{Pure bending of a circular cantilever beam subjected to an increasing end moment: Evaluation of the error in the displacement of the end point  caused by replacement of the \color{black}{consistent} theory by the \color{black}{simplified} one (normalized by the circle radius $R_0$).}
           \label{sometable}
 \end{table}
 \end{center} 
The deformed shapes in Fig.~\ref{fig:circular_unfolding}
and the error values in  Table~\ref{sometable}
are based on a highly accurate numerical solution with 1 element divided into 2048 integration segments. The dependence of the numerical error
on the number of integration segments is illustrated
in Table~\ref{tabCB}. The evaluated quantity $w$ is the
vertical displacement of the point initially located 
at $s=L/2$ caused by the applied moment $-M_1$ or $-2M_1$. 
At $M=-M_1$,
the exact value of this displacement is $w=-2\, R_0$, where $R_0=1$ in our example. At $M=-2M_1$, a highly accurate 
computation yields $w=-3.94612\, R_0$. 
 The table confirms
that asymptotically the error decreases in inverse proportion to the square of the number of integration segments. 

\begin{center}
 \begin{table}[htb]
     \centering
          \begin{tabular}{rlrrr}
\toprule
& \multicolumn{2}{c}{\textbf{$M = -M_1$}} & \multicolumn{2}{c}{\textbf{$M = -2M_1$}} \\
        \textbf{ Model} & \textbf{$w/R_0$} & \textbf{error $[\%]$} & \textbf{$w/R_0$} & \textbf{error $[\%]$} \\
\midrule
         Exact & -2 & -- & -3.946122 & --
         \\
           4 segments & -2.221453 &  11.07265 &  -4.383040 &          43.69177 \\
           8 segments & -2.052343 &   2.61715 & -4.049401 &         10.32787\\
          16 segments & -2.012914 &  0.64570 &  -3.971593 &         2.54707 \\
          32 segments & -2.003215 & 0.16075 & -3.952468 & 0.63457 \\
          64 segments & -2.000802 & 0.04010 &  -3.947707 & 0.15847\\
         128 segments & -2.000201 &  0.01005 &  -3.946518 &       0.03957\\
         256 segments & -2.000050 & 0.00249 & -3.946221   &    0.00987\\
         512 segments & -2.000012 &  0.00060 & -3.946147 & 0.00247\\
         1024 segments & -2.000003 &  0.00015 & -3.946128 & 0.00057\\
         2048 segments & -2.000001 &  0.00005 & -3.946124  &  0.00017\\
         \bottomrule
     \end{tabular}
      \caption{Pure bending of a circular cantilever beam subjected to end moment $-M_1$ or $-2M_1$: Evaluation of errors in vertical displacement caused by numerical integration along the beam element.}
      \label{tabCB}
 \end{table}
 \end{center}
         
\begin{figure}[h]
    \centering
    \begin{tabular}{ccc}
    (a) & (b) & (c)
    \\
    \includegraphics[width=0.25\linewidth]{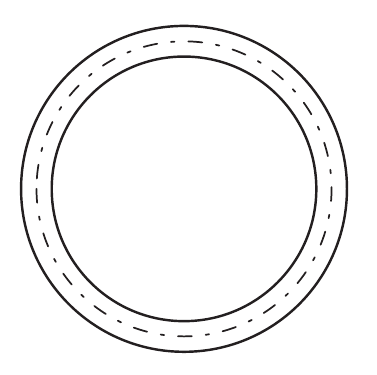}
    &  \includegraphics[width=0.25\linewidth]{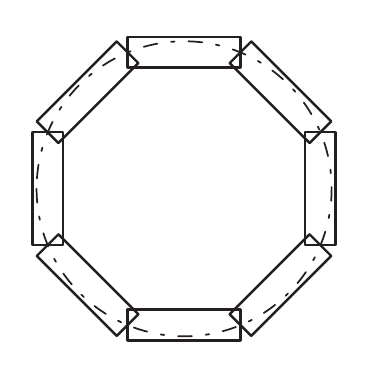} &
    \includegraphics[width=0.25\linewidth]{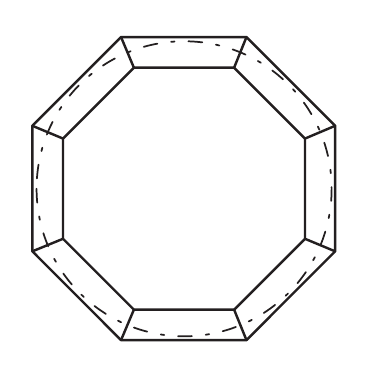}
    \end{tabular}
\caption{Geometry of a circular arch: (a) Exact, (b) approximated by straight beam elements with the same length of all longitudinal fibers, (c) approximated by straight beam elements with variable length of longitudinal fibers.}
\label{fig5}
\end{figure}

All results presented so far have been obtained
for a discretization of the entire circular cantilever
by 1 curved element. 
Consider now the effect of mesh refinement. For curved elements, the initial geometry is reproduced exactly and
no extra benefit would be gained by using
several elements, provided that the total number of integration segments would remain the same. This has already been
illustrated for the previous example
in Table~\ref{tabSR}.
On the other hand,
if the initial circular shape is approximated by
straight elements, the initial curvature of each 
element will be zero and the expressions for internal
forces will have the \color{black}{simplified} form (the information about initial curvature is transformed into jumps in the centerline slope between neighboring elements). Upon mesh
refinement, the initial centerline geometry approximated by 
a polygon will converge to a circle, but the 
numerical solutions obtained for the deformed shape
will converge to a limit that corresponds to the 
\color{black}{simplified} theory, which neglects the effect of initial
curvature on the relations between internal forces
and deformation variables. This paradox is related
to the fact that if the curved beam is approximated
by straight elements that connect nodes located
at the curved centerline, the total length of the 
polygonal approximation properly converges
to the length of the curved centerline, and the
length of fibers located at a given nonzero distance from
the centerline converges to the same limit,
which is however different from the actual length
of those fibers on the curved beam. 
This fact is schematically illustrated in Fig.~\ref{fig5}, which shows the circular arch (Fig.~\ref{fig5}a) and
its approximation by eight straight elements of the standard type (Fig.~\ref{fig5}b). 
A gradual change of the inclination angle of
individual sections is replaced by jumps at
element interfaces. The arrangement of rectangular domains
representing individual elements leads to gaps
and overlaps, the area of which does not vanish in the limit as the element size tends to zero. 

To get the proper limit
even with straight elements, one would need to abandon the idea that individual sections are initially perpendicular
to the centerline and 
consider non-parallel end sections 
as shown in Fig.~\ref{fig5}c. In this way, the information on the variable fiber length could
be incorporated into the model, but 
instead of this artificial adjustment
it is better to use fully consistent curved elements.

%\clearpage
%%%%%%%%%%%%%%%%%%%%%%%%%%%%%%%%%%%%%
\subsection{Asymmetric circular arch, complete load-displacement curve}
%%%%%%%%%%%%%%%%%%%%%%%%%%%%%%%%%%%%%

 The third example is representative of an arch instability after large deflections.
 The structural setup is similar to the arch analyzed
 in Section~\ref{sec:4.1} but the supports are
 not symmetric and the dimensions are different.
 This particular structure was investigated by several authors \cite{Noor1981, simo1986, wood1977} and the solution based on Euler's nonlinear theory of the inextensible curved elastica was evaluated by DaDeppo and Schmidt \cite{dadeppo1975}, who reported a value of the maximum load equal to $8.97\, EI/R^2$. The theory they used was exact in the sense that no restrictions were imposed on the magnitudes of deflections \cite{dadeppo1974}. The nonlinear boundary-value problem of the inextensible elastica was solved numerically to a high degree of accuracy, which was confirmed by comparisons with certain exact solutions of other problems previously derived by the same authors \cite{dadeppo1969}. 

\begin{figure}[h!]
    \centering
    \begin{tabular}{ccc}
    (a) & (b) & (c) 
    \\
    \includegraphics[width=0.3\linewidth]{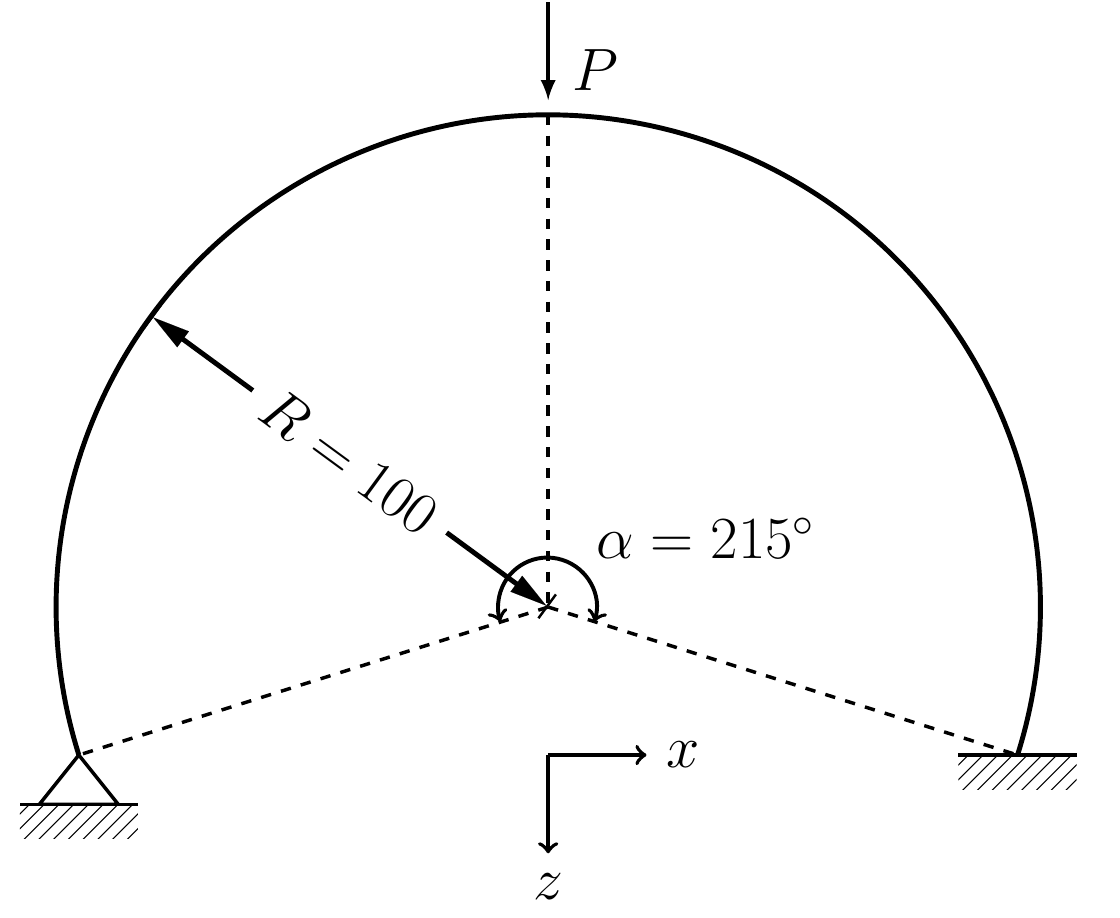}
    &  \includegraphics[width=0.3\linewidth]{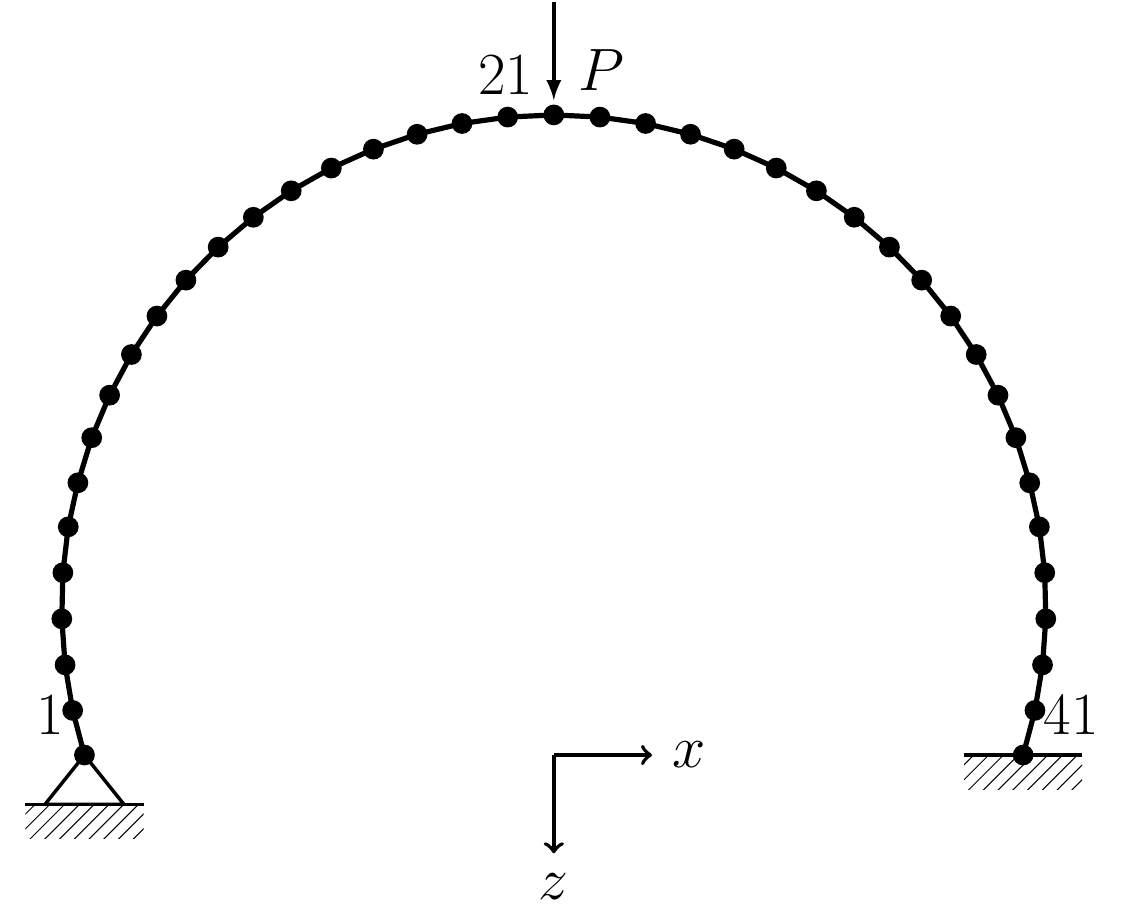}
    & \includegraphics[width=0.3\linewidth]{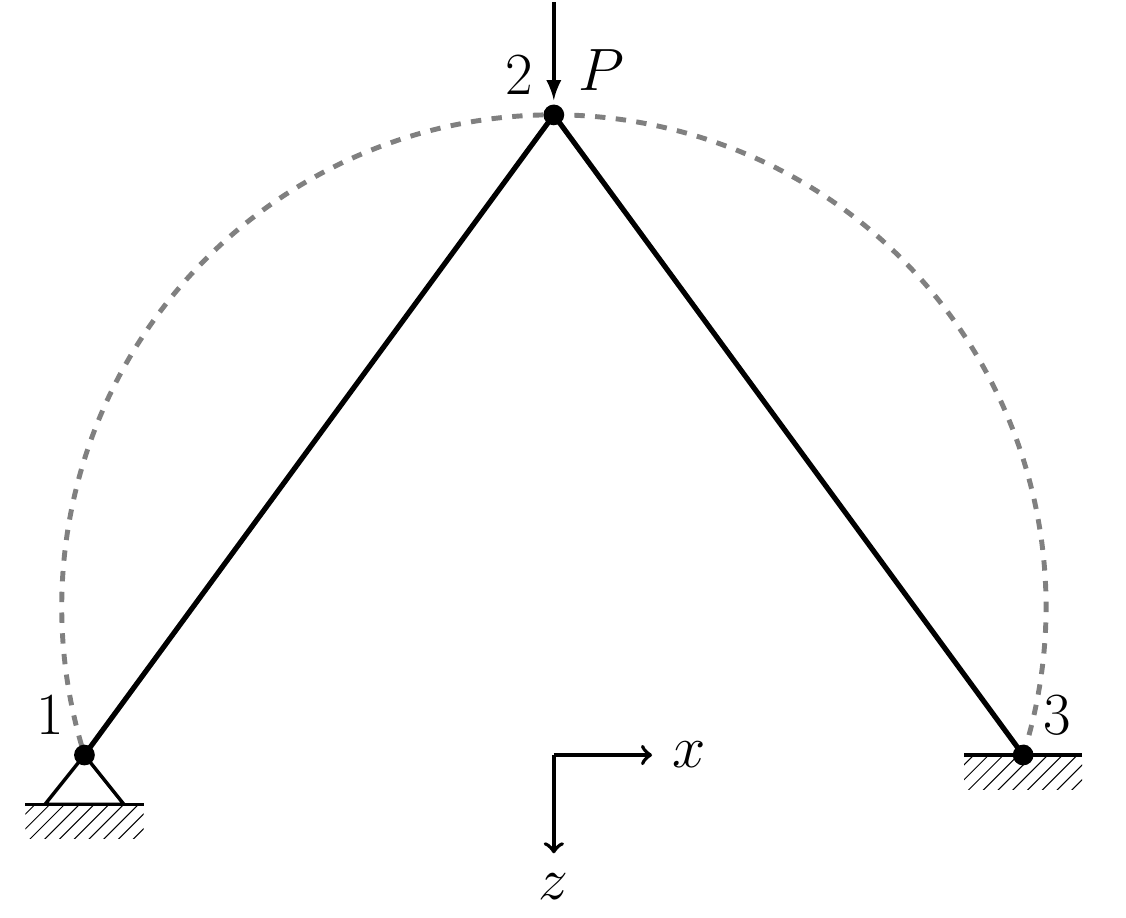}
    \end{tabular}
\caption{Asymmetric circular arch: (a) Geometry, (b)  forty-element mesh used by Simo and Vu-Quoc \cite{simo1986}, (c) two-element mesh used by the present curved beam formulation.}
\label{fig:circular_arch}
\end{figure}

\begin{figure}[h!]
    \centering
    \begin{tabular}{cc}
    (a) & (b) 
    \\
    \includegraphics[width=0.48\linewidth]{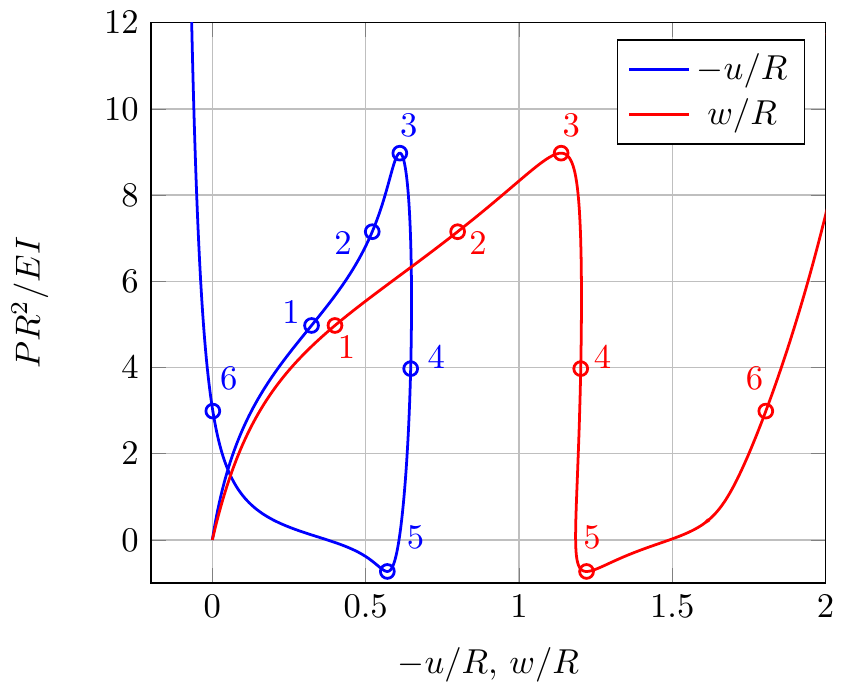}
    &  \includegraphics[width=0.48\linewidth]{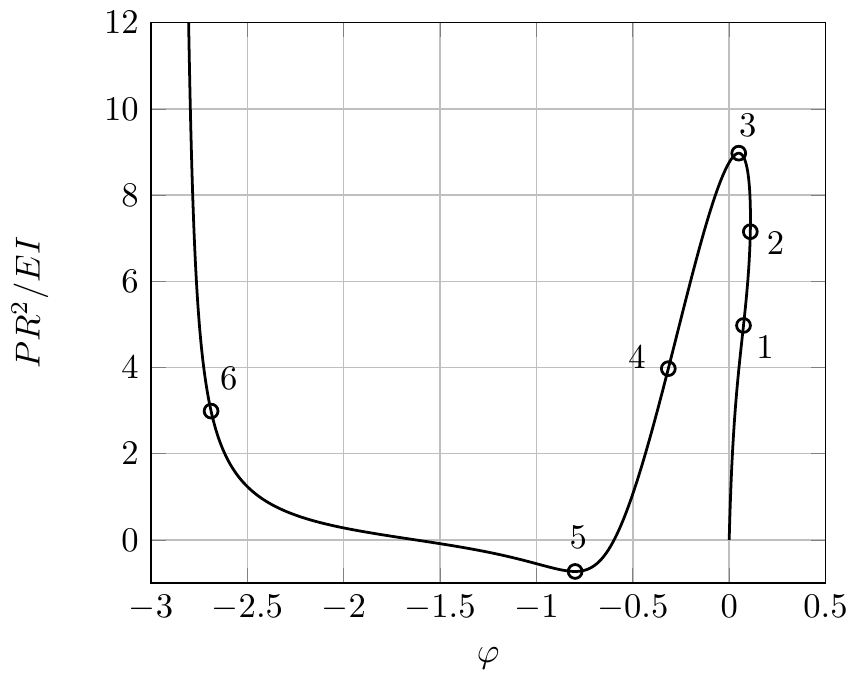}
    \end{tabular}
    \caption{Asymmetric circular arch:  (a) Load-displacement curves and (b) load-rotation curve for the top of the arch (node 2 in Fig.~\ref{fig:circular_arch}c)}
    \label{archcurves}
\end{figure}

\begin{figure}[h!]
    \centering
    \begin{tabular}{c}
    \includegraphics[width=0.6\linewidth]{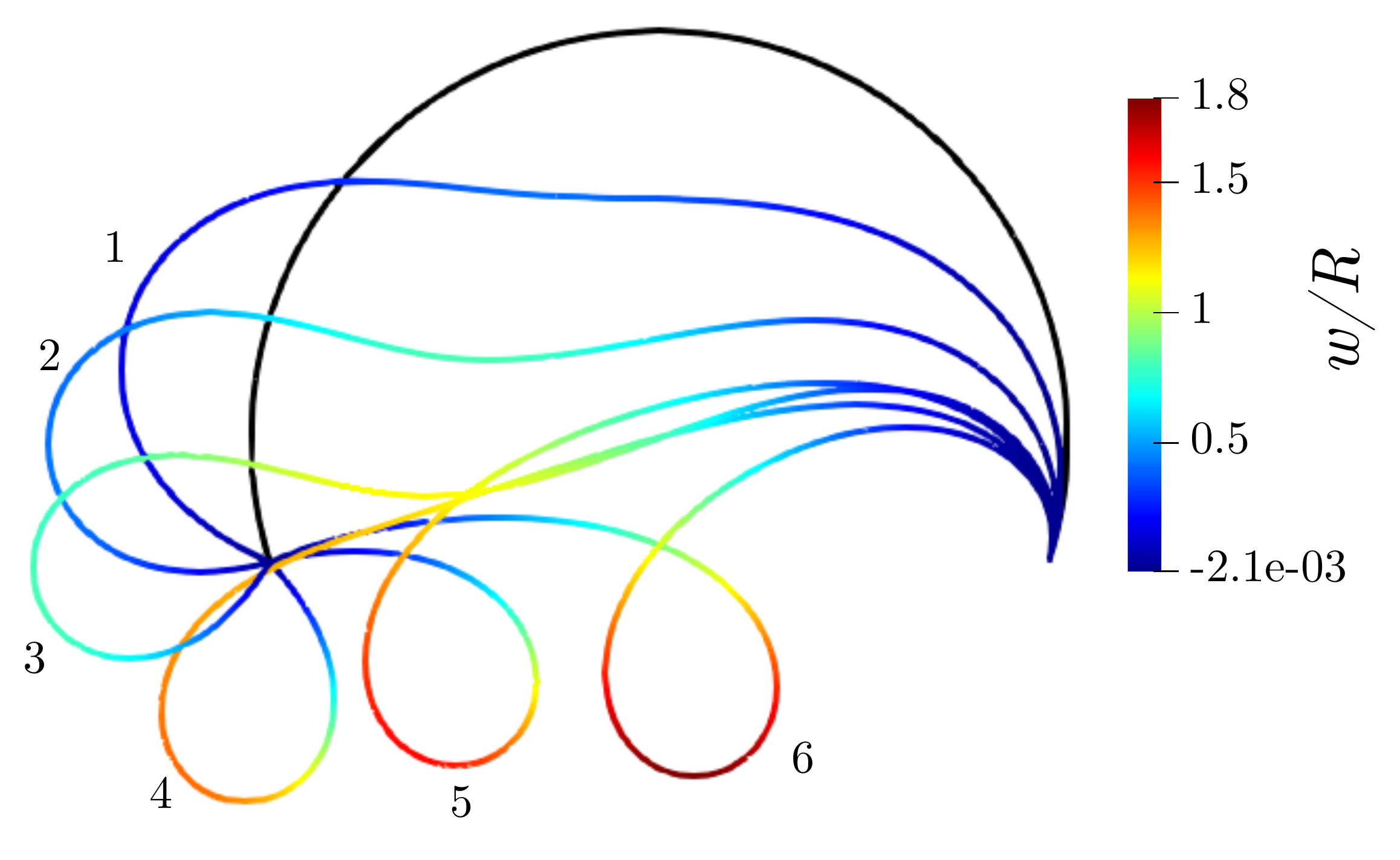}
    \end{tabular}
\caption{Asymmetric circular arch: Deformed shapes and normalized vertical displacement along the arch for the six states labeled by numbers 1--6 in Fig.~\ref{archcurves}. }
\label{fig:circular_arch_def}
\end{figure}

 The arch is circular with one boundary hinged and the other clamped, and it is loaded by a vertical concentrated load applied at the top, as shown in Fig.~\ref{fig:circular_arch}a. 
 %\textcolor{red}{The cross section is rectangular with a depth-to-width ratio $h/b=2.289$.} The elastic modulus is set to $E=1.0 \times 10^6$, 
 In calculations presented in the literature,
 the structure was usually considered as very slender,
 so that the solution obtained for the axially
 incompressible case with neglected shear distortion
 could be used as a reference. However, the specific
 values of parameters are in some cases hard to find.
 
 Wood and Zienkiewicz~\cite{wood1977} obtained a buckling load of $9.24\, EI/R^2$ using a mesh consisting of sixteen 2D six-noded paralinear elements (linear approximation in the direction of thickness and quadratic in the longitudinal direction), and one 2D three-noded linear element near the hinged support. Overall, their mesh had 67 nodes with 127 global unknowns, and they analyzed
 an arch of depth $h_s=1$, which corresponds
 to $h_s\kappa_0=0.01$. 
 
 Simo and Vu-Quoc \cite{simo1986} performed their analysis with a mesh consisting of forty straight beam elements (Fig.~\ref{fig:circular_arch}b) and found a maximum load of $9.0528\, EI/R^2$. 
 A similar result was obtained with straight elements by Ibrahimbegovi\'{c} \cite{IBRAHIMBEGOVIC1995},
 who also  reported a great improvement with 
 twenty 3-node curved elements.
 The resulting maximum load of $8.973\, EI/R^2$ was very close to the reference
 value published by DaDeppo and Schmidt.

In our analysis, we
have used the minimum number of elements required to discretize the structure, which results in a two-element mesh shown in Fig.~\ref{fig:circular_arch}c, with only four global unknowns---two displacements and the rotation at the loaded node plus the rotation at the left support. The analysis has been performed under indirect displacement control, with the load iteratively adjusted such that the rotation at the left support increases by prescribed increments.
Direct displacement control based on prescribed increments of the vertical displacement under the applied force is not advisable because the corresponding load-displacement diagram, plotted by the red line in Fig.~\ref{archcurves}a,
exhibits a section with a very steep drop and even 
a slight snapback (see the path 3-4-5 in the figure). 

The simulation has been done  with the same sectional
stiffnesses as in \cite{IBRAHIMBEGOVIC1995},
namely
$EA=10^8$ and $EI=10^6$, which correspond to a section
of depth $h_s=\sqrt{0.12}\approx 0.3464$ and thus to 
$h_s\kappa_0\approx 0.003464$. 
The numerical solution naturally depends on the number of segments used for integration of the governing equations on the element level. As seen in Table~\ref{tab:circarch}, the accuracy of the present
simulation is higher than in \cite{wood1977} if 10 integration segments are used
and higher than in \cite{simo1986}
if 20 integration segments are used. 

Further refinement of the integration grid 
leads again to quadratic convergence. The error is calculated by considering a highly accurate limit value of 8.972922, obtained by refinement until the resulting maximum load stabilizes up to seven valid digits. This limit is in good agreement with the truncated value of 8.97 reported by DaDeppo and Schmidt \cite{dadeppo1975}, and it perfectly agrees
with the value of 8.973 reported by Ibrahimbegovi\'{c} \cite{IBRAHIMBEGOVIC1995}.
One should note that \cite{IBRAHIMBEGOVIC1995}
used shear-flexible elements and the actual
limit value would be in that case slightly lower.

In the simulations reported in Table~\ref{tab:circarch}, the \color{black}{simplified} relations
between internal forces and deformation variables
have been used. Since, for the given input data,
$h_s\kappa_0\ll 1$, the results obtained with the
\color{black}{consistent} relations would be only slightly different.
The maximum load calculated with very high accuracy
would change from  8.972922 to  8.972950 if the \color{black}{consistent}
relations are used.

The load-displacement curves plotted in Fig.~\ref{archcurves} indicate that the arch
exhibits a snap-through behavior after reaching the
peak load at the state marked by label 3. The last physically reasonable numerical solution is obtained at the state marked by label 4.
The simulation can be continued without problems
but the deformed structure passes across the left support (see Fig.~\ref{fig:circular_arch_def}). This aspect was investigated by Simo et al.~\cite{simo1986b}, who showed that a contact constraint condition on the left support needs to be introduced to obtain a more realistic solution. Here, we have not considered contact activation and the subsequent stiffening effect in the structure because these phenomena are out of scope of the present study.
 
\begin{center}
\begin{table}[htb]
     \centering
          \begin{tabular}{rcll}
\hline\\[-4mm]
        \textbf{Model} & \textbf{N$^\circ$ global unknowns} & \textbf{$P_{cr}R^2/EI$} & \textbf{error $[\%]$}  \\[1mm]
\hline\\[-4mm]
DaDeppo and Schmidt \cite{dadeppo1975}  &  & 8.97 & \\
Wood and Zienkiewicz \cite{wood1977}  & 127 & 9.24 &  \\ %2.976
Simo and Vu-Quoc \cite{simo1986}  & 117 & 9.0528 & \\ % 0.890
Ibrahimbegovi\'{c} \cite{IBRAHIMBEGOVIC1995} & & 8.973 & \\
         present approach, 10 segments  & 4 & 8.735135  &  2.650\\
         20 segments  & 4 &  8.912875 & 0.669 \\
         40 segments  & 4 & 8.957865 & 0.168 \\
         80 segments  & 4 & 8.969161 & 0.0420 \\
         160 segments  & 4 & 8.971979 & 0.0105 \\
         320 segments  & 4 & 8.972686 &  0.00263\\
         640 segments  & 4 & 8.972863
          & 0.00066 \\
    $\to\infty$ & 4 & 8.972922 & 0\\
    \hline
     \end{tabular}
      \caption{Asymmetric circular arch: Evaluation of errors in maximum load caused by numerical integration along the beam element and comparison with results from the literature.} 
\end{table}
\label{tab:circarch}
\end{center}

\subsection{Parabolic arch}
%%%%%%%%%%%%%%%%%%%%%%%%%%%%

In the previous three examples, the initial
shape was supposed to be circular and its
analytical description was based on formulae
(\ref{circle1})--(\ref{circle3}). Let us now
consider a parabolic arch. In the simplest case, 
when the left end of the parabolic beam element
is located at the apex of the parabola, 
functions $u_{s0}$ and $w_{s0}$ that characterize the initial undeformed shape must satisfy the equation
\beq \label{eq163}
w_{s0} = \frac{a}{2}\,(x+u_{s0})^2
\eeq 
where $a$ is a given geometric parameter. These functions
are also constrained by the condition
\beq \label{eq164}
(1+u'_{s0})^2 + w'^2_{s0} = 1
\eeq 
which follows from the fact that differential segments in the fictitious straight configuration have the same length as in the initial undeformed configuration. 
Taking the derivative of (\ref{eq163}) and substituting into (\ref{eq164}),
we obtain after rearrangement
\beq 
\sqrt{1+a^2(x+u_{s0})^2}\,(1+u'_{s0}) = 1
\eeq 
and integration after separation of variables
leads to
\beq \label{eq166}
a(x+u_{s0})\sqrt{1+a^2(x+u_{s0})^2}+{\rm arcsinh} (a(x+u_{s0}))=2ax
\eeq 
This equation implicitly defines function $u_{s0}(x)$ but the closed-form expression for this function is not available. Still,
equation (\ref{eq166}) could be solved numerically for 
each prescribed value of $x_i$ and the corresponding $u_{s0}(x_i)$ could be evaluated.
In this way, the analytical description of the
initial shape would be replaced by a precomputed table of
values of $x_i$ and $u_{s0}(x_i)$ for all  points of the integration grid. It is then easy to
evaluate the corresponding $w_{s0}(x_i)$
from (\ref{eq163}) and the ``initial rotations $\varphi_{0}(x_i)$ from
\beq\label{eq167} 
\varphi_{0}(x) = -\arctan a(x+u_{s0})
\eeq

The procedure described above can be used to ensure uniform spacing of the grid points,
which was assumed in the algorithms described in Section~\ref{sec:algo}. As an alternative, one can 
prescribe uniform spacing in projection of the
curved beam element onto the local $x$ axis,
which coincides with the tangent to the centerline constructed at
the left end. In this case, the prescribed values
are $x_i+u_{s0}(x_i)=ih_p$, $i=0,1,2,\ldots N$,
where $h_p=L_p/N$ is the projected integration segment length derived from the projected element length, $L_p$.
The corresponding values of $x_i$ are then
directly obtained from (\ref{eq166}) by evaluating the
expression on the left-hand side and dividing by $2a$. Afterwards, $u_{s0}(x_i)=ih_p-x_i$ can be
determined and the corresponding values $w_{s0}(x_i)$
and $\varphi_{0}(x_i)$ are calculated as usual
from (\ref{eq163}) and (\ref{eq167}). 
All this can be done ``on the fly'' and there
is no need to solve nonlinear equations and store a precomputed table of values. In the algorithms,
the constant segment length $h$ needs to be 
replaced by a variable length $h_i$, computed
for each integration segment separately
as the Euclidean distance between points with
coordinates $((i-1)h_p,w_{s0}(x_{i-1}))$
and $(ih_p,w_{s0}(x_{i}))$. In theory, this
may slightly disturb the quadratic convergence 
rate when the number of segments increases,
since the integration of curvature is no longer
based on a central difference scheme.
However, no problems have been observed in the specific calculations to be reported next.

For these calculations, 
we adopt an example of a parabolic arch from \cite{borkovic2022geometrically} and investigate two selected geometries. One, characterized by $H=0.25$~m, corresponds to a shallow arch, and the other, with $H=0.5$~m, to a deeper one. The cross section is a square with $b_s=h_s=0.2$ m and the elastic modulus is $E = 10^4$ MPa.  

It is well known that the load-displacement curves of shallow arches typically exhibit snap-through followed by an unstable
branch and eventually by restoration of stable equilibrium, while the behavior of deep arches is more complex, with the so-called looping of the load-displacement curve~\cite{sabir1973large}. The aim of this example is to demonstrate that the proposed formulation is able to capture such phenomena. The arch is simply supported and loaded by a concentrated force acting at midspan. Symmetry is exploited and only one half of the arch is simulated,
which means that a possible bifurcation into a nonsymmetric shape is ignored.

\begin{figure}[h!]
\centering
\includegraphics[width=0.4\linewidth]{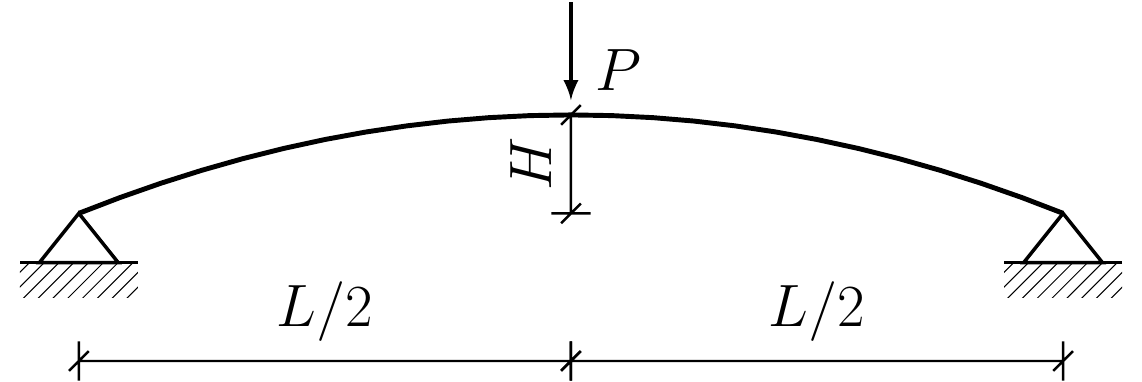}
\caption{Parabolic arch: Geometry and loading, with $L = 10$~m, $H = 0.25$ m for the shallow arch and $H = 0.5$ m for the deeper arch.}
\label{fig:parabolic_arch}
\end{figure}

\begin{figure}[h!]
    \centering
    \begin{tabular}{cc}
        (a) & (b) 
        \\
        \includegraphics[width=0.4\linewidth]{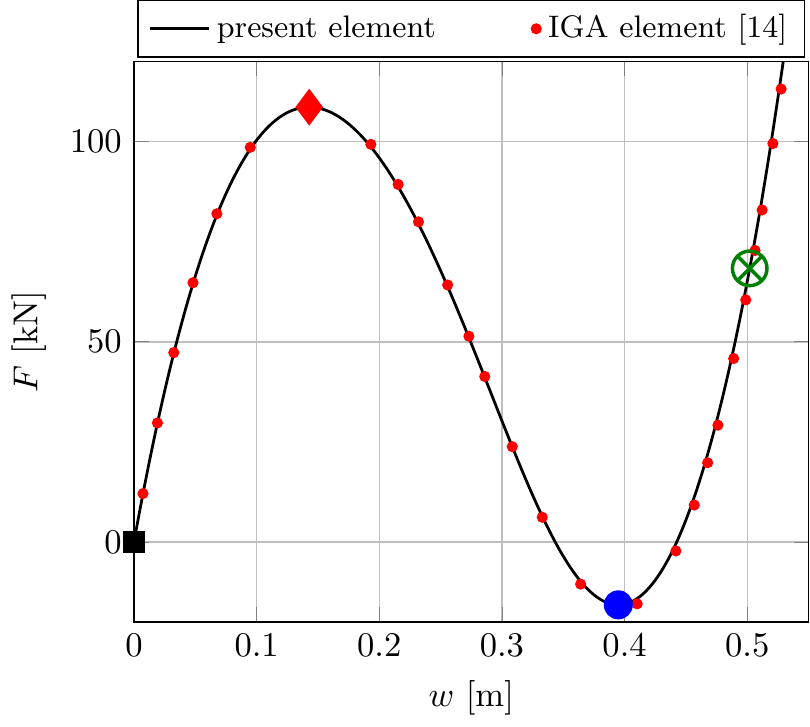} &  \includegraphics[width=0.4\linewidth]{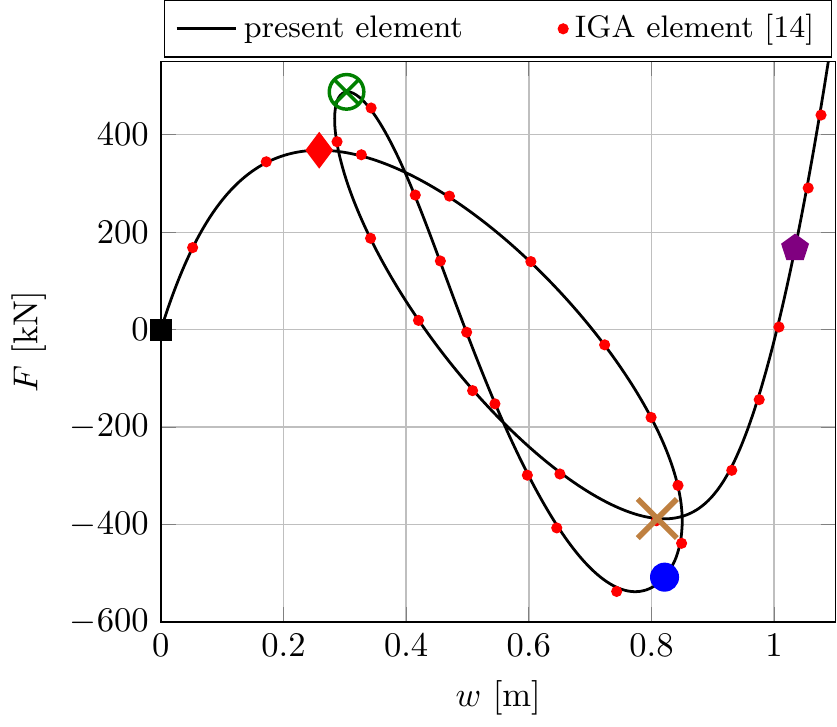}
        \\
        (c) & (d) 
        \\
        \includegraphics[width=0.4\linewidth]{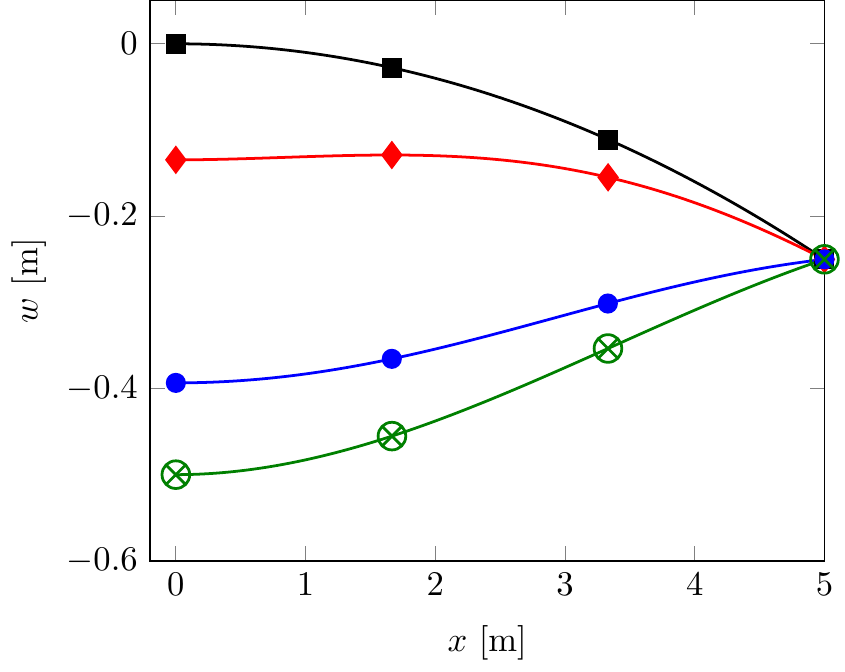} &  \includegraphics[width=0.4\linewidth]{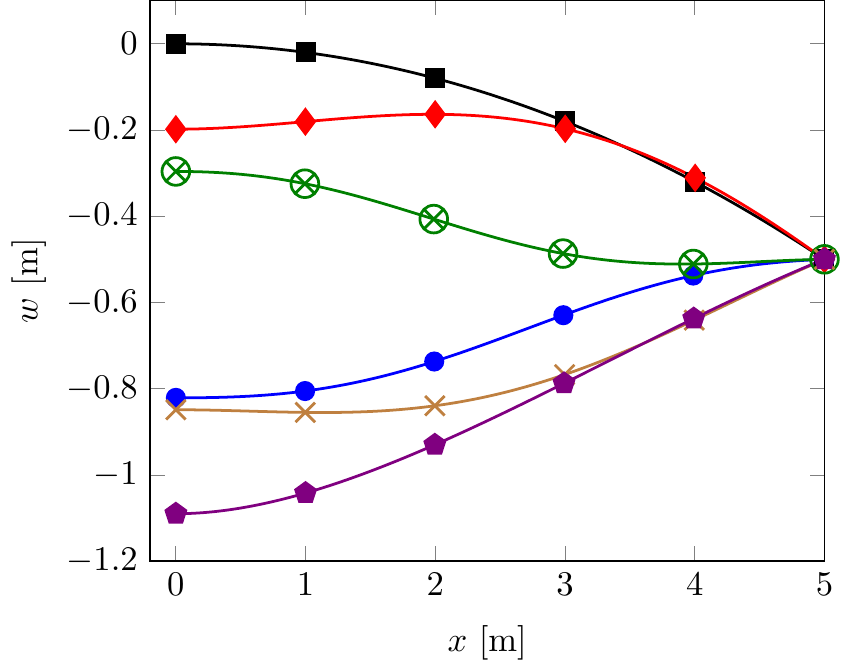}
    \end{tabular}
    \caption{Parabolic arch: Load-displacement diagrams (a)  for the shallow arch and (b) for the deeper arch, and deformed shapes (c)  for the shallow arch and (d) for the deeper arch.} 
    \label{fig:parabolicArch_results}
\end{figure}

The response of the shallow arch is described by the
diagrams in Fig.~\ref{fig:parabolicArch_results}a,c.
The load-displacement curve (Fig.~\ref{fig:parabolicArch_results}a)
shows excellent agreement with the results
reported in 
\cite{borkovic2022geometrically}, where isogeometric 
analysis was used.
To illustrate the deformation process,
the deformed shapes corresponding to
four selected states that are marked by special symbols in the load-displacement curve
are presented in
Fig.~\ref{fig:parabolicArch_results}c: 
the initial state (black),
the state at peak load and onset of snap-through instability (red), the state at the end of the
unstable branch (blue), and a stable state
attained at displacement $w=0.5$~m (green). 

In a similar fashion, the response of the deeper arch is described by the
diagrams in Fig.~\ref{fig:parabolicArch_results}b,d.
In contrast to the shallow arch, the vertical displacement under the load does not increase monotonically and the snapback  phenomenon is observed
in addition to the snap-through. The simulation cannot
be performed under direct displacement control and
an arc-length technique is needed. Still, the
complicated load-displacement diagram can be captured
and the results nicely agree with those from 
\cite{borkovic2022geometrically}.

%%%%%%%%%%%%%%%%%%%%%%%%%%%%%%%%
\subsection{Logarithmic spiral}
%%%%%%%%%%%%%%%%%%%%%%%%%%%%%%%%

The last example presents an extreme case of a highly curved beam, which has the initial shape of 
a planar spiral. It would not be so easy to properly 
approximate such a ``structure'' by straight elements
or by \color{black}{shallow} curved elements. Using the present approach, the whole spiral can be represented
by one single element and its shape can be taken
into account precisely, independently of the total
number of loops. 

\begin{figure}[t!]
\centering
\includegraphics[width=0.5\linewidth]{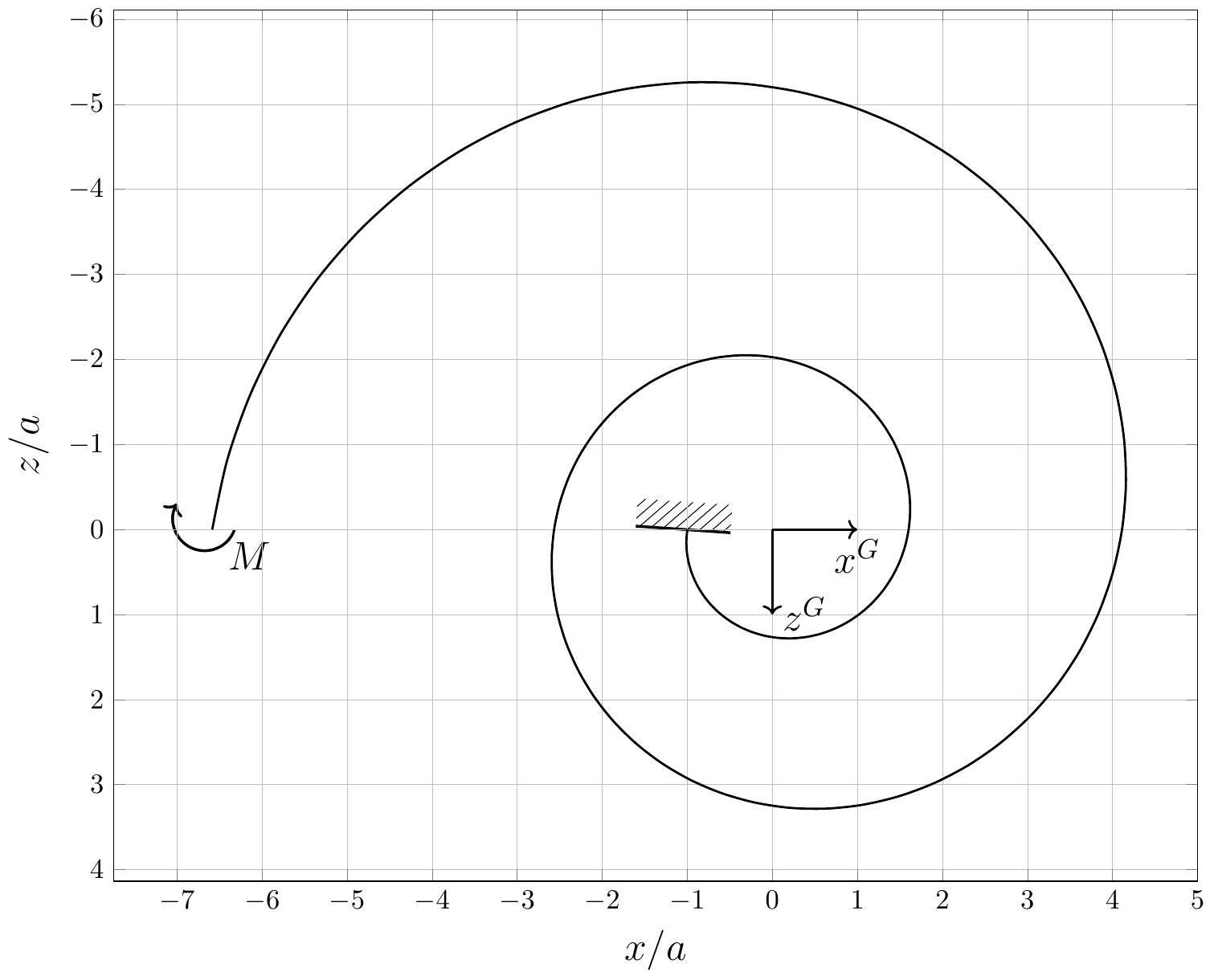}
\caption{Logarithmic spiral: Geometric shape, support and loading, and the choice of global
coordinate axes}
\label{fig:logarithmic_spiral}
\end{figure}

The spiral considered in this example
is a logarithmic one, and in polar coordinates $(r, \theta)$ it is
described by 
\begin{equation}\label{eq:logspir}
    r  = a\, {\rm e}^{b\theta}
\end{equation}
where $a$ and $b$ are positive parameters. 
Fig.~\ref{fig:logarithmic_spiral} depicts
the shape of the spiral obtained
for $b = 0.15$ and $\theta \in[0,4\pi]$.
Parameter $a$ sets the length scale and
determines the distance of the clamped
end of the spiral (at $\theta=0$) 
from the pole. 

For the logarithmic spiral, it is possible to derive 
closed-form expressions for functions that describe
the initial shape with respect to local Cartesian
coordinates with the origin at the left end
of the spiral and with the local $x$ axis in the
tangential direction. The details of the derivation
are provided in Appendix~\ref{appB}. The resulting
expressions read
\bea\label{spiral1} 
\varphi_0(x) &=& \frac{\ln\left(1+cx\right)}{b}\\
\label{spiral2} 
u_{s0}(x) &=& a\left((1+cx)\sin(\varphi_0(x)+\varphi^*) - \sin\varphi^*\right)-x\\
\label{spiral3} 
w_{s0}(x) &=& a\left((1+cx)\cos(\varphi_0(x)+\varphi^*)-\cos\varphi^*\right)
\eea 
where
\beq \label{eq:c}
c = \frac{b}{a\sqrt{1+b^2}}
\eeq 
and
\beq \label{eq:phistar}
\varphi^* = \arctan b
\eeq 
are auxiliary parameters, introduced only to simplify notation. Formulae (\ref{spiral1})--(\ref{spiral3}) can be understood as a generalized form of the formulae
that characterize a circular geometry.
Indeed, by setting $a=R$ and $b\to 0$,
we can reduce (\ref{spiral1})--(\ref{spiral3})
to
(\ref{circle1})--(\ref{circle3}). Since $b\to 0$
leads to $c\to 0$, the expression on the right-hand side of (\ref{spiral1}) tends to a fraction 0/0
but the limit, $x/R$, can be properly evaluated using
L'Hospital's rule. Formulae (\ref{spiral2})--(\ref{spiral3}) are then 
reduced simply by setting 
$a=R$, $c=0$, $\varphi^*=0$ and 
$\varphi_0(x)$ by $x/R$.

\begin{figure}[ht!]
    \centering
    \begin{tabular}{c}
        \includegraphics[width=0.5\linewidth]{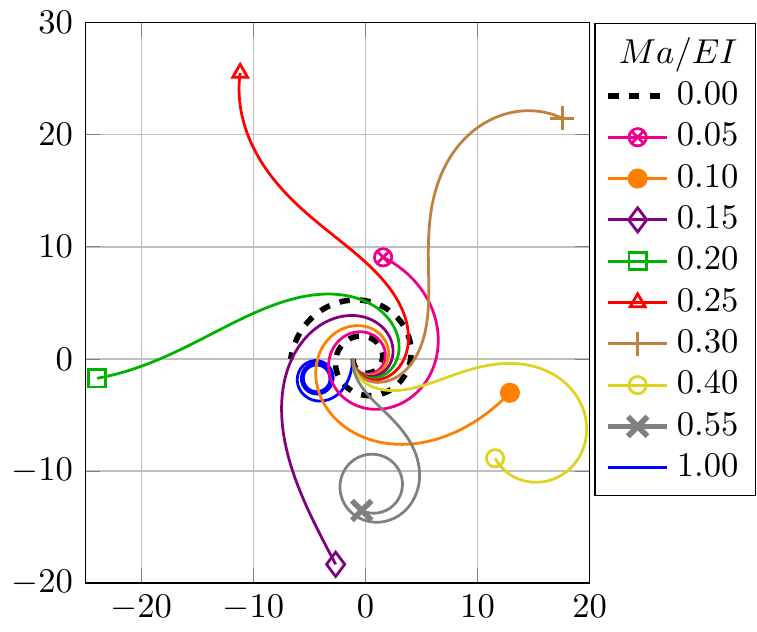}   
    \end{tabular}
    \caption{Initial undeformed shape of the spiral (black) and selected deformed shapes for applied moment increasing up to $M=EI/a$, plotted in the space of dimensionless coordinates normalized by $a$.} 
    \label{fig:logSpiral_results_all}
\end{figure}

Similarly to Example \ref{sec:unfolding}, one side of the curved beam is clamped and the other is loaded by an increasing end moment unfolding the spiral, see Fig. \ref{fig:logarithmic_spiral}.
%The following geometric and material parameters are used, $a = 3$, $b = 0.15$, $\theta \in(0,4\pi)$, $E = 1 \times 10^4$. Square cross section with $h = 0.1$.
Parameter $b$ that controls the shape of the spiral
is in this example set to 0.15 and the polar angle
$\theta$ varies from 0 to $4\pi$, which means that
the spiral has initially two full loops. 
The only other parameter that matters is the ratio
$h_s/a=1/30$, which leads to a dimensionless
parameter $EAa^2/EI=1080$.

\begin{figure}[ht!]
    \centering
    \begin{tabular}{c}
         \includegraphics[width=0.5\linewidth]{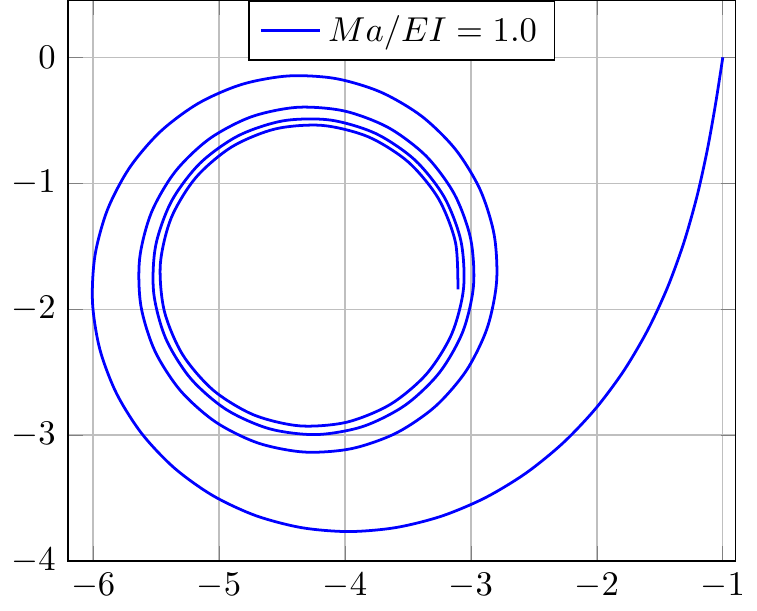} 
    \end{tabular}
    \caption{Deformed shape of the final configuration.} 
    \label{fig:logSpiral_results}
\end{figure}

Selected deformed shapes (and the initial undeformed
shape in black) for a sequence of applied moments
ranging from 0 to $EI/a$ 
are shown in Fig.~\ref{fig:logSpiral_results_all},
and the final state, reached at $M=EI/a$, is reproduced in detail in Fig.~\ref{fig:logSpiral_results}.
The spatial coordinates used
in Figs.~\ref{fig:logarithmic_spiral}--\ref{fig:logSpiral_results} have their origin at the pole 
of the spiral and are normalized by constant $a$.
In the dimensionless coordinates,
the clamped section is located at $(-1,0)$.

%\subsection{Lattice metamaterial composed of circular-arc curved beam elements}
%Example taken from \cite{fu2020design}.
%\begin{figure}[h!]
%\centering
%\includegraphics[width=0.4\linewidth,angle =90]{figures/metamat.pdf}
%\caption{Design of 2D arch-type cell}
%\label{fig:metamat_arch}
%\end{figure}

%%%%%%%%%%%%%%%%%%%%%%%%%%%%%%%%%%%%%%%%%%%%%%%%%%%%
\subsection{A zig-zag beam (two-dimensional spring)}
\label{sec:springexample}
%%%%%%%%%%%%%%%%%%%%%%%%%%%%%%%%%%%%%%%%%%%%%%%%%%%%

\color{black}{
One advantage of the proposed formulation is that it can easily handle
not only beams with a smooth curved centerline, but also beams
with kinks, i.e., with discontinuities in the centerline slope. 
An illustrative example is the zig-zag beam depicted in the top part
of Fig.~\ref{fig:zigzagbeam_results}a, which can be considered as a two-dimensional form of a spring.
This beam
consists of ten straight
segments, with a right angle between neighboring segments. 
During the initial phase of loading by an increasing horizontal force $P$ at the left
support, the spring is compressed and its ``macroscopic axis'' remains straight.
However, since the zig-zag centerline does not coincide with that fictitious
axis, individual segments experience bending and, from the macroscopic point
of view, the spring is relatively flexible in tension or compression. 
The overall deformed shape is plotted in the bottom part of Fig.~\ref{fig:zigzagbeam_results}a in black.
The vertical displacement
of the traced point marked by a special symbol remains very small and the horizontal
displacement increases almost proportionally to the applied force; see the first steep
part of the load-displacement diagrams in Fig.~\ref{fig:zigzagbeam_results}b.
When a critical force is reached, the spring starts buckling and the vertical
displacement increases dramatically (solid curve in Fig.~\ref{fig:zigzagbeam_results}b).
The deformed shape evolves as indicated by images in green and blue
in Fig.~\ref{fig:zigzagbeam_results}a until the ends of the spring meet (red image).

This highly nonlinear process has been simulated using one single element. 
The results are presented in Fig.~\ref{fig:zigzagbeam_results}b in
terms of the dimensionless force $PL^2/EI$ and dimensionless displacements
$u/L$ and $w/L$. The axial sectional stiffness was selected such that
$EAL^2/EI=1000$, which corresponds to a not too stocky beam, with individual
physical segments deforming mainly by bending. The fact that the centerline
consists of straight segments with kinks does not lead to any 
changes in the algorithm presented in Section~\ref{sec:algo1}. It is sufficient to make sure
that each kink point is at the same time one of the points of the computational
grid. In other words, each physical segment is divided into an integer number
of computational segments. The simulation results are presented here for 200
computational segments per regular full-length physical segments (100  computational
segments in the first and last physical segment), i.e., with NIS=1800,
but it has been checked that the load-displacement curves would remain visually almost the same 
even with NIS=90. It has also been verified that the same results are obtained
if the zig-zag beam is divided into 10 straight elements. 

\begin{figure}[ht!]
    \centering
    \begin{tabular}{cc}
    (a) & (b)
    \\
    \includegraphics[width=0.40\linewidth]{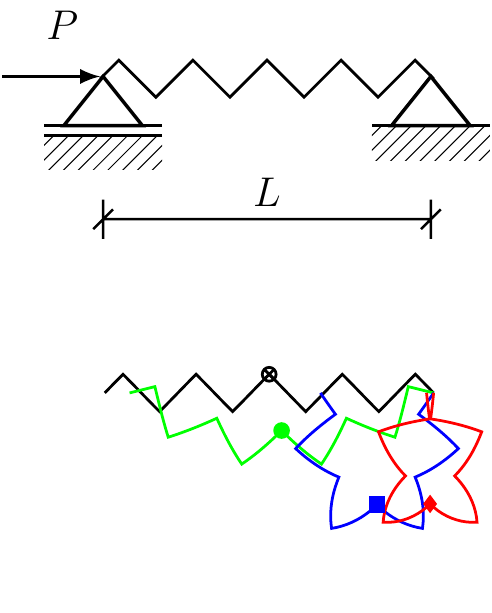} 
    &
\includegraphics[width=0.55\linewidth]{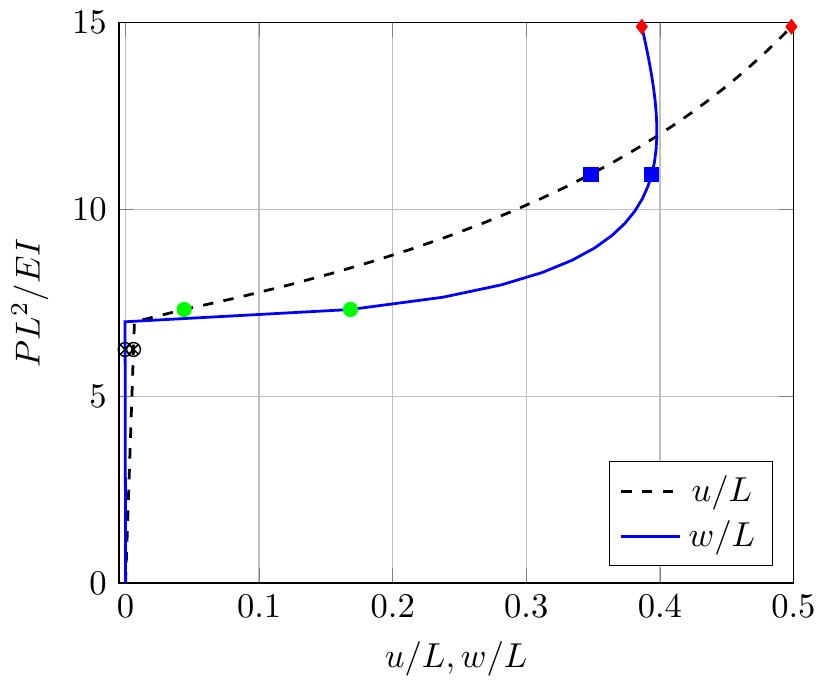}  
    \end{tabular}
    \caption{\color{black}{Zig-zag beam: (a) initial geometry and loading (top) and deformed shapes (bottom), (b) load-displacement diagrams for the horizontal and vertical displacements of the selected point at beam center} }
    \label{fig:zigzagbeam_results}
\end{figure}

The critical force can be estimated based on analogy with a standard
straight beam that has an equivalent stiffness. For the linearized model,
one can easily evaluate the displacement caused by a unit force applied
along the macroscopic axis and also the end rotations caused by two unit moments
applied at the opposite ends with opposite orientations. This is then set equal
to the displacement $L/\overline{EA}$ and rotation $L/\overline{EI}$ that would be produced
on a standard beam with axial sectional stiffness $\overline{EA}$ and flexural sectional
stiffness $\overline{EI}$. For the present input data, the resulting equivalent stiffnesses 
are $\overline{EA}=486\sqrt{2}EI/L^2$ and  $\overline{EI}=EI/\sqrt{2}$. A simple estimate
of the critical load is obtained using the standard Euler formula for a simply
supported beam,
\beq 
P_{E} = \frac{\overline{EI}\pi^2}{L^2} = \frac{EI\pi^2}{\sqrt{2}\,L^2}\approx 6.9789 \frac{EI}{L^2}
\eeq 
This would be exact for an axially incompressible beam. An improved estimate that
takes into account the axial compressibility is constructed using formula (112)
from our previous paper \cite{JLMRH21} (the more accurate version of the formula is used):
\beq  
%P_{cr} = \frac{\overline{EI}\pi^2}{L^2}\left(1+\frac{\overline{EI}\pi^2}{\overline{EA}L^2}\right) = \frac{EI\pi^2}{\sqrt{2}\,L^2}\left(1+\frac{EI\pi^2L^2}{\sqrt{2}\cdot 486\cdot \sqrt{2}\,EIL^2}\right)\approx 7.0497 \frac{EI}{L^2}
P_{cr} = \frac{\overline{EA}}{2}\left(1-\sqrt{1-\frac{4\overline{EI}\pi^2}{\overline{EA}L^2}}\right) = \frac{243\sqrt{2}\,EI}{L^2}\left(1-\sqrt{1-\frac{\pi^2}{243}}\right) \approx 7.0512\frac{EI}{L^2}
\eeq 
The critical force evaluated by a highly accurate numerical simulation 
turns out to be $7.0514EI/L^2$.

The reason why the algorithm from Section~\ref{sec:algo1} does not need to be modified is that the discontinuity
in the initial slope, i.e., in function $\varphi_0$, does not affect the evaluation
of $\varphi_0(x_{i-1/2})$ in the algorithmic steps described by equations
(\ref{e60}) and (\ref{e63})--(\ref{e64}), provided that the midpoints of
integration segments do not coincide with the kink points on the centerline.
The initial curvature $\kappa_0$ is considered as zero and $\varphi_0$ is constant
in each physical segment. For the present geometry, $\varphi_0=0$ in odd physical
segments and $\varphi_0=-\pi/2$ in even physical segments. 
Functions $u_0$ and $w_0$
are also easy to describe. The accuracy depends on the number of computational
segments per physical segments. The whole spring with a zig-zag centerline
can be handled by one element and the number of global degrees of freedom
remains low (equal to 2 if the loading process is controlled by applied horizontal displacement at the left support) and independent of the number of physical segments in the spring.
Since the function that describes the distribution of bending moments is not smooth
(for small changes of the initial geometry, it is piecewise linear), it would be
very hard to approximate the resulting rotation function by a polynomial.
Therefore, the approach of Saje~\cite{Saje1991}, described in Section~\ref{sec:treat}, would in this case require discretization
into elements that correspond to individual physical segments, leading to
many global degrees of freedom. Of course, the example is somewhat artificial,
but it is not totally meaningless and illustrates the high flexibility of our approach.
}

%%%%%%%%%%%%%%%%%%%%%%%%%%%%%%%%%%%%%%%%%%%%%%%%%%%%
%\newpage
\section{Concluding remarks}
%%%%%%%%%%%%%%%%%%%%%%%%%%%%%%%%%%%%%%%%%%%%%%%%%%%%

As illustrated by the examples, the formulation developed in this paper can describe curved elastic beams under large displacements and rotations with high accuracy. In summary, the main idea is that the equilibrium equations are used in their integrated form (\ref{e149x})--(\ref{e150x}), and they are combined 
with the geometrically exact kinematic relations (\ref{mj142x})--(\ref{mj144x})
and sectional equations \color{black}{(\ref{mj140})--(\ref{mj141})}. 
The resulting set of three first-order ordinary differential equations is then numerically approximated by an explicit finite difference scheme and the boundary value problem is converted to an initial value problem using the shooting method. 
On the global (structural) level, the governing equations are assembled in the same way as for a standard two-noded beam element with six degrees of freedom. In this sense, it plays the same role as traditional finite elements in the context of structural analysis.  The advantage is that accuracy of the numerical approximation can be conveniently increased by refining the integration scheme on the element level while the number of global degrees of freedom is kept constant.

The specific formulation presented in this paper is based on a number of simplifying
assumptions, some of which could be generalized.
From the kinematic point of view, the formulation has been developed for curved planar beams assuming the validity of
the Navier-Bernoulli hypothesis.
An extension to 
shear-flexible beams would be relatively straightforward. 
One would need to include the shear force
among the internal forces and link it to the
sectional shear distortion. This would not
substantially affect the structure of the
governing equations. A much more challenging task would be an extension to three dimensions,
which would require a major change of the 
descriptor that characterizes the sectional rotation. 

The initial shape of the curved beam that can be considered by the present formulation is virtually arbitrary. It is
described by functions that relate the
position of each centerline point and the inclination of the corresponding section 
to the arc-length coordinate. For some shapes,
most notably for the circular shape, such
functions are available in closed form.
If this is not the case, a table of values
of these functions can be precomputed numerically. Alternatively, for flat elements, one can use the projection onto a straight line as a replacement of the arc-length
coordinate, with the slight drawback that the
integration segments along the centerline
then do not have a constant length.

Since accuracy can be efficiently improved
by increasing the number of integration segments into which the element is divided,
the size of the element does not need to be
reduced and the number of global degrees of 
freedom can be kept low and independent of the refinement level. There is of course a limitation stemming from the presently used 
assumption that the loading is applied
exclusively at joints that connect the elements while individual element are not loaded
at their intermediate sections. However,
this assumption has been introduced only for
simplicity and it
could easily be removed. In the integrated
form of the equilibrium equations, the effect
of loads applied directly on the element
can be incorporated. This is conceptually easy
and the extended implementation only requires a more refined 
specification and processing of the input data. 

\color{black}{A very attractive feature of the present formulation is that it does not suffer from
membrane locking. This is documented in
examples in Sections~\ref{sec:4.1}--\ref{sec:ex431}. Convergence to the exact solution is found to be very regular, the results are meaningful even for relatively coarse discretizations, and
no parasitic oscillations of internal forces
are detected.}

From the constitutive point of view, the presented simple version of the method is based on the assumption of linear elasticity. 
The expressions for internal forces (normal force and bending moment) as functions of the
sectional deformation variables (axial stretch or strain and curvature) are then linear and
easily invertible. Still, for a curved beam,
it is interesting to note that the sectional
equations are coupled, in contrast to the case
of a straight beam. This phenomenon and its 
consequences are \color{black}{illustrated}  by
an example in Section~\ref{sec:unfolding}
\color{black}{and analyzed in detail in Appendix~\ref{appA}}.
The difference as compared to the \color{black}{simplified}
uncoupled equations is negligible for very
slender beams but it may play an important
role if, for instance, a thick-walled cylinder
is analyzed using a circular beam model.

The constitutive description can be generalized to nonlinear elasticity \color{black}{(as outlined in Appendix~\ref{app:a2})} and
even to inelastic behavior, e.g., to plasticity.
Of course, the computational demands are then
increased, because the inverted form of the
relation between internal forces and deformation variables can hardly be described
analytically and numerical evaluation,
involving an iterative solution of a set of two nonlinear algebraic equations, needs
to be adopted. It is fair to admit that
the method would fail if this relation
becomes non-invertible, e.g., due to 
softening in the moment-curvature diagram. 
Such cases would lead to localization
phenomena and would need to be handled
by adaptively introducing generalized inelastic hinges.

\color{black}{Certain theoretical aspects that are not necessary for proper understanding of the present approach but could be of general interest are discussed in detail in Appendix~\ref{appA}.
In particular, the appendix presents a rigorous
analysis of the structure of sectional equations
for curved beams. The present approach, with 
sectional equations consistently derived from
a uniaxial hyperelastic stress-strain law, 
is compared to the reduced form of the equations
used for description of shells by Simo and Fox\cite{Simo1989}, which were derived from 
a potential postulated directly in terms of the
deformation variables on the sectional level.
It is shown that such a potential cannot have
an arbitrary form and that, if it is derived
consistently, the definition of the effective
membrane stress resultant used by Simo and Fox\cite{Simo1989} needs to be modified.
}

\section*{Acknowledgments}
The authors are grateful for the support of the Czech Science Foundation (project No.\ 19-26143X). The authors would also like to thank A. Borkovi{\'c} for providing data for the load-displacement curves of parabolic arches calculated by the IGA formulation \cite{borkovic2022geometrically}.

\subsection*{Conflict of interest}
The authors declare no potential conflict of interests.

\bibliography{bibliography}%

\appendix

\color{black}{
\section{Structure of sectional equations for curved beams}
\label{appA}

\subsection{Brief summary of the origin of sectional equations adopted in this paper}

In general, sectional equations link the internal forces
to variables that characterize the deformation of an infinitesimal
beam segment. In our case (planar Euler-Bernoulli beam), the relevant internal forces are the
normal force, $N$, and the bending moment, $M$, which have a clear
physical meaning. On the other hand, the deformation variables
can be selected in different ways, but they should characterize the
processes of axial stretching and bending. 
Perhaps the most natural choice is to consider the centerline stretch
$\lambda_s$, defined in (\ref{eq:lambdas}), 
and the curvature $\kappa$, defined as the derivative 
of the 
sectional rotation $\varphi$ with respect to the initial centerline length, i.e., as $\kappa=\varphi'$. Using these variables,
one can conveniently describe the distribution of normal stretch
$\lambda$ across the section by formula (\ref{eq:lambda}), which also contains
the initial curvature, $\kappa_0$. 

It is worth noting that formula (\ref{eq:lambda}) directly follows from the basic kinematic
assumptions that are behind equations (\ref{equ0})--(\ref{eqw0}) and (\ref{equ})--(\ref{eqw}). The stretch, $\lambda$, is understood here
as the ratio of the current and initial lengths of an infinitesimal
fiber parallel to the centerline. Since shear deformation is neglected
and individual material fibers parallel to the centerline are 
expected to be in a state of uniaxial stress, it is perfectly justified
to assume that the density of strain energy (per unit initial volume),
${\cal E}_{int}$,
is a function of the stretch. 
The resulting expression (\ref{eq:deltaEint}) for the strain energy variation shows that, at the sectional level, the normal force is work-conjugate with
the centerline stretch, and the bending moment is work-conjugate 
with the curvature (understood in the present sense, i.e.,
as the derivative of sectional rotation with respect to the initial
centerline length). The stress $\sigma$ that appears in the 
integral definitions of internal forces
as stress resultants, eqs.~(\ref{eq:25})--(\ref{eq:26}), 
is understood as the quantity that is work-conjugate with the
stretch at the fiber level.

In the special case of a quadratic expression for strain energy density
(in terms of the Biot strain $\eps=\lambda-1$), the corresponding material law
that links the stress to the stretch (or to the Biot strain) is linear,
which then leads to a linear form of the sectional equations (\ref{eq:n})--(\ref{eq:m}).  In contrast to what would be
obtained for an initially straight beam, these equations are
coupled, in the sense that the normal force depends not only on the axial strain
but also on the change of curvature, and the bending moment depends
not only on the change of curvature but also on the axial strain.

\subsection{Extension to other hyperelastic material laws}\label{app:a2}

For general hyperelasticity, the strain energy density
does not need to be quadratic in terms of Biot strain.
The description of stretch distribution across the section by formula (\ref{eq:lambda}) as well as the definitions of internal forces
as stress resultants given by (\ref{eq:25})--(\ref{eq:26}) remain valid,
independently of the constitutive model. One only needs to adjust
the material law that links the stress to the stretch, and then
derive the corresponding sectional equations. Often, the integrals in
(\ref{eq:25})--(\ref{eq:26}) would have to be evaluated numerically.

\subsubsection{St.~Venant-Kirchhoff model}

It is instructive to look at one of the special cases that permit
analytical evaluation. Let us consider the 
St.~Venant-Kirchhoff model, which postulates a linear relationship
between the Green-Lagrange strain and its work-conjugate stress,
which is the second Piola-Kirchhoff stress. This model is obtained
by defining the strain energy density as a quadratic function
of the Green-Lagrange strain $\eps_{GL}=(\lambda^2-1)/2$.
From
\beq \label{a:1}
{\cal E}_{int} = \half E \eps_{GL}^2 = \frac{E}{8}  (\lambda^2-1)^2
\eeq 
we get 
\beq\label{e:a2} 
\sigma = \frac{\dd{\cal E}_{int}}{\dd\lambda} = \frac{E}{2}  (\lambda^2-1)\lambda = \frac{E}{2} (\lambda^3-\lambda) 
\eeq 
in which $\sigma$ is the first Piola-Kirchhoff stress (note that the second Piola-Kirchhoff stress 
would be obtained by differentiating ${\cal E}_{int}$ with respect to $\eps_{GL}$ and would be given by $E\eps_{GL}$). 
The 
internal forces are then evaluated as
\bea \nonumber
N&=&\int_A \sigma\,\dA = \frac{E}{2}\int_A \left(\frac{(\lambda_s+z\kappa)^3}{(1+z\kappa_0)^3}-\frac{\lambda_s+z\kappa}{1+z\kappa_0}\right)\,\dA
= \\
\nonumber
&=&\frac{EA_3}{2} \left(2\eps_s+3\eps_s^2+\eps_s^3\right) +\frac{ES_3}{2}\left(\kappa_0\eps_s(4+3\eps_s)+\left(2+6\eps_s+3\eps_s^2\right)\Delta\kappa\right)+\\
&&+\frac{EI_3}{2}\left(\left(2\kappa_0^2+6\kappa_0\Delta\kappa+3\Delta\kappa^2\right)\eps_s+4\kappa_0\Delta\kappa+3\Delta\kappa^2\right) + \frac{EJ_3}{2}\left(2\kappa_0^2\Delta\kappa+3\kappa_0\Delta\kappa^2+\Delta\kappa^3\right)
\label{e:a3}
\\
\nonumber
M&=&\int_A z\sigma\,\dA = \frac{E}{2}\int_A \left(\frac{z(\lambda_s+z\kappa)^3}{(1+z\kappa_0)^3}-\frac{z\lambda_s+z^2\kappa}{1+z\kappa_0}\right)\,\dA = \\
\nonumber
&=& \frac{ES_3}{2} \left(2\eps_s+3\eps_s^2+\eps_s^3\right) +\frac{EI_3}{2}\left(\kappa_0\eps_s(4+3\eps_s)+\left(2+6\eps_s+3\eps_s^2\right)\Delta\kappa\right)+\\
&&+\frac{EJ_3}{2}\left(\left(2\kappa_0^2+6\kappa_0\Delta\kappa+3\Delta\kappa^2\right)\eps_s+4\kappa_0\Delta\kappa+3\Delta\kappa^2\right) +\frac{EK_3}{2}\left(2\kappa_0^2\Delta\kappa+3\kappa_0\Delta\kappa^2+\Delta\kappa^3\right)
\label{e:a4}
\eea 
in which
\bea \label{eq:a3}
A_3 &=& \int_A \frac{1}{(1+z\kappa_0)^3}\,\dA \\
S_3 &=& \int_A \frac{z}{(1+z\kappa_0)^3}\,\dA \\
I_3 &=& \int_A \frac{z^2}{(1+z\kappa_0)^3}\,\dA \\
J_3 &=& \int_A \frac{z^3}{(1+z\kappa_0)^3}\,\dA \\
K_3 &=& \int_A \frac{z^4}{(1+z\kappa_0)^3}\,\dA \label{eq:k3}
\eea 
are modified sectional characteristics. These five quantities
are linked by the relations $A_3+3\kappa_0S_3+3\kappa_0^2I_3+\kappa_0^3J_3=A$
and $S_3+3\kappa_0I_3+3\kappa_0^2J_3+\kappa_0^3K_3=0$,
and so only three of them are independent. For a straight beam,
they reduce to $A_3=A$, $S_3=0$, $I_3=I$, $J_3=\int_A z^3\dA$ and $K_3=\int_A z^4\dA$. 

The internal forces are now nonlinear functions of the 
(Biot) axial strain, $\eps_s$, and the curvature change,
$\Delta\kappa$. Inversion of the nonlinear sectional equations (\ref{e:a3})--(\ref{e:a4})
in closed form is not possible, but the deformation variables
that correspond to a given combination of internal forces $N$ and $M$
can, at least within a certain range, be evaluated numerically
by the Newton-Raphson method. 

The main reason why we present the explicit form of these more complicated sectional equations (\ref{e:a3})--(\ref{e:a4}) is that they can now be compared
with the linear sectional equations (\ref{eq:n})--(\ref{eq:m}). The common feature
is that, in both cases, the equations are coupled. For instance,
the normal force depends not only on the axial strain but also
on the change of curvature. For the linear sectional equations,
it is possible to find a transformed quantity, $N+\kappa_0 M$,
which is linked exclusively to the centerline strain and independent
of the curvature change. However, this special property is closely related
to the choice of the constitutive model and cannot be considered
as general. To see that, it is sufficient to realize that
\beq \label{e:a10}
N+\kappa_0 M = \int_A\sigma\,\dA + \kappa_0\int_A z\sigma\,\dA = 
\int_A (1+z\kappa_0)\,\sigma\,\dA
\eeq
If the stress-strain law (\ref{eq:stress-strain}) is used, the resulting stress
distribution is given by (\ref{eq:29}), with $1+z\kappa_0$ in the denominator,
and expression (\ref{e:a10}) yields a very simple result:
\beq 
\int_A (1+z\kappa_0)\sigma\,\dA = \int_A (1+z\kappa_0)\,E\,\frac{\eps_s+z\Delta\kappa}{1+z\kappa_0}\,\dA =
E\int_A(\eps_s+z\Delta\kappa)\,\dA = EA\eps_s
\eeq
In contrast to that, for the stress-strain law given by (\ref{e:a2}),
the integrand does not simplify that much, and the result still depends on $\Delta\kappa$:
\bea
\int_A (1+z\kappa_0)\sigma\,\dA &=& 
 \int_A  (1+z\kappa_0)\,\frac{E}{2}\left(\frac{(\lambda_s+z\kappa)^3}{(1+z\kappa_0)^3}-\frac{\lambda_s+z\kappa}{1+z\kappa_0}\right)\,\dA
 = \frac{E}{2} \int_A  \left(\frac{(\lambda_s+z\kappa)^3}{(1+z\kappa_0)^2}-\lambda_s-z\kappa\right)\,\dA = \\
 &=& \frac{E}{2} \int_A  \frac{(\lambda_s+z(\kappa_0+\Delta\kappa))^3}{(1+z\kappa_0)^2}\,\dA -\frac{EA}{2}\lambda_s
\eea
One can also verify that if the expression on the right-hand side
of (\ref{e:a4}) is multiplied by $\kappa_0$ and added to the expression
on the right-hand side of (\ref{e:a3}), terms with $\Delta\kappa$
do not cancel out. This fact will be used later on in the discussion
of the differences between the present modeling approach and 
the framework used by Simo and Fox\cite{Simo1989} for shells.

It is interesting to note that if all nonlinear terms 
in equations (\ref{e:a3})--(\ref{e:a4}) are neglected, the equations
reduce to
\bea 
N&=& \left(EA_3+2\kappa_0ES_3+\kappa_0^2EI_3\right)\eps_s
+ \left(ES_3+2\kappa_0EI_3+\kappa_0^2EJ_3\right)\Delta\kappa
\label{e:a3x}
\\
M&=&\left(ES_3+2\kappa_0EI_3+\kappa_0^2EJ_3\right)\eps_s
+ \left(EI_3+2\kappa_0EJ_3+\kappa_0^2EK_3\right)\Delta\kappa
\label{e:a4x}
\eea 
This form of sectional equations linearized around the initial state is equivalent to the linear sectional
equations (\ref{eq:n})--(\ref{eq:m}). To see that, it is sufficient to realize
that $A_3+2\kappa_0S_3+\kappa_0^2I_3=A_{\kappa_0}$,
$S_3+2\kappa_0I_3+\kappa_0^2J_3=S_{\kappa_0}$ and
$I_3+2\kappa_0J_3+\kappa_0^2K_3=I_{\kappa_0}$,
as follows from definitions (\ref{eq:a3})--(\ref{eq:k3}) and (\ref{e:A0})--(\ref{e:I0}). In the next subsection, we will show that
the same linearized form  is obtained for sectional equations derived from any uniaxial hyperelastic stress-strain law, i.e., not only from the presently considered  St.~Venant-Kirchhoff law.

\subsubsection{General hyperelasticity}

A certain drawback of uniaxial stress-strain laws that follow
from the assumed quadratic expression for strain energy density in terms
of Biot strain or Green-Lagrange strain is that they fail to describe
the extreme compression limit, $\lambda\to 0^+$. The energy density
remains finite and the stress as well. For the St.~Venant-Kirchhoff model, the first Piola-Kirchhoff stress even tends to zero in this limit. 
This deficiency is not so critical for applications to slender beams,
which usually buckle in compression and large compressive strains
are not really attained. If needed, a more realistic behavior
in extreme compression can be obtained if the strain energy
density is taken as quadratic in terms of the logarithmic strain,
i.e., if we set
\beq 
{\cal E}_{int} = \half E\, (\ln\lambda)^2
\eeq 
The first Piola-Kirchhoff stress is then given by
\beq 
\sigma = \frac{\dd{\cal E}_{int}}{\dd\lambda} = E\,\frac{\ln\lambda}{\lambda}
\eeq 
This tends to minus infinity as $\lambda\to 0^+$, i.e., in extreme compression. On the other hand,
in tension the first Piola-Kirchhoff stress attains its maximum
at $\lambda={\rm e}$ and afterwards decreases to zero. 

Of course, one could further improve the material description by constructing
a more appropriate (and presumably more complicated) expression for
strain energy density, but it is also possible to postulate
directly the function $\strf$ that links the first Piola-Kirchhoff stress
to the stretch. For such a general elastic material law
\beq 
\sigma = \strf(\lambda)
\eeq 
the corresponding sectional equations read
\bea 
N &=& \int_A \strf\left(\frac{\lambda_s+z\kappa}{1+z\kappa_0}\right)\dx = \int_A \strf\left(1+\frac{\eps_s+z\,\Delta\kappa}{1+z\kappa_0}\right)\dx
\\
M &=& \int_A z\,\strf\left(\frac{\lambda_s+z\kappa}{1+z\kappa_0}\right)\dx = \int_A z\,\strf\left(1+\frac{\eps_s+z\,\Delta\kappa}{1+z\kappa_0}\right)\dx
\eea 
 Differentiating the internal forces
with respect to the deformation variables $\eps_s$ and $\Delta\kappa$,
we obtain the tangent sectional stiffnesses
\bea 
\frac{\partial N}{\partial \eps_s} = \int_A \frac{1}{1+z\kappa_0}\strf_{,\lambda}\left(1+\frac{\eps_s+z\,\Delta\kappa}{1+z\kappa_0}\right)\dx
\hskip 5mm && \hskip 5mm 
\frac{\partial N}{\partial \Delta\kappa} = \int_A \frac{z}{1+z\kappa_0}\strf_{,\lambda}\left(1+\frac{\eps_s+z\,\Delta\kappa}{1+z\kappa_0}\right)\dx
\\
\frac{\partial M}{\partial \eps_s} = \int_A \frac{z}{1+z\kappa_0}\strf_{,\lambda}\left(1+\frac{\eps_s+z\,\Delta\kappa}{1+z\kappa_0}\right)\dx
\hskip 5mm && \hskip 5mm 
\frac{\partial M}{\partial \Delta\kappa} = \int_A \frac{z^2}{1+z\kappa_0}\strf_{,\lambda}\left(1+\frac{\eps_s+z\,\Delta\kappa}{1+z\kappa_0}\right)\dx
\eea 
Here, $\strf_{,\lambda}\equiv \dd\strf/\dd\lambda$ is the tangent
material modulus. Its value at $\lambda=1$ is the standard elastic
modulus, $E$.

The main point to be made here is that no matter which specific form
of the material law is used, the coupling effect is always present,
already in the small-strain range.
Indeed, the cross-coupling stiffness, $\partial N/\partial \Delta\kappa \equiv \partial M/\partial\eps_s$, vanishes only exceptionally, e.g., when
$\kappa_0=0$ and $\strf_{,\lambda}$ is constant, which corresponds
to a straight beam and linear material law (\ref{eq:stress-strain}).
In the initial undeformed state, we have $\eps_s=0$ and $\Delta\kappa=0$,
and the cross-coupling stiffness is
\beq 
\frac{\partial N(0,0)}{\partial \Delta\kappa}  = \frac{\partial M(0,0)}{\partial \eps_s} =\int_A \frac{z}{1+z\kappa_0}\strf_{,\lambda}\left(1\right)\dx = \int_A \frac{Ez}{1+z\kappa_0}\dx = ES_{\kappa 0}
\eeq 
In a similar fashion, one gets ${\partial N(0,0)}/{\partial \eps_s}  = EA_{\kappa 0}$ and ${\partial M(0,0)}/{\partial \Delta\kappa}  = EI_{\kappa 0}$. This means that even if the material law is nonlinear, 
the sectional equations linearized  around the initial state
always have the form (\ref{eq:n})--(\ref{eq:m}) and thus are coupled,
provided that the beam is curved.

Material nonlinearity cannot eliminate the coupling, and it can even
introduce coupling for straight beams. For $\kappa_0=0$ (but a general
deformation state), the
cross-coupling sectional stiffness reduces to
\beq \label{a:22}
\frac{\partial N(\eps_s,\Delta\kappa)}{\partial \Delta\kappa}  = \frac{\partial M(\eps_s,\Delta\kappa)}{\partial \eps_s} =\int_A z\,\strf_{,\lambda}\left(1+\eps_s+z\,\Delta\kappa\right)\dx
\eeq 
This vanishes if the tangent modulus $\strf_{,\lambda}=$~const., but not in general. 
For instance, for the St.~Venant-Kirchhoff model we have
$\strf_{,\lambda}(\lambda)=E(3\lambda^2-1)/2$, and substitution into (\ref{a:22}) leads to 
\beq 
\int_A z\,\strf_{,\lambda}\left(1+\eps_s+z\,\Delta\kappa\right)\dx
= \frac{E}{2}\int_A z\, [3(1+\eps_s+z\,\Delta\kappa)^2-1]\,\dA
= \frac{E}{2} 3\cdot 2\cdot (1+\eps_s)\Delta\kappa\int_Az^2\dA = 3EI(1+\eps_s)\Delta\kappa
\eeq 
Here, the coupling effect vanishes as long as $\Delta\kappa=0$,
i.e., the beam remains straight.  However, as the beam gets curved,
the coupling effect gradually builds up. 

\subsection{Comparison with the framework used by Simo and Fox for shells}

Simo and Fox\cite{Simo1989} proposed a framework for geometrically exact modeling of shells
based on systematic tensorial  description of basic quantities and
governing equations. Among other developments, they introduced the 
concept of the {\it effective stress resultants}, which were
obtained from the specific normal and shear forces by adding a certain linear 
combination of specific moments. The main objective was to construct
internal forces that can be linked by sectional equations
exclusively to the in-plane deformation variables, with the effect
of bending eliminated. It is interesting to check how the Simo-Fox 
approach would reduce to the case of a two-dimensional Euler-Bernoulli
beam studied here, and whether the effective membrane stress resultant
corresponds to our $N+\kappa_0 M$, which was found to be dependent 
exclusively on the axial strain in the special case of linear sectional
equations (\ref{eq:n})--(\ref{eq:m}).

Simo and Fox\cite{Simo1989} worked with a general shape of the
shell midsurface and described it using curvilinear coordinates 
and the corresponding surface convected frame, plus an additional
director orthogonal frame. For comparison with the two-dimensional
curved beam, it is sufficient to consider a cylindrical surface.
Moreover, we assume that the section remains perpendicular to the
deformed centerline, and so the director vector of the corresponding
shell remains perpendicular to the deformed midsurface. The first curvilinear coordinate can be selected as our $x$, measured as the distance along the initial centerline, and the second one as the
out-of-plane coordinate $y$. The surface convected frame is then
orthogonal, initially even orthonormal; the first base vector
stretches with the centerline and its norm is $\lambda_s$, the second 
remains a unit out-of-plane vector, and the third coincides with
the unit director vector perpendicular to the midsurface.

Considering the curved beam (with a rectangular cross section) 
as a cylindrical shell strip of width $b_s$ and using the special
choice of curvilinear coordinates described above, we can reduce
the expression for stress power presented in eq.~(5.1) in Simo and Fox\cite{Simo1989} to
\beq \label{a:24}
{\cal W} = \int_0^L b_s(n^{11} \lambda_s\dot{\lambda}_s + \tilde{m}^{11}\lambda_s\dot{\kappa})\lambda_s\, \dx
\eeq 
where
\bea \label{a:25}
n^{11} &=& \frac{N}{b_s\lambda_s^2} \\ \label{a:26}
\tilde{m}^{11} &=& \frac{M}{b_s\lambda_s^2}
\eea 
are stress resultant components used by Simo and Fox\cite{Simo1989},
evaluated from their equations (4.7), (4.11) and (4.22).
Fractions $N/b_s$
and $M/b_s$ are identified as the specific normal force and specific
bending moment, and the additional scaling by $1/\lambda_s^2$ is related
to the fact that the first base vector is not normalized and its
norm is $\lambda_s$. Substituting (\ref{a:25})--(\ref{a:26}) into
(\ref{a:24}), we can rewrite the stress power expression as
\beq \label{a:27}
{\cal W} = \int_0^L b_s\left(\frac{N}{b_s\lambda_s^2} \lambda_s\dot{\lambda}_s + \frac{M}{b_s\lambda_s^2}\lambda_s\dot{\kappa}\right)\lambda_s\, \dx =
\int_0^L (N \dot{\lambda}_s + M\dot{\kappa})\, \dx
\eeq
This is consistent with our previously derived equation (\ref{eq:deltaEint}), which expressed virtual work instead of stress
power, and so the rates were replaced by virtual changes. 
%Another formal difference is that $\varphi'$ was denoted as $\kappa$. 

In the present paper, we have decided to define one of the
deformation variables as $\kappa=\varphi'$, because this quantity
is work-conjugate with the bending moment, as seen in (\ref{eq:deltaEint}). We refer to this variable as the ``curvature'',
even though it is equal to the reciprocal value of the radius
of curvature only in the initial state, where $\kappa_0=\varphi_0'=1/R_0$.
In the deformed state, if the centerline segment is on a circle
of radius $R$, has length $\lambda_s \,\dx$ and the difference
in rotation of the right and left end is $\dd\varphi$,
elementary geometry considerations lead to the relation $R\,\dd\varphi=\lambda_s \,\dx$, from which $R\varphi'=\lambda_s$.
The true curvature is thus given by $k=1/R=\varphi'/\lambda_s$.
However, Simo and Fox\cite{Simo1989} used yet another measure
of curvature, defined as 
\beq \label{eq:kap11}
\kappa_{11}=\lambda_s\varphi'=\lambda_s\kappa
\eeq 
combined with a measure of membrane strain,
defined as 
\beq \label{eq:eps11}
\eps_{11}=(\lambda_s^2-1)/2
\eeq 
which is easily
recognized as the Green-Lagrange strain. 
The rates of the deformation variables $\eps_{11}$ and $\kappa_{11}$
are 
\bea 
\dot{\eps}_{11} &=& \lambda_s\dot{\lambda}_s
\\
\dot{\kappa}_{11} &=& \dot{\lambda_s}\kappa + \lambda_s \dot{\kappa}
\eea 
and so the expression in parentheses inside the integral in (\ref{a:24})
can be transformed as follows:
\beq 
n^{11} \lambda_s\dot{\lambda}_s + \tilde{m}^{11}\lambda_s\dot{\kappa} = n^{11} \dot{\eps}_{11} + \tilde{m}^{11}(\dot{\kappa}_{11}-\kappa\dot{\lambda}_s) = 
n^{11} \dot{\eps}_{11} + \tilde{m}^{11}(\dot{\kappa}_{11}-\dot{\eps}_{11}\kappa/\lambda_s) = (n^{11}- \tilde{m}^{11}\kappa/\lambda_s)\,\dot{\eps}_{11} + \tilde{m}^{11}\dot{\kappa}_{11}
\label{a:30}
\eeq
Consequently, if we define the effective membrane stress resultant
\beq \label{a:31}
\tilde{n}^{11} = n^{11}- \frac{\kappa}{\lambda_s}\tilde{m}^{11}
\eeq 
then the stress power can be expressed as
\beq \label{a:32} 
{\cal W}  = \int_0^L b_s(\tilde{n}^{11}\dot{\eps}_{11} + \tilde{m}^{11}\dot{\kappa}_{11})\lambda_s\, \dx
\eeq 
This means that quantities $\tilde{n}^{11}$ and $\tilde{m}^{11}$ are
work-conjugate with deformation measures $\eps_{11}$ and $\kappa_{11}$.
In fact, since the curvature does not vanish in the initial state,
it is preferable to use the curvature increment $\Delta\kappa_{11}=\kappa_{11}-\kappa_{0,11}$
as the deformation variable (this quantity was denoted as $\rho_{11}$ in Simo and Fox\cite{Simo1989}, but we prefer $\Delta\kappa_{11}$, to avoid confusion with the density).
Of course, by $\kappa_{0,11}$ we mean the value of $\kappa_{11}$ in the initial state.
Note that the factor $\kappa/\lambda_s$ in (\ref{a:31}) represents the true
curvature, $k$. 

The constitutive description used by Simo and Fox\cite{Simo1989} was based on the assumption
that there exists a stored energy function $\psi$ that depends on the deformation variables,
in our case only on $\eps_{11}$ and $\Delta\kappa_{11}$, and represents the strain energy
per unit mass. Since the deformation variables already characterize the strain state
in the whole elementary segment, the strain energy of the shell would be evaluated by
integrating over the midsurface, and for the 2D beam this reduces to integration over
the centerline:
\beq\label{a:33} 
E_{int} = \int_0^L b_sh_s\rho_0\psi\,\dx
\eeq 
The product $b_sh_s$ would in general correspond to the sectional area, but since we consider
a beam model that represents a strip of a cylindrical shell, we insert right away
the expression valid for a rectangular section. Symbol $\rho_0$ represents the initial mass density
(per unit volume), and so the product $b_sh_s\rho_0$ is the mass per unit length of the
centerline. Differentiation of (\ref{a:33}) with respect to time 
leads to the rate of the beam strain energy 
\beq 
\dot{E}_{int} = \int_0^L b_sh_s\rho_0\left(\frac{\partial\psi}{\partial\eps_{11}}\dot{\eps}_{11} +\frac{\partial\psi}{\partial\Delta\kappa_{11}}\Delta\dot{\kappa}_{11}\right) \,\dx
\eeq 
and when this is compared with (\ref{a:32}), the constitutive equations are obtained
in the form
\bea\label{a:35} 
\tilde{n}^{11} &=& \frac{h_s\rho_0}{\lambda_s}\frac{\partial\psi}{\partial\eps_{11}}\\
\tilde{m}^{11} &=& \frac{h_s\rho_0}{\lambda_s}\frac{\partial\psi}{\partial\Delta\kappa_{11}}
\label{a:36}
\eea 
In eq.~(5.18) in Simo and Fox\cite{Simo1989}, these relations were written as
\bea \label{a:37}
\tilde{n}^{11} &=& \bar{\rho}\frac{\partial\psi}{\partial\eps_{11}}\\
\tilde{m}^{11} &=& \bar{\rho}\frac{\partial\psi}{\partial\Delta\kappa_{11}}\label{a:38}
\eea 
where $\bar{\rho}$ is the mass density per unit surface area in the deformed state,
which can be for a beam segment evaluated as
\beq 
\bar{\rho} = \frac{\int_A \rho_0(1+z\kappa_0)\,\dA\,\dx }{b_s \lambda_s\,\dx} = \frac{ \rho_0 A\,\dx }{b_s \lambda_s\,\dx} = \frac{h_s\rho_0}{\lambda_s}
\eeq 
This confirms that equations (\ref{a:35})--(\ref{a:36}) and (\ref{a:37})--(\ref{a:38})
are indeed equivalent. 

An important difference between our approach and the framework used by Simo and Fox\cite{Simo1989} is that we start from constitutive equations at the material point level
and construct the sectional equations by incorporating the kinematic assumptions
that link strain to the sectional deformation variables, while Simo and Fox\cite{Simo1989}
start on the sectional level and postulate the expression for stored energy density $\psi$
directly in terms of the sectional deformation variables. The resulting sectional
equations are equivalent only if the expression for $\psi$ is consistently derived
instead of arbitrarily postulated. Indeed, if we consider the strain energy density
(per unit volume) as a given function ${\cal E}_{int}$ of the stretch and combine this
with the kinematic assumptions that lead to the distribution of stretch described by (\ref{eq:lambda}),
substitution of (\ref{eq:lambda}) into  (\ref{eq:strainenergy}) gives 
\beq \label{eq:strainenergy2}
E_{int} = \int_0^L \int_A (1+z\kappa_0)\,
\Eint\left(\frac{\lambda_s+z\kappa}{1+z\kappa_0}\right) \,\dA\,\dx
\eeq 
This is equivalent to  (\ref{a:33}), provided that function $\psi$ is defined as
\beq \label{eq:psi}
\psi = \frac{1}{b_sh_s\rho_0}\int_A (1+z\kappa_0)\,
\Eint\left(\frac{\lambda_s+z\kappa}{1+z\kappa_0}\right) \,\dA\,
=\frac{1}{b_sh_s}\int_A (1+z\kappa_0)\,
\psi_0\left(\frac{\lambda_s+z\kappa}{1+z\kappa_0}\right) \,\dA\,
\eeq 
in which $\psi_0={\cal E}_{int}/\rho_0$ is introduced only for convenience
and represents the specific strain energy (i.e., energy per unit mass) of our original
material model. 
The expression on the right-hand side of (\ref{eq:psi}) can be interpreted as
averaging of the specific strain energy over the section, however, instead of uniform
averaging, a certain weight function dependent on the initial curvature is used.
Effectively, we average over the volume of the curved elementary segment.

In (\ref{eq:psi}), function $\psi$ is presented as dependent on deformation
variables $\lambda_s$ and $\kappa$, but it is no problem to transform the expression
into a function of any other two suitable chosen sectional deformation variables,
such as our $\eps_s$ and $\Delta\kappa$, or $\eps_{11}$ and $\Delta\kappa_{11}$ used
by Simo and Fox\cite{Simo1989}. For instance, if we adopt the model based on quadratic
energy density in terms of Biot strain, function $\psi_0(\lambda)$ is given by
$(E/2\rho_0)(\lambda-1)^2$ and formula (\ref{eq:psi}) yields
\beq \label{eq:psi2}
\psi = \frac{1}{b_sh_s}\int_A (1+z\kappa_0)\,
\frac{E}{2\rho_0}\left(\frac{\lambda_s+z\kappa}{1+z\kappa_0}-1\right)^2 \,\dA\,
= \frac{E}{2\rho_0 b_sh_s}\int_A (1+z\kappa_0)\left(\frac{\eps_s+z\Delta\kappa}{1+z\kappa_0}\right)^2\,\dA =
\frac{E}{2\rho_0 b_sh_s}\left(A_{\kappa_0}\eps_s^2 + 2 S_{\kappa_0}\eps_s\Delta\kappa + I_{\kappa_0}\Delta\kappa^2\right)
\eeq 
Based on (\ref{eq:kap11})--(\ref{eq:eps11}), the deformation variables used by Simo and Fox\cite{Simo1989} are 
\bea\label{a:45} 
\eps_{11} &=& \half (\lambda_s^2-1) = \half(\eps_s^2+2\eps_s) \\
\Delta\kappa_{11} &=& \lambda_s\kappa-\kappa_0 = \Delta\kappa + (\kappa_0+\Delta\kappa)\eps_s 
\label{a:46}
\eea 
and by inversion we obtain
\bea \label{a:47}
\eps_s &=& \sqrt{1+2\eps_{11}}-1 \\
\Delta\kappa &=& \frac{\Delta\kappa_{11}+\kappa_0\left(1-\sqrt{1+2\eps_{11}}\right)}{\sqrt{1+2\eps_{11}}}\label{a:48}
\eea
Substitution of these expressions into (\ref{eq:psi2}) leads, after omission of constant terms, to 
\bea\nonumber 
\psi &=&
\frac{E}{\rho_0 b_sh_s}\left(A_{\kappa_0}\left(\eps_{11}-\sqrt{1+2\eps_{11}}\right) +  S_{\kappa_0}\left(\Delta\kappa_{11}-\frac{2\kappa_0(1+\eps_{11})+\Delta\kappa_{11}}{\sqrt{1+2\eps_{11}}}\right)\right.+\\
&&\left.+I_{\kappa_0}\frac{(\kappa_0+\Delta\kappa_{11})^2/2-\kappa_0(\kappa_0+\Delta\kappa_{11})\sqrt{1+2\eps_{11}}}{1+2\eps_{11}}\right)
\label{eq:psi3}
\eea
This function of $\eps_{11}$ and $\Delta\kappa_{11}$ can now be used in  (\ref{a:37})--(\ref{a:38}),
where $\bar{\rho}$ needs to be replaced by 
\beq 
\bar{\rho} = \frac{h_s\rho_0}{\lambda_s}=\frac{h_s\rho_0}{\sqrt{1+2\eps_{11}}}
\eeq 
The resulting expression
for the effective membrane stress resultant is
\beq 
\tilde{n}^{11} = \frac{E}{b_s}\left(A_{\kappa_0}\frac{\sqrt{1+2\eps_{11}}-1}{1+2\eps_{11}}
+S_{\kappa_0} \left(\frac{2\kappa_0(1+\eps_{11})+\Delta\kappa_{11}}{(1+2\eps_{11})^2}-\frac{2\kappa_0}{1+2\eps_{11}}\right) 
+I_{\kappa_0}\left(\frac{\kappa_0(\kappa_0+\Delta\kappa_{11})}{(1+2\eps_{11})^2}-\frac{(\kappa_0+\Delta\kappa_{11})^2}{(1+2\eps_{11})^{5/2}}\right)
\right)
\eeq 
A considerably simpler expression is obtained for the specific bending moment:
\beq\label{a:52} 
\tilde{m}^{11} = \frac{E}{b_s}\left(S_{\kappa_0}\frac{\sqrt{1+2\eps_{11}}-1}{1+2\eps_{11}} 
+I_{\kappa_0}\left(\frac{\kappa_0+\Delta\kappa_{11}}{(1+2\eps_{11})^{3/2}}-\frac{\kappa_0}{1+2\eps_{11}}\right)
\right)
\eeq 
Recalling that the actual bending moment is $M=b_s\lambda_s^2\tilde{m}^{11}$
and taking into account that deformation variables $\eps_{11}$ and $\Delta\kappa_{11}$
are linked to ``our'' variables $\eps_s$ and $\Delta\kappa$ by equations (\ref{a:45})--(\ref{a:46}), we can easily check that equation  (\ref{a:52}) exactly
corresponds to the simple linear law (\ref{eq:m}) for the bending moment $M$
in terms of $\eps_s$ and $\Delta\kappa$. For the effective membrane stress resultant
$\tilde{n}^{11}$, the equivalence is more difficult to check because $\tilde{n}^{11}$
corresponds to a combination of $N$ and $M$, namely to  $(\lambda_sN-\kappa M)/(b_s\lambda_s^3)$. The comparison becomes easier if the ``original'' (i.e., not effective)
membrane stress resultant is expressed first:
\beq 
n^{11} = \tilde{n}^{11} + \frac{\kappa}{\lambda_s}\tilde{m}^{11} = \tilde{n}^{11} + \frac{\kappa_0+\Delta\kappa_{11}}{1+2\eps_{11}}\tilde{m}^{11} =  \frac{E}{b_s}\left(A_{\kappa_0}\frac{\sqrt{1+2\eps_{11}}-1}{1+2\eps_{11}}
+S_{\kappa_0} \left(\frac{\kappa_0+\Delta\kappa_{11}}{(1+2\eps_{11})^{3/2}}-\frac{\kappa_0}{1+2\eps_{11}}\right) 
\right)
\eeq
The resulting expression has the same structure as (\ref{a:52}), only with 
$A_{\kappa_0}$ replaced by $S_{\kappa_0}$ and $S_{\kappa_0}$ replaced by $I_{\kappa_0}$,
and so the equivalence to (\ref{eq:n}) is easy to verify.

We have shown that, for a given uniaxial hyperelastic stress-strain law, one can derive
the corresponding stored energy density function $\psi$ dependent on the sectional
deformation variables such that the sectional equations (\ref{a:37})--(\ref{a:38}) used by Simo and Fox\cite{Simo1989}
become equivalent to our sectional equations (\ref{eq:n})--(\ref{eq:m}).
The opposite is not true, i.e., not all choices of $\psi$ are consistent 
with a uniaxial material law. To see that, let us think of how the 
%procedure leading from
%(\ref{eq:strainenergy2}) to 
transformation performed in (\ref{eq:psi}) can be inverted, i.e., how the functional dependence of the strain energy density
${\cal E}_{int}$ on the stretch can be identified if the dependence of $\psi$ on the sectional
deformation variables is prescribed. 
To be specific, let us consider that $\psi=\hat{\psi}(\lambda_s,\kappa)$, even though the proposed
method would work for other
choices of independent deformation variables as well.
The key idea is that if $\lambda_s$ and $\kappa$ are
selected such that $\kappa=\lambda_s\kappa_0$, then the fraction in (\ref{eq:psi}) that
represents the stretch becomes independent of $z$, and the corresponding value of ${\cal E}_{int}$ can be taken out of the integral. The integral of $1+z\kappa_0$ then gives the sectional area, $A=b_sh_s$,
and the first equality in (\ref{eq:psi}) can be transformed into
\beq\label{a:eint} 
{\cal E}_{int} (\lambda_s) = \rho_0 \hat{\psi}(\lambda_s, \lambda_s\kappa_0)
\eeq 
This means that function ${\cal E}_{int}$ of one variable (the stretch) is uniquely determined
by the values of function $\psi$ on one single line in the plane of variables $\lambda_s$
and $\kappa$. Once ${\cal E}_{int}$ is extracted in this way, the values of $\hat{\psi}$
outside that special line are uniquely determined by (\ref{eq:psi}), and so they cannot
be chosen arbitrarily. Of course, by far not all possible choices of $\hat{\psi}$ satisfy this
constraint. 

Let us now return attention to the approach used by Simo and Fox\cite{Simo1989}.
The reason why they introduced the effective membrane stress resultant $\tilde{n}^{11}$
was that if $\eps_{11}$ and $\Delta\kappa_{11}$
are selected as the sectional deformation variables, then $\tilde{n}^{11}$ is
the static quantity that is work-conjugate with $\eps_{11}$. For stored energy density 
 $\psi$ considered as a function of $\eps_{11}$ and $\Delta\kappa_{11}$, the sectional
 equations have the form (\ref{a:37})--(\ref{a:38}). As the simplest formulation,
 Simo and Fox\cite{Simo1989} suggested to use 
 linear and decoupled sectional
 equations, which would be in the present notation written as
 \bea\label{a:57} 
 \tilde{n}^{11} &=& \rho Eh_s\eps_{11} \\
 \tilde{m}^{11} &=& \frac{\rho Eh_s^3}{12}\Delta\kappa_{11}
 \label{a:58} 
 \eea 
 This form of the equations would be obtained by transcribing the original 
 equation (5.19) from Simo and Fox\cite{Simo1989}, taking into account their
 equation (5.20) and setting the Poisson ratio to zero, to eliminate the
 difference between the beam and the cylindrical shell strip. However, equations
 (\ref{a:57})--(\ref{a:58}) are not dimensionally correct, because $\eps_{11}$
 is dimensionless, $E$ is the Young modulus in Pa, $h_s$ is the shell thickness
 (or beam depth) in m, $\tilde{n}^{11}$ is a force per unit width, in N/m,
 and $\rho$ is supposed to be the mass density per unit volume, in kg/m$^3$.
 The derivation of these equations was not presented in detail, only briefly described
 as an approximation based on an asymptotic expansion, and the corresponding
 choice of stored energy potential $\psi$ was not specified. One can only guess that
 the objective was to construct the equations in the simplest and decoupled form,
 which could be achieved by using the quadratic potential
 \beq \label{a:psi}
 \psi(\eps_{11},\Delta\kappa_{11}) = \half C_1 \eps_{11}^2 + \half C_2 \Delta\kappa_{11}^2
 \eeq 
 with the meaning of constants $C_1$ and $C_2$ yet to be identified. The corresponding
 sectional equations (\ref{a:37})--(\ref{a:38}) would read 
  \bea\label{a:59} 
 \tilde{n}^{11} &=& \bar{\rho}C_1\eps_{11} = h_s\rho_0 C_1\frac{\eps_{11}}{\sqrt{1+2\eps_{11}}} \\
 \tilde{m}^{11} &=& \bar{\rho}C_2\Delta\kappa_{11}=h_s\rho_0 C_2 \frac{\Delta\kappa_{11}}{\sqrt{1+2\eps_{11}}}
 \label{a:60} 
 \eea
 In this way, we would obtain an expression for $\tilde{n}^{11}$ exclusively in terms
 of $\eps_{11}$ (and given constants), while $\tilde{m}^{11}$ would depend not only
 on $\Delta\kappa_{11}$ but also slightly on $\eps_{11}$, even though the potential
 $\psi$ did not contain a mixed term with the product $\eps_{11}\Delta\kappa_{11}$. 
 
 Equations (\ref{a:59})--(\ref{a:60}) are nonlinear
 and  (\ref{a:60}) contains both $\eps_{11}$ and $\Delta\kappa_{11}$. This could be fixed by considering
 static quantities $\hat{n}=\lambda_s\tilde{n}^{11}$
 and $\hat{m}=\lambda_s\tilde{m}^{11}$ instead of
 $\tilde{n}^{11}$
 and $\tilde{m}^{11}$. The resulting sectional equations would
 be linear and decoupled. A more fundamental problem is
 related to the specific choice of quadratic potential
 (\ref{a:psi}), which is not consistent with any 
 hyperelastic uniaxial stress-strain law on the 
 material level. To see that, let us first exploit
 formula (\ref{a:eint}) and construct the strain energy density
 expression that {\it could be} (but actually is not)
 at the origin of the assumed quadratic potential
 (\ref{a:psi}):
 \beq\label{a:eintx} 
{\cal E}_{int} (\lambda_s) = \rho_0 \hat{\psi}(\lambda_s, \lambda_s\kappa_0) = \rho_0\psi\left(\half\left(\lambda_s^2-1\right),\kappa_0\left(\lambda_s^2-1\right)\right) = 
%\rho_0\half C_1 \frac{1}{4}\left(\lambda_s^2-1\right)^2 + \rho_0\half C_2 \kappa_0^2 \left(\lambda_s^2-1\right)^2 = 
\left(\frac{\rho_0C_1}{8}+\frac{\rho_0C_2\kappa_0^2}{2}\right)\left(\lambda_s^2-1\right)^2
\eeq 
Recalling (\ref{a:1}), we realize that this is the strain energy density of the St.~Venant-Kirchhoff material
with elastic modulus
\beq \label{a:61}
E = \rho_0C_1 + 4 \rho_0C_2\kappa_0^2
\eeq 
So far we have shown that if the quadratic potential 
(\ref{a:psi}) is derivable from a material-level strain energy 
expression, then the expression would need to have
the form (\ref{a:1}), with $E$ given by (\ref{a:61}).
However, for this particular strain energy density, 
formula (\ref{eq:psi}) gives
\bea\nonumber 
\hat{\psi}(\lambda_s,\kappa) &=& \frac{1}{b_sh_s\rho_0}\int_A (1+z\kappa_0)\,
\Eint\left(\frac{\lambda_s+z\kappa}{1+z\kappa_0}\right) \,\dA=\frac{E}{8b_sh_s\rho_0}\int_A (1+z\kappa_0)\,
\left(\left(\frac{\lambda_s+z\kappa}{1+z\kappa_0}\right)^2-1\right)^2 \,\dA=\\ \nonumber
&=& \frac{E}{8b_sh_s\rho_0}\int_A \frac{\left(\left(\lambda_s+z\kappa\right)^2-\left(1+z\kappa_0\right)^2\right)^2}{(1+z\kappa_0)^3} \,\dA
= \frac{E}{8b_sh_s\rho_0}\int_A \frac{\left(\lambda_s^2-1+2(\lambda_s\kappa-\kappa_0)z+(\kappa^2-\kappa_0^2)z^2\right)^2}{(1+z\kappa_0)^3} \,\dA= \\ \nonumber
&=& 
\frac{E}{8b_sh_s\rho_0}\left(A_3\left(\lambda_{s}^2-1\right)^2 +4S_3\left(\lambda_{s}^2-1\right)(\lambda_s\kappa-\kappa_0)+
2I_3 \left(2(\lambda_s\kappa-\kappa_0)^2+\left(\lambda_{s}^2-1\right)\left(\kappa^2-\kappa_{0}^2\right) \right) + \right.\\
&&\left.+4J_3(\lambda_s\kappa-\kappa_0)\left(\kappa^2-\kappa_{0}^2\right)+K_3\left(\kappa^2-\kappa_{0}^2\right)^2
\right)
\label{eq:psi4}
\eea
%Using expression (\ref{a:61}) for the elastic modulus and 
Replacing $\lambda_s$ by $\sqrt{1+2\eps_{11}}$ and $\kappa$
by $(\kappa_0+\Delta\kappa_{11})/\sqrt{1+2\eps_{11}}$,
which are expressions that easily follow from (\ref{a:47})--(\ref{a:48}),
we obtain 
\bea\nonumber
\psi \left(\eps_{11},\Delta\kappa_{11} \right) &=& \frac{E}{8\rho_0b_sh_s}\left(A_3 4\eps_{11}^2 +8S_3\eps_{11}\Delta\kappa_{11}+
2I_3 \left(2\Delta\kappa_{11}^2+2\eps_{11}\frac{\Delta\kappa_{11}^2+2\kappa_0\Delta\kappa_{11}-2\kappa_0^2\eps_{11}}{1+2\eps_{11}} \right) + \right.\\
&&\left.+4J_3\Delta\kappa_{11}\frac{\Delta\kappa_{11}^2+2\kappa_0\Delta\kappa_{11}-2\kappa_0^2\eps_{11}}{1+2\eps_{11}}+K_3\frac{\left(\Delta\kappa_{11}^2+2\kappa_0\Delta\kappa_{11}-2\kappa_0^2\eps_{11}\right)^2}{\left(1+2\eps_{11}\right)^2} \right)
\label{eq:psi5}
\eea 
Here, $A_3$ to $K_3$ are modified sectional characteristics defined in (\ref{eq:a3})--(\ref{eq:k3}).
The resulting functional expression for $\psi$ contains not only terms proportional to
the squares of $\eps_{11}$ and $\Delta\kappa_{11}$ but also
many terms with other powers, including mixed terms that
depend on $\eps_{11}$ and $\Delta\kappa_{11}$ simultaneously. 
Therefore, the simple expression (\ref{a:psi}) can be considered only
as a convenient approximation, but it is not consistent
with any hyperelastic law on the fiber level.

If the complete expression (\ref{eq:psi5}) is substituted into (\ref{a:35})--(\ref{a:36}), the resulting sectional equations are
written in terms of internal forces $\tilde{n}^{11}$ and $\tilde{m}^{11}$ and deformation
variables $\eps_{11}$ and $\Delta\kappa_{11}$,
but they are
equivalent with equations (\ref{e:a3})--(\ref{e:a4}) written in terms of internal forces $N$ and $M$ and 
deformation variables $\eps_s$ and $\Delta\kappa$. For this consistent model,
the effective membrane stress
resultant $\tilde{n}^{11}$ depends not only 
on $\eps_{11}$ but also on $\Delta\kappa_{11}$,
in contrast to what was assumed in Simo and Fox\cite{Simo1989}. Even if the consistently
derived potential (\ref{eq:psi5}) is approximated by a quadratic expression, the mixed terms still persist and no decoupling is achieved. The quadratic approximation 
obtained by neglecting terms of third and higher
order in $\eps_{11}$ and $\Delta\kappa_{11}$
reads
\bea\nonumber
\psi_Q \left(\eps_{11},\Delta\kappa_{11} \right) &=& \frac{E}{8\rho_0b_sh_s}\left(A_3 4\eps_{11}^2 +8S_3\eps_{11}\Delta\kappa_{11}+
2I_3 \left(2\Delta\kappa_{11}^2+2\eps_{11}\left(2\kappa_0\Delta\kappa_{11}-2\kappa_0^2\eps_{11}\right) \right) + \right.\\
&&\left.+4J_3\Delta\kappa_{11}\left(2\kappa_0\Delta\kappa_{11}-2\kappa_0^2\eps_{11}\right)+K_3\left(2\kappa_0\Delta\kappa_{11}-2\kappa_0^2\eps_{11}\right)^2\right)=
\nonumber\\
&=&
\frac{E}{2\rho_0b_sh_s}\left(
\left(A_3-2\kappa_0^2I_3+\kappa_0^4K_3\right) \eps_{11}^2
+2\left(S_3+\kappa_0I_3-\kappa_0^2J_3-\kappa_0^3K_3\right)\eps_{11}\Delta\kappa_{11}
+\left(I_3+2\kappa_0J_3+\kappa_0^2K_3\right) \Delta\kappa_{11}^2
 \right)=\nonumber\\
 &=&
\frac{E}{2\rho_0b_sh_s}\left(
\left(A+4\kappa_0^2I_{\kappa_0}\right) \eps_{11}^2
+4S_{\kappa_0}\eps_{11}\Delta\kappa_{11}
+I_{\kappa_0} \Delta\kappa_{11}^2
 \right)
\label{eq:psiQ}
\eea 
and the corresponding approximated sectional equations (\ref{a:35})--(\ref{a:36}) are
\bea\label{a:35x} 
\tilde{n}^{11} &\approx& \frac{h_s\rho_0}{\lambda_s}\frac{\partial\psi_Q}{\partial\eps_{11}} \approx 
\frac{E}{b_s}\left(
\left(A+4\kappa_0^2I_{\kappa_0}\right) \eps_{11}
+2S_{\kappa_0}\Delta\kappa_{11}
 \right)
\\
\tilde{m}^{11} &\approx& \frac{h_s\rho_0}{\lambda_s}\frac{\partial\psi_Q}{\partial\Delta\kappa_{11}} \approx 
\frac{E}{b_s}\left(
2S_{\kappa_0}\eps_{11}
+I_{\kappa_0} \Delta\kappa_{11}
 \right)
\label{a:36x}
\eea 
Here we have replaced $1/\lambda_s=1/(1+\eps_s)$ by 1,
because $\eps_s$ is of the same order as $\eps_{11}$, which is tacitly assumed to be small 
compared to 1 (note that $\partial\psi_Q/\partial\eps_{11}$ is in fact a linear approximation of $\partial\psi/\partial\eps_{11}$, which is applicable if the deformation variables are small). In the same spirit, 
we can use the linearized approximations
\bea 
\eps_{11} &=& \half\left(\lambda_s^2-1\right) =
\half\left(2\eps_s+\eps_s^2\right)\approx \eps_s
\\
\Delta\kappa_{11} &=& \lambda_s\kappa-\kappa_0 =
(1+\eps_s)(\kappa_0+\Delta\kappa)-\kappa_0\approx
\Delta\kappa+\kappa_0\eps_s
\eea 
and evaluate the transformation from the specific to the total internal forces:
\bea 
M &=& b_s\lambda_s^2 \tilde{m}^{11} \approx b_s \tilde{m}^{11} \approx 
2ES_{\kappa_0}\eps_{11}
+EI_{\kappa_0} \Delta\kappa_{11} \approx 
2ES_{\kappa_0}\eps_s
+EI_{\kappa_0}\left(\Delta\kappa+\kappa_0\eps_s\right) = ES_{\kappa_0}\eps_s
+EI_{\kappa_0}\Delta\kappa
\label{a:71}
\\
N &=& b_s\lambda_s^2 n^{11} = b_s\lambda_s^2 \left(\tilde{n}^{11}+\frac{\kappa}{\lambda_s}\tilde{m}^{11}\right) \approx b_s\left(\tilde{n}^{11}+\kappa_0\tilde{m}^{11}\right)\approx 
\left(EA+4\kappa_0^2EI_{\kappa_0}\right) \eps_{11}
+2ES_{\kappa_0}\Delta\kappa_{11}+ 2\kappa_0ES_{\kappa_0}\eps_{11}+\kappa_0EI_{\kappa_0} \Delta\kappa_{11}=\nonumber
\\
&=& 
\left(EA+2\kappa_0^2EI_{\kappa_0}\right) \eps_{11}
+ES_{\kappa_0}\Delta\kappa_{11}
 \approx  \left(EA+2\kappa_0^2EI_{\kappa_0}\right) \eps_s
+ES_{\kappa_0}\left(\Delta\kappa+\kappa_0\eps_s\right)
= \left(EA+\kappa_0^2EI_{\kappa_0}\right)\eps_s+ES_{\kappa_0}\Delta\kappa
\label{a:72}
\eea 
All the foregoing approximations have been based on the assumptions of small deformations ($\eps_s\ll 1$, $\Delta\kappa\ll 1/h_s$) but nothing special has been assumed regarding the initial curvature, $\kappa_0$ (see Section~\ref{sec:over} for extended analysis).
The resulting equations (\ref{a:71})--(\ref{a:72}) are identical with
equations (\ref{e:a3x})--(\ref{e:a4x}) obtained by linearization of 
equations (\ref{eq:n})--(\ref{eq:m}).

In summary, we have shown that the effective membrane stress resultant $\tilde{n}^{11}$
in the form introduced by Simo and Fox does
not really help to decouple the membrane and bending effects for an initially curved beam
(considered here as a special case of a shell). The simple form of quadratic potential (\ref{a:psi}) is not consistent with any
model systematically derived from  a hyperelastic uniaxial material law governing the response of each fiber, not even as
an approximation valid in the small-strain range. If the quadratic approximation of
the consistently derived potential is 
properly derived, the resulting formula (\ref{eq:psiQ})
contains a mixed term, which is then responsible for the coupling effect 
demonstrated in sectional equations (\ref{a:35})--(\ref{a:36}).
This effect is reflected by a mixed sectional
stiffness $2ES_{\kappa_0}$, which is in fact
the double of the mixed stiffness in
sectional equations (\ref{e:a3x})--(\ref{e:a4x}) or (\ref{eq:n})--(\ref{eq:m}), written in terms
of the standard (not effective) stress resultants. 

\subsection{Alternative definitions of effective force}\label{sec:alt}

The effective membrane stress resultant was defined by Simo and Fox such that it corresponds to the quantity work-conjugate with the membrane strain measure.
However, this definition is also affected by
the specific choice of the deformation variable
characterizing flexural effects. For instance,
if we select $\eps_s$ and $\Delta\kappa$ as the
sectional deformation variables, the stress power
per unit initial length of the beam segment
is given by $N\dot{\eps}_s+M\Delta\dot{\kappa}$, and so the force that provides work on increments
of $\eps_s$ at constant $\Delta\kappa$ is simply
the normal force, $N$. The shell model
of Simo and Fox is based on deformation variables
which, for a beam, reduce to $\eps_{11}$ and 
$\kappa_{11}$, defined in (\ref{eq:eps11}) and (\ref{eq:kap11}), and the
stress power is then given by (\ref{a:32}). 
The force that provides work on increments
of $\eps_{11}$ at constant $\kappa_{11}$ is
in fact not directly $\tilde{n}^{11}$ defined
in (\ref{a:31}) but the product $\tilde{n}^{11}\lambda_s$
(the additional multiplication by $b_s$ is simple scaling by
a constant, which only transforms the specific
internal force, i.e., force per unit width, into the internal
force at the level of the whole section of a given
width).
Clearly, other choices of sectional deformation variables are possible and they may lead to other
work-conjugate internal forces. 

One natural choice would be to select the flexural deformation measure such that it remains
equal to zero if all fibers are stretched uniformly. A uniformly stretched straight beam
would of course remain straight. For a curved beam segment whose centerline is initially
a circular arc of radius $R_0$ that subtends
an angle $\dd\varphi_0=\dx/R_0$, uniform stretching of all
fibers leads to a centerline still located on
a circle of radius $R_0$ but subtending an
angle $\dd\varphi=\lambda_s\dd\varphi_0$.
We have defined the curvature as $\kappa=\varphi'=\dd\varphi/\dx$, which means
that the initial curvature $\kappa_0=\dd\varphi_0/\dx$ is equal to $1/R_0$
but the curvature in a deformed state is in general not equal to the reciprocal value of
the current radius of curvature. This is fine
as long as the meaning of $\kappa$ is properly
interpreted. In the case of uniform stretching
of all fibers, we get $\kappa=\dd\varphi/\dx=\lambda_s\dd\varphi_0/\dx=\lambda_s\kappa_0$. This is why we need to
consider the combination of $\lambda_s$ and
$\kappa=\lambda_s\kappa_0$ when constructing
the state of uniform strain energy density,
which is exploited, e.g., in (\ref{a:eint}). 
It is more natural to define the ``true''
curvature as $k=1/R$ where $R$ is the current
radius of inertia. In general, $R\,\dd\varphi=\lambda_s\dx$, and so $k=1/R=\varphi'/\lambda_s=\kappa/\lambda_s$,
from which $\kappa=k\lambda_s$
and $\dot\kappa=\dot{k}\lambda_s+k\dot{\lambda}_s$. The stress power per unit initial length is then
\beq 
N\dot{\eps}_s+M\dot\kappa = (N+kM)\dot{\eps}_s +\lambda_s M\dot{k}
\eeq 
The initial value of $k$ is $k_0=\kappa_0=1/R_0$.
If we define potential 
\beq 
\psi(\lambda_s,\Delta k) = \frac{1}{b_sh_s\rho_0}
\int_A (1+z\kappa_0){\cal E}_{int}\left(\frac{\lambda_s+z(\kappa_0+\Delta k)\lambda_s}{1+z\kappa_0}
\right)\dA = \frac{1}{b_sh_s\rho_0}
\int_A (1+z\kappa_0){\cal E}_{int}\left(\lambda_s\left(1+\frac{z\Delta k}{1+z\kappa_0}\right)
\right)\dA
\eeq
as function of
$\lambda_s$ and $\Delta k=k-\kappa_0$, the sectional
equations resulting from this choice of deformation measures read
\bea 
N+kM &=& b_sh_s\rho_0\frac{\partial\psi}{\partial\lambda_s} \\
\lambda_sM &=& b_sh_s\rho_0\frac{\partial\psi}{\partial\Delta k}
\eea
The advantage of this formulation is that,
for given $\psi(\eps_s,\Delta k)$, the strain
energy density is easily identified as
\beq 
{\cal E}_{int}(\lambda_s)=\rho_0\psi(\lambda_s,0)
\eeq 
For instance, for the material model with strain energy density given by (\ref{ss27}), we obtain
\beq \label{a:78}
\psi(\lambda_s,\Delta k) = \frac{E}{2\rho_0}(\lambda_s-1)^2 +  \frac{EI_{\kappa_0}}{2\rho_0b_sh_s}\lambda_s^2\Delta k^2
\eeq 
and the corresponding sectional equations read
\bea 
N+kM &=& EA(\lambda_s-1) +  EI_{\kappa_0}\lambda_s\Delta k^2 \\
\lambda_sM &=&   EI_{\kappa_0}\lambda_s^2\Delta k
\eea
The effective normal force, $N+kM$, depends
not only on $\lambda_s$ but also on $\Delta k$,
and a full decoupling effect is not achieved. 
However, using the small-strain assumptions,
we can construct the quadratic approximation
\beq\label{a:81} 
\psi_Q(\eps_s,\Delta k) = \frac{E}{2\rho_0}\eps_s^2 +  \frac{EI_{\kappa_0}}{2\rho_0b_sh_s}\Delta k^2
\eeq 
and the corresponding linearized sectional equations
\bea 
N+kM &=& EA\eps_s \\
M &=&   EI_{\kappa_0}\Delta k
\eea
are indeed decoupled. The main point here is that the
quadratic approximation (\ref{a:81}) does not contain
the mixed term with the product $\eps_s\Delta k$.
For other uniaxial hyperelastic models combined
with the present choice of deformation variables,
the exact potential is different from
(\ref{a:78}) and introduces slight coupling, but the 
quadratic approximation of that potential 
always has the form (\ref{a:81}) and leads to decoupled linearized sectional equations.
In this sense, $N+kM$ is a meaningful choice
of the effective normal force, but it is not the only one.

To complete the picture, let us consider yet
another choice of deformation variables,
namely $\lambda_s$  combined with
\beq\label{a:84} 
\tilde\kappa = \kappa - \kappa_0\eps_s 
\eeq 
or, equivalently, $\eps_s$ combined with
\beq 
\Delta\tilde\kappa = \tilde\kappa-\kappa_0 = \kappa-\kappa_0\lambda_s = \Delta\kappa - \kappa_0\eps_s = \Delta k\, \lambda_s
\eeq 
The stress power per unit initial length can be expressed as
\beq 
N\dot{\eps}_s+M\Delta\dot\kappa = N\dot{\eps}_s+M(\Delta\dot{\tilde{\kappa}}+\kappa_0\dot{\eps}_s)=
(N+\kappa_0M)\dot{\eps}_s +M\Delta\dot{\tilde{\kappa}}
\eeq 
and so the forces work-conjugate with $\eps_s$
and $\Delta\tilde{\kappa}$ are $N+\kappa_0M$ and $M$. This time, for the material model with strain energy density given by (\ref{ss27}), the exact potential
\beq 
\psi(\lambda_s,\Delta k) = \frac{E}{2\rho_0}(\lambda_s-1)^2 +  \frac{EI_{\kappa_0}}{2\rho_0b_sh_s}\Delta\tilde{\kappa}^2
\eeq 
is already quadratic in terms of $\eps_s=\lambda_s-1$ and $\Delta\tilde{\kappa}$,
and moreover does not contain the mixed term. 
The corresponding sectional equations
\bea 
N+\kappa_0M &=& EA\eps_s \\
M &=& EI_{\kappa_0}\Delta\tilde{\kappa}
\eea 
are then linear and decoupled, provided that
$N+\kappa_0M$ is considered as the effective
normal force. 
Linearity and decoupling for arbitrarily large
deformation are special properties that are
achieved only for one particular uniaxial hyperelastic
material law.
For other  laws, the exact sectional equations
would still be nonlinear and coupled, but 
their linearized form, applicable under the
small-strain assumption, would be decoupled.
Note that if the strains are considered as small,
the difference between effective normal forces
defined as $N+kM$ or as $N+\kappa_0M$ is negligible.

\subsection{Overview of formulations and discussion}\label{sec:over}

The formulations discussed above are summarized in Table~\ref{tab:formulations}.
Each of them is characterized by a specific choice of two variables that
characterize the deformation of an infinitesimal segment. The stress power density (per unit length of the undeformed centerline) can be expressed in terms of rates
of these variables and the corresponding conjugate variables (generalized internal forces) can be identified. 

\begin{table}
    \centering
    \begin{tabular}{|c|c|c|c|}
    \hline
    formulation &  deformation variables     &  stress power density   &  conjugate forces \\
    \hline
    1a -- present & $\lambda_s$     &    $N\dot{\lambda}_s+M\dot\kappa$ &  $N$ \\
    &$\kappa$   &  &  $M$ \\
    \hline
    1b -- present &  $\eps_s$     &    $N\dot{\eps}_s+M\Delta\dot\kappa$ &  $N$ \\
    &$\Delta\kappa$   &  &  $M$ \\
     \hline
    2 -- Simo and Fox &  $\eps_{11}$     &    $b_s\lambda_s\left(\tilde{n}^{11}\dot{\eps}_{11}+\tilde{m}^{11}\dot{\kappa}_{11}\right)$ &   $b_s\lambda_s\tilde{n}^{11}\equiv N/\lambda_s-\kappa M/\lambda_s^2$ \\
    &$\kappa_{11}$   &  &   $b_s\lambda_s\tilde{m}^{11}\equiv M/\lambda_s$ \\
    \hline
    3a -- alternative &  $\lambda_s$     &    $(N+kM)\dot{\lambda}_s+\lambda_sM\dot{k}$ &   $N+kM$ \\
    &$k$   &  &   $\lambda_sM$ \\
    \hline
    4b -- alternative &  $\eps_s$     &    $(N+\kappa_0M)\dot{\eps}_s+M\Delta\dot{\tilde{\kappa}}$ &   $N+\kappa_0M$ \\
    &$\Delta\tilde{\kappa}$   &  &   $M$ \\
    \hline
     \end{tabular}
    \caption{Various formulations that differ by the choice of sectional deformation variables and the corresponding conjugate forces}
    \label{tab:formulations}
\end{table}

Formulations 1a and 1b are equivalent versions of the approach used in the present paper.
The difference is only formal -- 1a uses variables $\lambda_s$ and $\kappa$,
which do not vanish in the initial state, while formulation 1b uses $\eps_1=\lambda_s-1$
and $\Delta\kappa=\kappa-\kappa_0$, which are true deformation variables that are
initially equal to zero. The rates are the same for both formulations, and therefore the conjugate forces
are also the same and they correspond to the standard normal force, $N$, and bending moment, $M$. 

Formulation 2 is a reduced version of the approach used by Simo and Fox, adapted
to the case of a planar curved beam. The original model for shells uses certain tensorial components which, for the beam, reduce to the Green-Lagrange strain
at the centerline, $\eps_{11}=(\lambda_s^2-1)/2$, and a curvature measure defined as $\kappa_{11}=\lambda_s\kappa$. The conjugate forces at the beam section level
turn out to be $N/\lambda_s-\kappa M/\lambda_s^2$ and $M/\lambda_s$. 

In Section~\ref{sec:alt}, we have also outlined two alternative formulations,
which are presented in Table~\ref{tab:formulations} under labels 3a and 4b. 
Formulation 3a combines the centerline stretch, $\lambda_s$, with the ``true
curvature'', $k$, which is equal to the reciprocal value of the current radius of curvature. These variables could be easily replaced by deformation variables
$\eps_s$ and $\Delta k=k-\kappa_0$, which would lead to formulation 3b, with the
same conjugate forces as for formulation 3a. Finally, formulation 4b is written
directly in terms of deformation variables that vanish in the initial state,
$\eps_s$ and $\Delta\tilde\kappa$, but it could also be presented as formulation
4a using $\lambda_s$ and $\tilde{\kappa}$. The special definition of 
the ``quasi-curvature'' $\tilde{\kappa}$ given in (\ref{a:84}) leads to conjugate forces
$N+\kappa_0M$ and $M$, which have a certain potential advantage.

Each formulation provides a specific expression for the distribution of stretch
across the section and, when combined with a suitable uniaxial hyperelastic stress-strain law, leads to a potential written in terms of the selected deformation
variables. The sectional equations are then obtained by setting the conjugate forces
equal to the partial derivatives of that potential with respect to the deformation
variables. If the uniaxial law is given, all formulations obtained in this way
are equivalent, just written in terms of different variables. Their mathematical
form is in some particular cases simpler. For instance, for the law based
on a linear relation between the Biot strain and its work-conjugate stress,
formulation 1 gives linear sectional equations (for arbitrarily large deformations),
and formulation 4 gives sectional equations that are not only linear, but also
fully decoupled, still for arbitrarily large deformations. However, this would not
be the case if formulation 4 is combined with another material law, e.g., the
St.~Venant-Kirchhoff law. 

If the deformation is assumed to be small,  the sectional equations can be linearized
around the initial state.
Since the distribution of Biot strain across the section is given by
$\eps=\lambda-1=(\eps_s+z\Delta\kappa)/(1+z\kappa_0)$ where $z$ ranges from 
$-h_s/2$ to $h_s/2$, the small-strain assumption is in terms of the deformation
variables written as $\eps_z\ll 1$ and $\Delta\kappa\ll1/h_s$. For other
types of deformation variables this means that $\eps_{11}\ll 1$, $\Delta\kappa_{11}\ll 1/h_s$, $\Delta k\ll1/h_s$
and $\Delta\tilde\kappa\ll 1/h_s$.
The resulting form of sectional equations becomes independent
of the specific choice of the uniaxial law (provided that the elastic modulus is fixed), but it still depends on the choice of deformation variables. Consistent linearization
leads to coupled linear equations for formulations 1 and 2, and to two independent
equations for formulations 3 and 4 (in fact, in the small-strain limit,  formulations
3 and 4 become equivalent). For formulation 2, the consistently derived quadratic
approximation of the potential $\psi$ based on the assumptions of small 
deformation variables ($\eps_{11}\ll 1$ and $\Delta\kappa_{11}\ll 1/h_s$) with no
other simplifying assumptions is given by (\ref{eq:psiQ}) and a mixed term with the product
$\eps_{11}\Delta\kappa_{11}$ is present. 
The resulting linear sectional equations (\ref{a:35x})--(\ref{a:36x}) contain a coupling sectional stiffness 
$2ES_{\kappa_0}$, which is in fact ``twice as strong'' as the coupling stiffness
in equations (\ref{eq:n})--(\ref{eq:m}) for our formulation 1. This means that transformation
from formulation 1 to formulation 2 does not eliminate the coupling (makes it
even stronger) and the interpretation of the effective membrane stress resultant
$\tilde{n}^{11}$ as a quantity related exclusively to the membrane strain and independent of the curvature is in general not justified.

All previous considerations have been done without any restrictions on the initial
curvature, except for the mild assumption that $1\pm h_s\kappa_0/2$ remains positive
and not ``too small'' (in geometric terms, the initial center of curvature of the centerline is outside the beam). It is instructive to rewrite the quadratic potential (\ref{eq:psiQ}) as
\beq
\psi_Q \left(\eps_{11},\Delta\kappa_{11} \right) =
\frac{EA}{2\rho_0b_sh_s}\left(
\left(1+4\kappa_0^2i_{\kappa_0}^2\right) \eps_{11}^2
-4\frac{\kappa_0i_{\kappa_0}^2}{h_s}\eps_{11}(h_s\Delta\kappa_{11})
+\frac{i_{\kappa_0}^2}{h_s^2} (h_s\Delta\kappa_{11})^2
 \right)
\label{eq:psiQ1}
\eeq
where $i_{\kappa_0}=\sqrt{I_{\kappa_0}/A}$ is the modified radius of inertia
of the section, which is comparable to the sectional depth, $h_s$. We have
exploited here the general relation $S_{\kappa_0}=-\kappa_0I_{\kappa_0}$ and
transformed the curvature change, $\Delta\kappa_{11}$, to a small dimensionless
variable $h_s\Delta\kappa_{11}$.  One could now make an additional assumption that
the initial radius of curvature is much larger than the beam depth, i.e., 
$\kappa_0\ll 1/h_s$. This leads to $\kappa_0i_{\kappa_0}\ll 1$ and the factor
that multiplies the mixed term in (\ref{eq:psiQ1}) becomes much smaller
than the factors that multiply the other two terms. The quadratic potential
can then be legitimately approximated by
\beq
\psi_Q \left(\eps_{11},\Delta\kappa_{11} \right) =
\frac{EA}{2\rho_0b_sh_s}\left(
\eps_{11}^2
+\frac{i_{\kappa_0}^2}{h_s^2} (h_s\Delta\kappa_{11})^2
 \right)
\label{eq:psiQ2}
\eeq
So, if the initial curvature is small, one can indeed postulate the
potential in the form (\ref{a:psi}) and obtain decoupled sectional equations. 
However, under the same assumptions, one could use analogous arguments to 
simplify the quadratic potential (\ref{eq:psi2}) that is written in terms of $\eps_s$
and $\Delta\kappa$ and corresponds to formulation 1, and the resulting equations
would be directly the traditional simplified sectional equations (\ref{eq:ns})--(\ref{eq:ms}).
This means that, if the initial curvature is small, there is no need to transform the standard normal force
into an effective one. 
And if the initial curvature is not small, the proper definition of the effective
normal force should be based on formulation 3 or 4.
}

%%%%%%%%%%%%%%%%%%%%%%%%%%%%%%%%%%%%%%%%%%%%%%%%%%%%%%%%%%%%%%%%%%%%%%%%%%%%%%%%%%%%%%%%%
\section{Logarithmic spiral: Description of the initial shape}
\label{appB}
The logarithmic spiral is easily defined in polar coordinates by equation (\ref{eq:logspir}) but the algorithm developed in Section~\ref{sec:3} works with local Cartesian coordinates
aligned with the curved element. The initial
deviation from a straight shape needs to be 
described by functions that depend on the 
coordinate measured as the arc length along the
undeformed centerline.
The purpose
of this appendix is to show the derivation of these functions.

\begin{figure}[h!]
\centering
\begin{tabular}{c}
\includegraphics[width=0.3\linewidth]{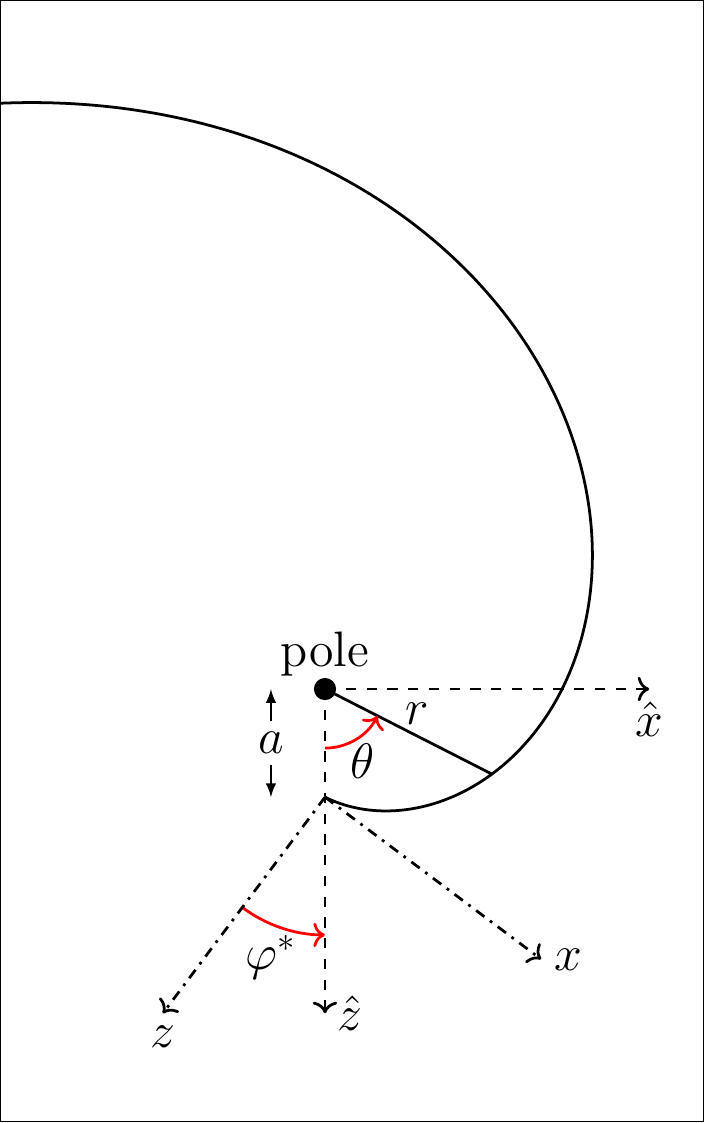}
\end{tabular}
\caption{Geometry of the logarithmic spiral
and position of the auxiliary Cartesian axes
$\hat{x}$ and $\hat{z}$ and of the local
Cartesian axes $x$ and $z$ aligned with the
left end of the element (illustrative plot with $b=0.5$, leading to a faster increase of the radial coordinate than for $b=0.15$ considered in the example).}
\label{fig:LogarithmicSpural_Appendix}
\end{figure}

In the first step, we can easily express 
Cartesian coordinates with respect to an auxiliary
coordinate system with the origin placed at the
pole of the polar coordinates and with axis $\hat{z}$ considered as the axis from which
the polar angle is measured counterclockwise;
see Fig.~\ref{fig:LogarithmicSpural_Appendix}. Since the 
radial coordinate is a given exponential function
of the polar angle, the Cartesian coordinates
can be expressed as unique functions of the polar angle:
\bea \label{eq:a1}
\hat{x} &=& r\sin\theta = a{\rm e}^{b\theta}\sin\theta \\
\label{eq:a2}
\hat{z} &=& r\cos\theta = a{\rm e}^{b\theta}\cos\theta
\eea 

The final objective is to use the arc-length coordinate $s$ as the independent variable,
and so we need to find the link between $\theta$
and $s$. The differential of the arc length is
expressed as
\bea \nonumber
{\rm d}s &=& \sqrt{({\rm d}\hat{x})^2+({\rm d}\hat{z})^2} = \\ \nonumber
&=& \sqrt{(ab{\rm e}^{b\theta}\sin\theta+a{\rm e}^{b\theta}\cos\theta)^2+(ab{\rm e}^{b\theta}\cos\theta-a{\rm e}^{b\theta}\sin\theta)^2}\,{\rm d}\theta = 
\\
&=& a\sqrt{1+b^2}\,{\rm e}^{b\theta}\,{\rm d}\theta
\eea
and integration with initial condition $s=0$ at $\theta=0$ leads to
\beq\label{eq:a4} 
s = \frac{a\sqrt{1+b^2}}{b}\left({\rm e}^{b\theta}-1\right)
= \frac{1}{c}\left({\rm e}^{b\theta}-1\right)
\eeq 
where $c=b/(a\sqrt{1+b^2})$ is an auxiliary parameter introduced for convenience; see also (\ref{eq:c}).
By inversion of (\ref{eq:a4}) one easily gets
\beq\label{eq:a5}  
\theta(s) = \frac{\ln\left(1+cs\right)}{b}
\eeq 
and substitution back into (\ref{eq:a1})--(\ref{eq:a2}) yields
\bea \label{eq:a6}
\hat{x}(s) &=&  a(1+cs)\sin\theta(s) \\
\label{eq:a7}
\hat{z}(s) &=&  a(1+cs)\cos\theta(s)
\eea
It is also useful to express the angle $\hat{\varphi}$ by which the tangent to the
spiral deviates (counterclockwise) from the
$\hat{x}$ axis, because this at the same time determines the deviation of the cross section from
the $\hat{z}$ axis. From
\bea\nonumber 
\cos\hat{\varphi}(s) &=& \frac{{\rm d}\hat{x}(s)}{{\rm d}s} = ac\sin\theta(s) + a(1+cs)\cos\theta(s)\,\frac{c}{b(1+cs)}
=\\
&=&\frac{b}{\sqrt{1+b^2}}\sin\theta(s) + \frac{1}{\sqrt{1+b^2}}\cos\theta(s)
\eea
one can infer that
\beq\label{eq:a9} 
\hat{\varphi}(s) =  \theta(s) - \varphi^*
\eeq 
where $\varphi^*=\arctan b$, as defined in (\ref{eq:phistar}).
Interestingly, each section deviates from the radial
direction by the same angle, $\varphi^*$.

The final step consists in transformation of the
derived expressions from the auxiliary Cartesian
coordinate system aligned with the pole and starting
point of the spiral to the local Cartesian coordinate system used by the algorithm, which
as its origin at the starting point of the spiral,
i.e., at $\hat{x}=0$ and $\hat{z}=a$, and the 
$x$ axis is tangent to the spiral at that point.
This implies that $x$ and $z$ are rotated clockwise
by $\varphi^*$ with respect to $\hat{x}$ and $\hat{z}$, and the transformation equations
can be written as
\bea 
x &=& \hat{x}\cos\varphi^* +(\hat{z}-a)\sin\varphi^*
\\
z &=& -\hat{x}\sin\varphi^* +(\hat{z}-a)\cos\varphi^*
\eea 
Substitution from (\ref{eq:a6})--(\ref{eq:a7}) then gives 
\bea\label{eq:a12}
x(s) &=& a\left((1+cs)\sin(\theta(s)+\varphi^*) - \sin\varphi^*\right)\\
\label{eq:a13}
z(s) &=& a\left((1+cs)\cos(\theta(s)+\varphi^*)-\cos\varphi^*\right)
\eea 

Having derived the description of the initial spiral shape, we can proceed to the interpretation of the
results in the notation used in the main body of this paper.
In the present derivation, $s$ denotes the arc-length coordinate and $x(s)$ and $z(s)$ specify the local Cartesian coordinates of the point 
on the centerline at arc-length distance $s$ from
the left end. However, the theoretical considerations in Section~\ref{sec:2} as well as the numerical techniques described in Section~\ref{sec:3} use $x$
instead of $s$ and interpret this symbol as the
$x$ coordinate of the point in the fictitious straight configuration, which would arise if the centerline were unfolded without changing its length (see Section~\ref{sec:kinematic}). The position
of each centerline point in the initial curved
configuration is then described by the differences
$u_{s0}$ and $w_{s0}$ with respect to the straight
configuration. This means that on the left-hand
sides of (\ref{eq:a12})--(\ref{eq:a13}) we need
to replace $x(s)$ by $x+u_{s0}(x)$ and $z(s)$
by $w_{s0}(x)$, and on the right-hand sides $s$ by $x$. Furthermore, the initial deviation of a generic section from the direction aligned with the left-end section, denoted in Sections~\ref{sec:2}--\ref{sec:3} as $\varphi_0$, is given by $\hat\varphi+\varphi^*$ (because the present $xz$ coordinate system is rotated with respect to the $\hat{x}\hat{z}$ coordinate system by $\varphi^*$ clockwise), and therefore, in view of (\ref{eq:a9}),
$\varphi_0$ turns out to be equal to the polar angle $\theta$,
which is expressed by (\ref{eq:a5}). This leads to equation
(\ref{spiral1}), and we can also replace  the symbol
$\theta(s)$
on the right-hand sides of (\ref{eq:a12})--(\ref{eq:a13}) by $\varphi_0(x)$, which finally yields
equations (\ref{spiral2})--(\ref{spiral3}). 

\end{document}